%% file: paper.tex
\pdfoutput=1

\documentclass{sig-alternate}

\usepackage{times}
\usepackage{amsmath}
\usepackage{amsfonts}
\usepackage{amssymb}
\usepackage{array}
\usepackage{graphicx}
\usepackage{color}
\usepackage{url}
\usepackage[algo2e,ruled,vlined]{algorithm2e}
\usepackage{paralist}   
\usepackage{verbatim}
\usepackage{mathtools}
\usepackage{subfigure}

\input{dfn}

\newcolumntype{P}[1]{>{\centering\arraybackslash}p{#1}}
\makeatletter
\def\@copyrightspace{\relax}
\makeatother

\begin{document}

\title{ M3A: {\underline M}odel, {\underline M}eta{\underline M}odel, and {\underline A}nomaly Detection \\ in Web Searches}

\numberofauthors{6} 
\author{
\alignauthor
Da-Cheng Juan\titlenote{Dr.~Da-Cheng Juan is now with Google Inc.}\\
       \affaddr{Carnegie Mellon University}\\
       \affaddr{Pittsburgh, PA, U.S.A.}\\
       \email{dacheng@alumni.cmu.edu}
\alignauthor 
Neil Shah\\
       \affaddr{Carnegie Mellon University}\\
       \affaddr{Pittsburgh, PA, U.S.A.}\\
       \email{neilshah@cs.cmu.edu}
\alignauthor
Mingyu Tang\\
       \affaddr{Carnegie Mellon University}\\
       \affaddr{Pittsburgh, PA, U.S.A.}\\
       \email{mingyut@cmu.edu}
\and  
\alignauthor
Zhiliang Qian\\
       \affaddr{Hong Kong University of Science and Technology}\\
       \affaddr{Hong Kong, China}\\
       \email{qianzl@connect.ust.hk}
\alignauthor Diana Marculescu\\
       \affaddr{Carnegie Mellon University}\\
       \affaddr{Pittsburgh, PA, U.S.A.}\\
       \email{dianam@cmu.edu}
\alignauthor Christos Faloutsos\\
       \affaddr{Carnegie Mellon University}\\
       \affaddr{Pittsburgh, PA, U.S.A.}\\
       \email{christos@cs.cmu.edu}
}

\maketitle

\begin{abstract}
\input{abstract}
\end{abstract}

\section{Introduction} 
\label{sec:Introduction}
\input{introduction}

\section{Problem Definition} 
\label{sec:Problem Definition}
\input{prob_def}

\section{Single User Analysis: \mll} 
\label{sec:mll}
\input{mll}

\section{Group-level analysis: \cop} 
\label{sec:cop}
\input{cop}


\section{\mmm: Practitioners' Guide}
\label{sec:mmm at work}
\input{practice}


\section{Related Work} 
\label{sec:Related Work}
\input{Related_Work}

\section{Conclusion} 
\label{sec:Conclusion}
\input{Conclusion}



\bibliographystyle{abbrv}
\bibliography{BIB/myref}

\section*{Appendix}
\input{appendix}

\end{document}

%% file: dfn.tex
\newtheorem{observation}{Observation}
\newtheorem{problem}{Problem}

\newtheorem{definition}{Definition}
\newtheorem{lemma}{Lemma}

\newcommand{\hide}[1]{}

\newcommand{\bit}{\begin{compactitem}}
\newcommand{\eit}{\end{compactitem}}
\newcommand{\ben}{\begin{compactenum}}
\newcommand{\een}{\end{compactenum}}

\newcommand{\eg}{\textit{e.g.}}

\newcommand{\tsc}{\textsuperscript}

\newcommand{\mmm}{M3A\xspace}
\newcommand{\mll}{Camel-Log\xspace}
\newcommand{\cop}{Meta-Click\xspace}
\newcommand{\LL}{$\mathcal{LL}$\xspace}
\newcommand{\IN}{in}
\newcommand{\OFF}{off}
\newcommand{\Ratio}{R}
\newcommand{\ratio}{r}
\newcommand{\Mean}{M}
\newcommand{\mean}{m}
\newcommand{\Ken}{\eta}
\newcommand{\Copula}{C}
\newcommand{\pone}{Pattern discovery and interpretation}
\newcommand{\ptwo}{Behavioral modeling}
\newcommand{\pthree}{Anomaly detection}


%% file: abstract.tex
`Alice' is submitting one web search per five minutes, for three hours in a row$-$is it normal? How to detect abnormal search behaviors, among Alice and other users? Is there any distinct pattern in Alice's (or other users') search behavior?
We studied what is probably the largest, publicly available,
query log, containing more than 
{\em 30 million} queries from {\em 0.6 million} users.
In this paper, we present a novel, user-and group-level framework, {\bf \mmm}: {\bf M}odel, {\bf M}eta{\bf M}odel and {\bf A}nomaly detection.
For each user, we discover and explain a surprising, bi-modal pattern of the inter-arrival time (IAT) of landed queries (queries with user click-through).
Specifically, the model \mll is proposed to describe such an IAT distribution; we then notice the correlations among its parameters at the group level. Thus, we further propose the metamodel \cop, to capture and explain the two-dimensional, heavy-tail distribution of the parameters. Combining \mll and \cop, the proposed \mmm has the following strong points: (1) the accurate modeling of marginal IAT distribution, (2) quantitative interpretations, and (3) anomaly detection.

%% file: introduction.tex
\pdfoutput=1

\textit{`` `Alice' is submitting one web search per five minutes, for three hours in a row$-$is it normal?'' ''How to detect abnormal search behaviors, among Alice and other users?'' ``Is there any distinct pattern in Alice's (or other users') search behavior?''} These three questions serve as the motivations of this work.

Conventionally, each of Alice's queries is assumed (1) to be submitted independently and (2) to follow a constant rate $\lambda$, which results in a simple and elegant model, Poisson process (PP). PP generates independent and identically distributed (i.i.d.) inter-arrival time (IAT) that follows an (negative) exponential distribution \cite{fischer1993markov}. In reality, however, does PP accurately model her search behavior?

\begin{figure*}[tb]
\centering
\subfigure[Empirical IAT in lin. scale]{
    \includegraphics[width=0.23\textwidth]{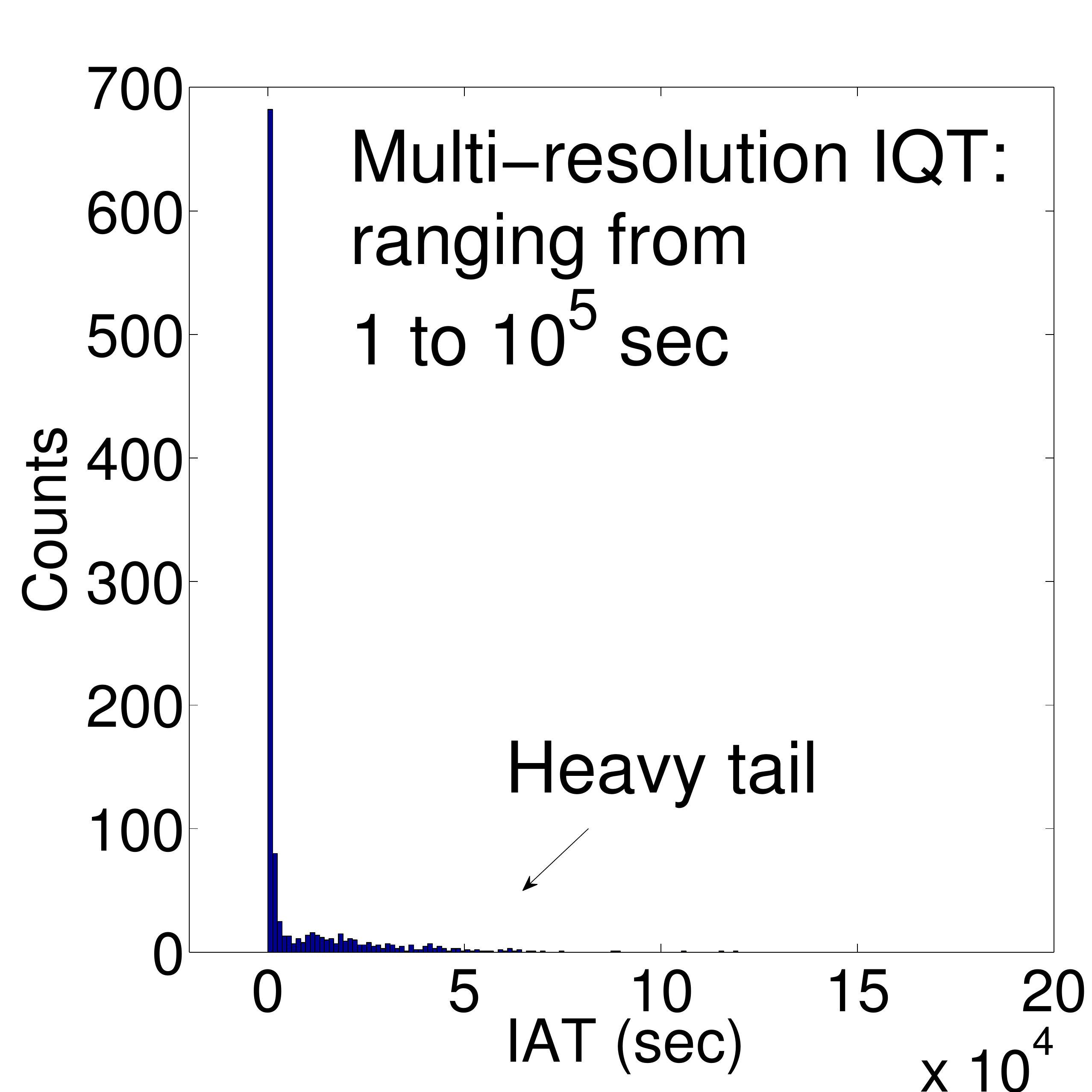}
}
\subfigure[Empirical IAT in log scale and ``\mll'' fit]{
    \includegraphics[width=0.23\textwidth]{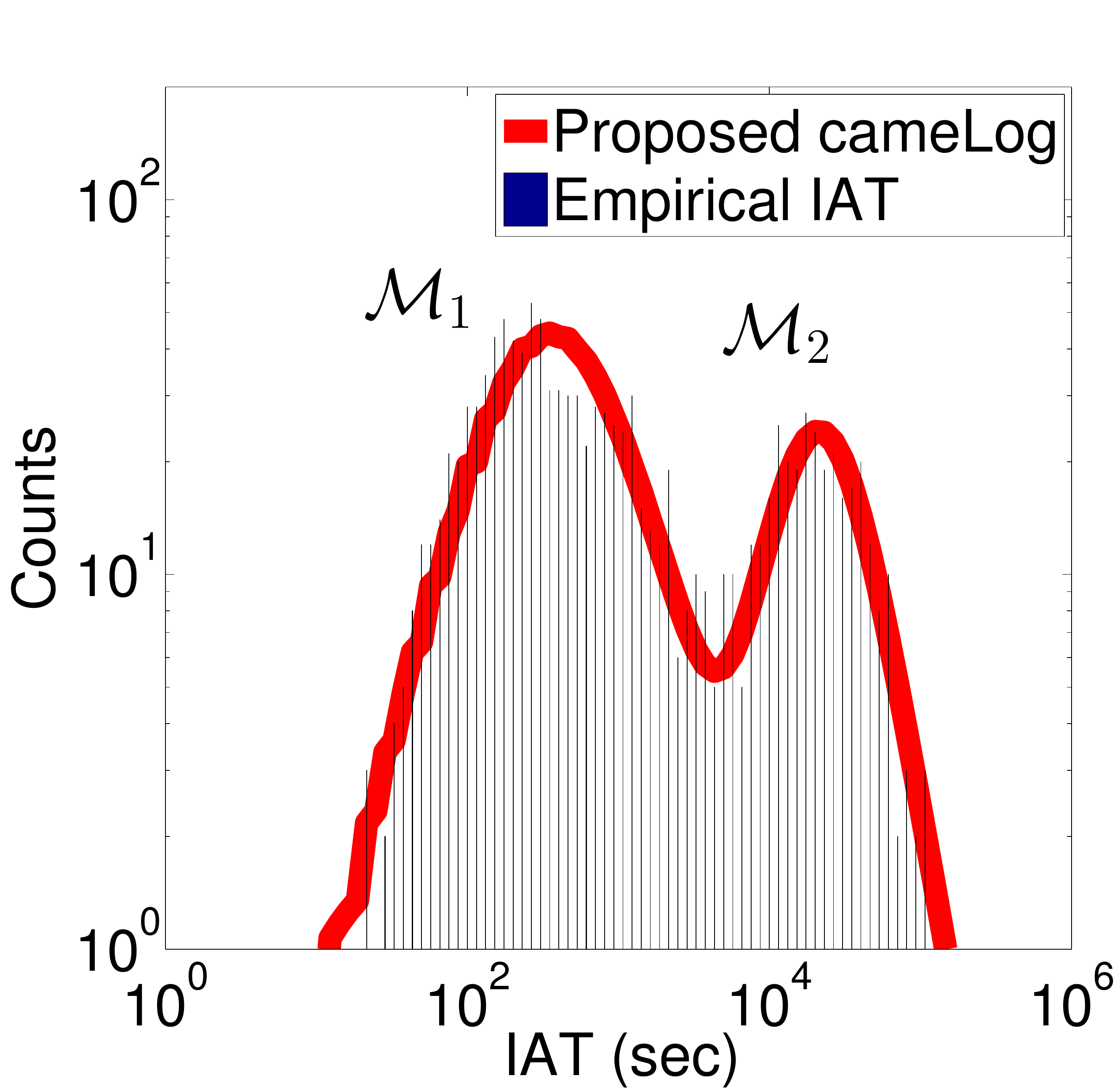}
}
\subfigure[Group-level analysis]{
    \includegraphics[width=0.23\textwidth]{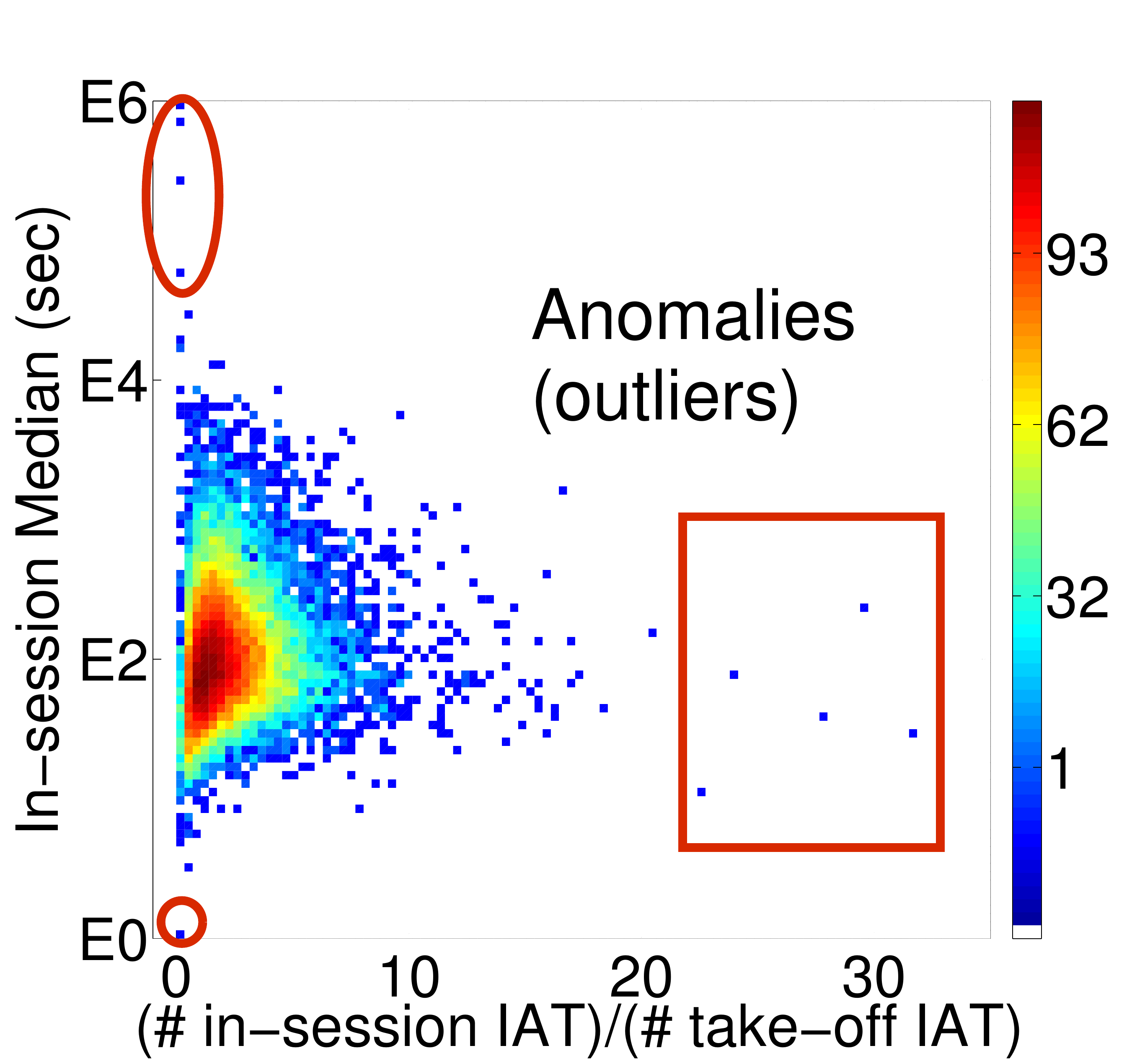}
}
\subfigure[Rank-weirdness plot]{
    \includegraphics[width=0.23\textwidth]{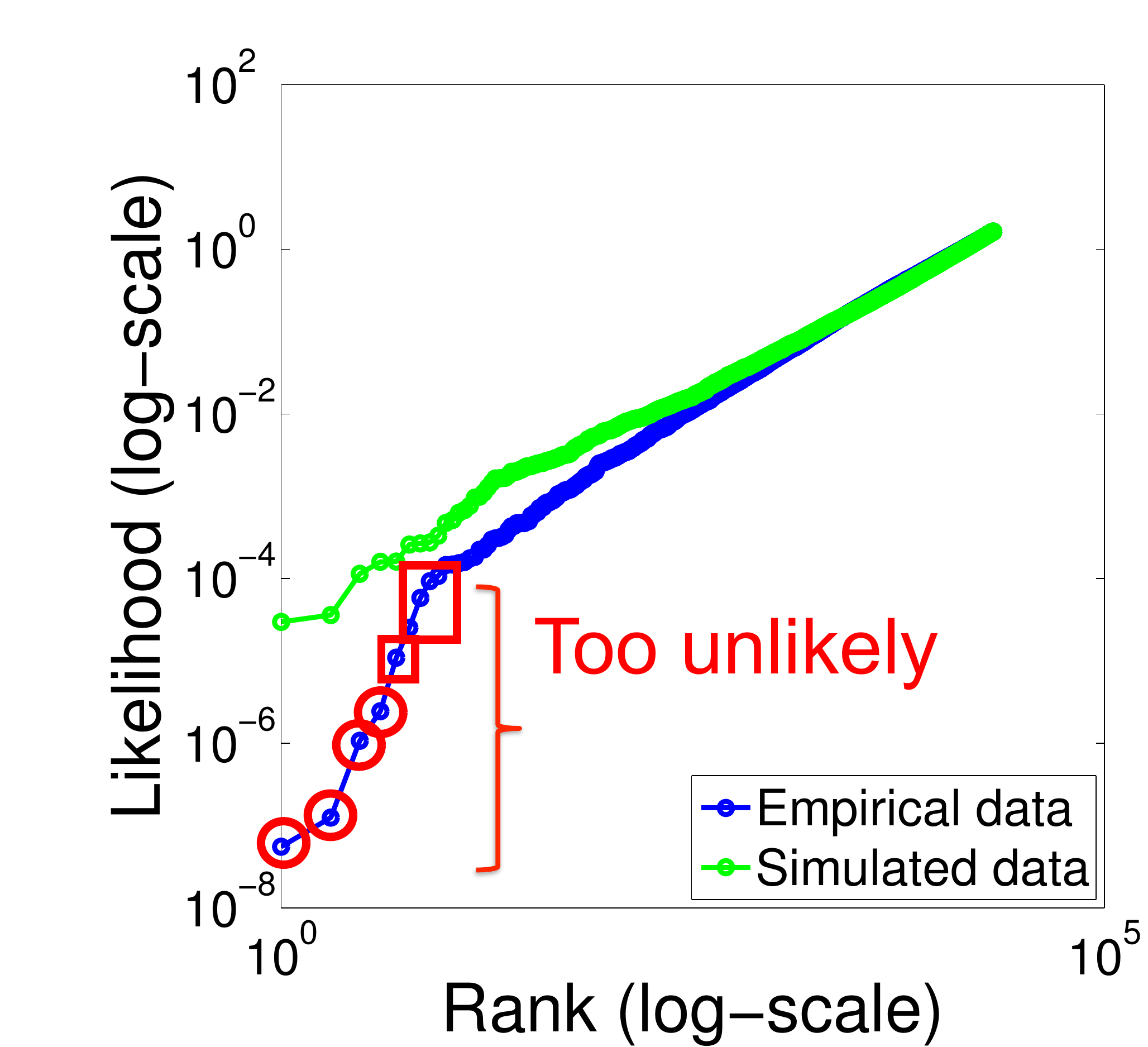}
}
\caption{\textit{Patterns and anomalies with \mmm}: (a) Histogram of inter-arrival time (IAT) for a single user in linear scale. No prevailing patterns are shown. (b) Logarithmic binning (equally-spaced in log-scale) of IAT with \mll fit. A \textit{bi-modal} distribution can be seen: $\mathcal{M}_1$ at $~$5 minutes (typical inter-query time) and $\mathcal{M}_2$ at hours (typical time between sessions). (c) illustrates group-level analysis with scatter plot of the ratio (in-session/take-off queries) vs. the median of in-session intervals. Anomalies are spotted: anomalies (circled by red) cannot be detected by using only the marginal PDF of X-variable, whereas anomalies (marked by the red rectangle) cannot be detected by using the Y-variable. (d) shows an automated way of spotting anomalies through \cop: the blue deviants (within red circles/boxes) correspond to the outliers (in circles/boxes) in (c).} 
\label{fig:typical_behavior}
\end{figure*}

To answer this question, we investigate a large, industrial query log that contains more than 30 million queries submitted by 0.6 million users. Figure~\ref{fig:typical_behavior} illustrates the histogram of a user's IAT. The temporal resolution is one second. As Figure \ref{fig:typical_behavior}(a) shows, this distribution has a ``heavy tail'' as opposed to an (negative) exponential distribution whose tail decays exponentially fast. In the logarithmic scale as Figure~\ref{fig:typical_behavior}(b) shows, surprisingly, \textit{two} distinct modes (denoted as $\mathcal{M}_1$ and $\mathcal{M}_2$) with approximately symmetric shapes can be seen. This distribution (or a mixture of distributions) clearly does not follow an (negative) exponential distribution, which has a strictly right-skewed shape in logarithmic scale and therefore cannot depict such shapes. This phenomenon suggests that the assumptions of PP rarely hold, since the arrival rate may change, or certain queries may be submitted depending on the previous queries. 

In this paper we aim at solving the following problems:
\bit
	\item {\textbf{P1: \pone.} Is there any pattern in the IAT on Alice's behalf?}
	\item {\textbf{P2: \ptwo.} How to characterize the marginal distribution of IAT?}
	\item {\textbf{P3: \pthree.} Given IAT from `Bob,' how to determine whether his behavior is abnormal from Alice and other users?}
\eit

The answers to the above questions are exactly the contributions brought by the proposed \mmm:
\bit
    \item{\textbf{A1: \pone.} One key observation of IAT is provided: a bi-modal ($\mathcal{M}_1$, $\mathcal{M}_2$) distribution with $\mathcal{M}_1$ referred as \textit{in-session} whereas $\mathcal{M}_2$ is referred as \textit{take-off} (\eg, sleep time) query.}
    \item{\textbf{A2: \ptwo.} Specifically, we propose:
	\bit
		\item ``\mll\footnote{The bi-modal distribution of a user's IAT is analogous to a baktrian \textbf{Camel}'s back, in \textbf{Log} scale.}'' to parametrically characterize Alice's (or any person's) IAT by mixing two heavy-tail distributions.
		\item ``\cop'' to describe the joint probability of two parameters of \mll by using a lesser-known tool of \textit{Copula}.
	\eit
	}	
	\item{\textbf{A3: \pthree.} \mll generates IAT with the same statistical properties as in the real data shown in Figure~\ref{fig:typical_behavior}(b), and \cop can detect abnormal users as in Figure~\ref{fig:typical_behavior}(c)(d).
	}
\eit

The remainder of this paper is organized as follows. Section \ref{sec:Problem Definition} provides the problem definition. Section \ref{sec:mll} details the user-level model \mll and Section \ref{sec:cop} details the group-level metamodel \cop. Section \ref{sec:mmm at work} provides the usage of \mmm. Section \ref{sec:Related Work} surveys the previous work. Finally, Section \ref{sec:Conclusion} concludes this paper.

%% file: prob_def.tex
\pdfoutput=1

In this work, we use a large-scale, industrial query log released by AOL \cite{pass2006picture}, which is essentially a Google query log since AOL searches are powered by Google \cite{bar2007position}. The basic statistics of this query log are provided here:
\bit
	\item Duration: three months, from March 1\tsc{st} to May 31\tsc{st}, 2006.
	\item 36 millions queries submitted from 657,000 users: 
		\bit
			\item 19 millions queries WITH click-through\\(referred as \textit{landed queries}).
			\item 17 millions queries WITHOUT click-through\\(referred as \textit{orphan queries}).
		\eit
	\item The temporal resolution is 1 second.
\eit

\subsection{Terminology and problem formulation}
\label{subsec:Terminology}
Table \ref{tbl:symbol_table} provides the symbols and the corresponding definitions used throughout this paper. By the convention in statistics, random variables are represented in upper-case (\eg, $\Mean$) and the corresponding values (\eg, $\mean$) are in lower-case.

\begin{table*}[tb]
	\centering
	\caption{Symbols and definitions}
	\begin{tabular}{|p{0.14\textwidth}||p{0.84\textwidth}|}	
	\hline 
	Symbol& Definition\\
	\hline\hline
	IAT	 & Inter-arrival time \\ \hline	
	$t_{i,j}$ & IAT between $j$\tsc{th} and $(j+1)$\tsc{th} query submitted by user $i$. \\ \hline
	$F_T(\cdot)$ & Cumulative distribution function (CDF) for: (a) the random variable $T$ or (b) the distribution $T$\\
	$f_T(\cdot)$ & Probability density function (PDF) for: (a) the random variable $T$ or (b) the distribution $T$ (\eg, $f_{\mathcal{LL}}$ is the PDF of log-logistic)\\ \hline
	\LL	 & Log-logistic distribution: a skewed (in linear scale), heavy-tail distribution\\ \hline
	\mll & Proposed mixture of two log-logistic distribution: modeling marginal IAT\\ \hline
	\cop & Proposed 2-d log-logistic distribution using Gumbel's copula: metamodeling the parameters of \mll\\	
	\hline\hline	
	\multicolumn{2}{|c|}{Symbols used by \mll} \\ \hline
	$\alpha_{\IN}$, $\beta_{\IN}$ & Parameters: median and shape of log-logistic distribution (for modeling in-session IAT) \\ \hline
	$\alpha_{\OFF}$, $\beta_{\OFF}$ & Parameters: median and shape of log-logistic distribution (for modeling take-off IAT) \\ \hline
	$\theta$ & Proportion parameter: $\theta \in$ [0,1] for in-session IAT, and $(1-\theta)$ for take-off IAT\\ \hline\hline
	\multicolumn{2}{|c|}{Symbols used by \cop} \\ \hline
	$\Ratio$ & Random variable representing the ratio of in-session and take-off IAT: $\Ratio \triangleq \theta/(1-\theta)$ \\ \hline
	$\Mean$ & Random variable representing the log-median of in-session IAT: $\Mean  \triangleq \log(\alpha_{\IN})$\\ \hline
	$\alpha_{\Ratio}$, $\beta_{\Ratio}$ & Hyper-parameters: median and shape of log-logistic distribution (for modeling $\Ratio$) \\ \hline		
	$\alpha_{\Mean}$, $\beta_{\Mean}$ & Hyper-parameters: median and shape of log-logistic distribution (for modeling $\Mean$) \\ \hline	
	$\Copula(\cdot,\cdot)$ & Copula: Joint CDF of two random variables considering their dependency $[0,1]\times[0,1]\rightarrow [0,1]$\\ \hline
	$\Ken$ & Parameter in Gumbel's copula that captures correlations between random variables $\Ratio$ and $\Mean$\\ \hline	
	\end{tabular}	
	\label{tbl:symbol_table}
\end{table*}

As mentioned in Section \ref{sec:Introduction}, we aim at solving the following three problems:
\begin{problem}[\pone]
Given each user ID and the time stamp of each query, find and interpret the most distinct pattern sufficient to characterize the IAT distribution of each user.
\end{problem}
\begin{problem}[\ptwo]
Given the pattern found in P1, design:
\ben
	\item A model (and a metamodel) that matches the statistical properties of the empirical data.
	\item The parameters (and the hyper-parameters).
\een
\end{problem}
\begin{problem}[\pthree]
Given:
\ben
	\item The model (and metamodel) from P2.
	\item The time stamp of each query from a user.
\een
Determine if her/his query behavior in terms of IAT is abnormal.
\end{problem}

\subsection{Observation on non-landed queries:\\ ``orphan queries''}
\label{subsec:Detecting obvious anomalies}

\begin{figure}[tb]
	\centering
	\includegraphics[width=0.8\linewidth]{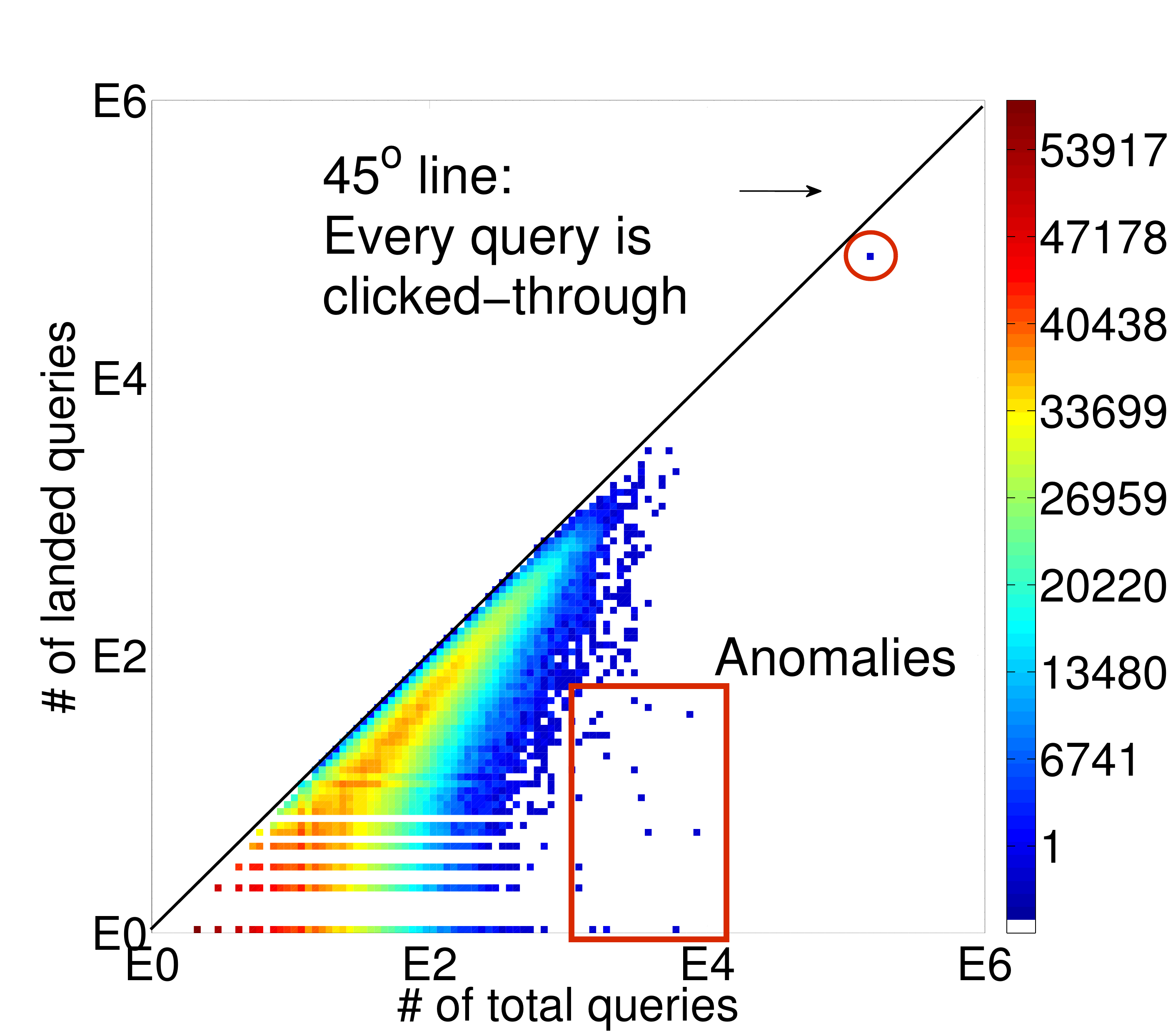}
    \caption{\textit{Orphan queries}. Queries without following through are suspicious (see the red box). One user (circled by red) has submitted $\approx$ 130,000 queries, with the longest IAT of only 20 minutes (no sleep time).}
    \label{fig:total_vs_landed}
\end{figure}

In Figure~\ref{fig:total_vs_landed}, notice that certain users (marked by the red rectangle) have submitted more than 1,000 queries but clicked through very few (less than 100, or even zero!) of them, resulting in abnormally-many of orphan queries. Another obvious evidence is: these orphan queries usually submitted (a) consecutively and (b) with the same keyword, leading to a clear robotic behavior. Therefore, we provide the following qualitative observation.

\begin{observation}[Orphan queries]
Users who have submitted many (usually more than 1,000) queries but clicked through very few (less than 100) of them are abnormal.
\end{observation}

Furthermore, one user (circled by red) in the upper-right corner of Figure~\ref{fig:total_vs_landed} has submitted more queries (by two order of magnitudes, $\approx$ 130,000) than typical users ($\approx$ hundreds to thousands), with the longest IAT of only 20 minutes (no sleep time). Clearly, this user is suspicious and therefore an anomaly.

After being able to detect obvious anomalies with orphan queries, we again ask the major motivating question (as mentioned in Section~\ref{sec:Introduction}): \textit{``How frequently does `Alice' submit a web query and click through the search results?''} Starting immediately, we ignore orphan queries and focus on the IAT of landed queries.

%% file: mll.tex
\pdfoutput=1

In this section, we first detail the proposed \mll distribution (Section~\ref{subsec: mll distribution}), provide validations (Section~\ref{subsec: mll validation}) and give comparisons with other well-known models (Section~\ref{subsec: mll comparison}). For convenience, we preview the mathematical form of \mll here:
\begin{eqnarray}
f_{\mll}(t) &=& \theta\cdot f_\mathcal{LL}(t;\alpha_{\IN},\beta_{\IN})+ \nonumber\\ 
			& & (1-\theta)\cdot f_\mathcal{LL}(t;\alpha_{\OFF},\beta_{\OFF}) \nonumber
\end{eqnarray}
\noindent where $t\ge 0$, $f_\mathcal{LL}(\cdot)$ stands for the probability density function (PDF) of log-logistic (\LL) distribution as shown in Eq\eqref{eq:PDF_LL}.

\subsection{\mll distribution}
\label{subsec: mll distribution}

\begin{figure}[tb]
\centering
\subfigure{\includegraphics[width=0.32\columnwidth]{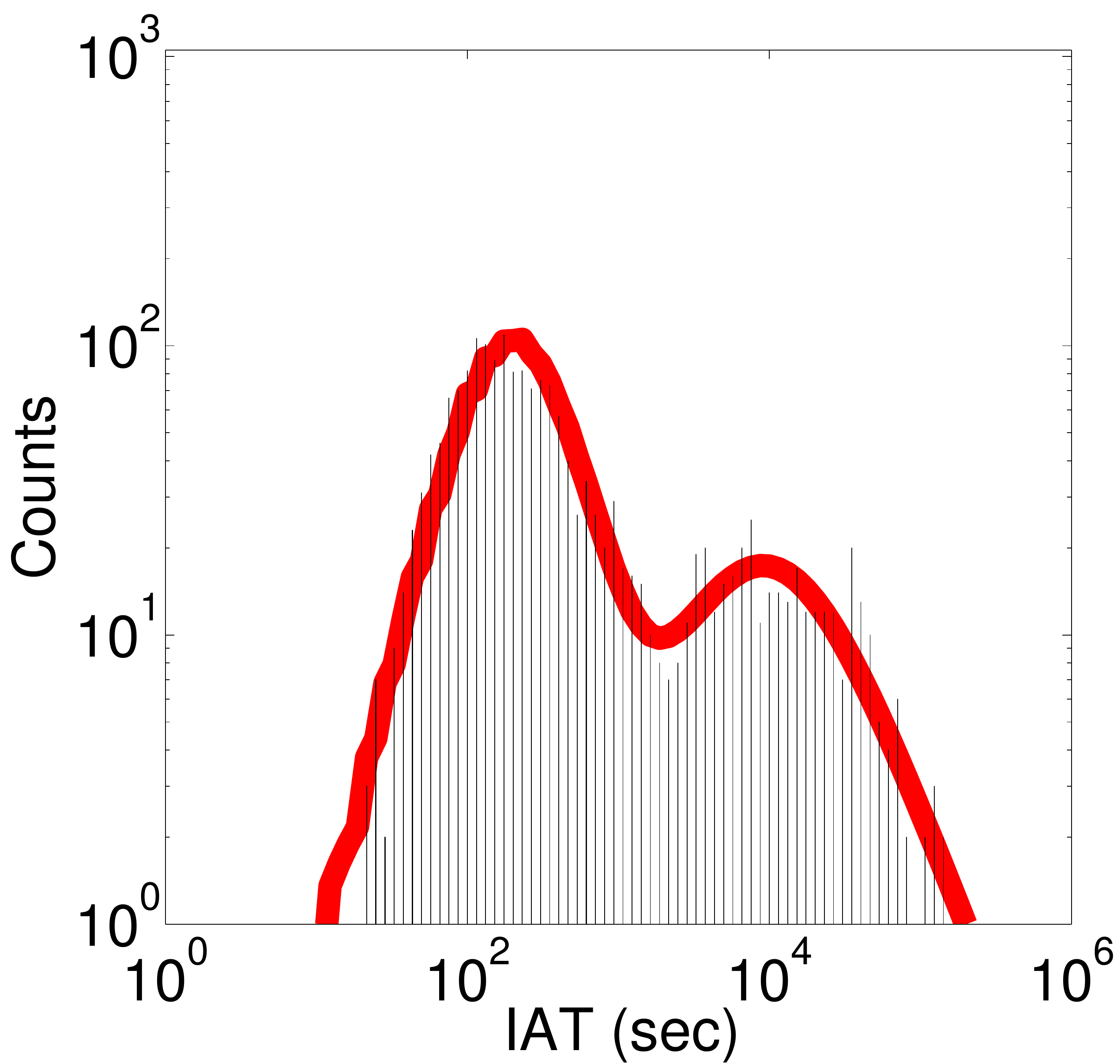}}
\subfigure{\includegraphics[width=0.32\columnwidth]{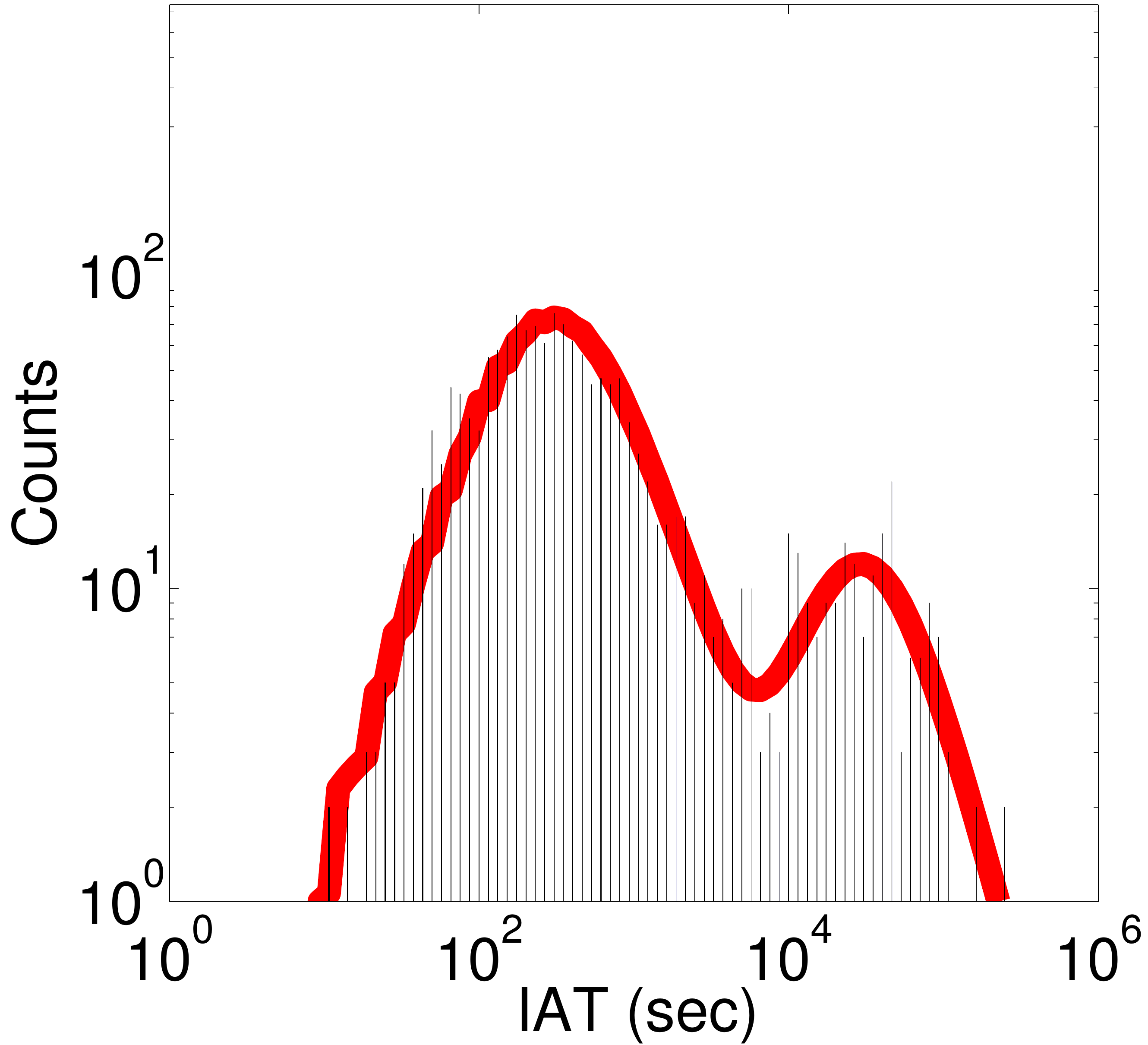}}
\subfigure{\includegraphics[width=0.32\columnwidth]{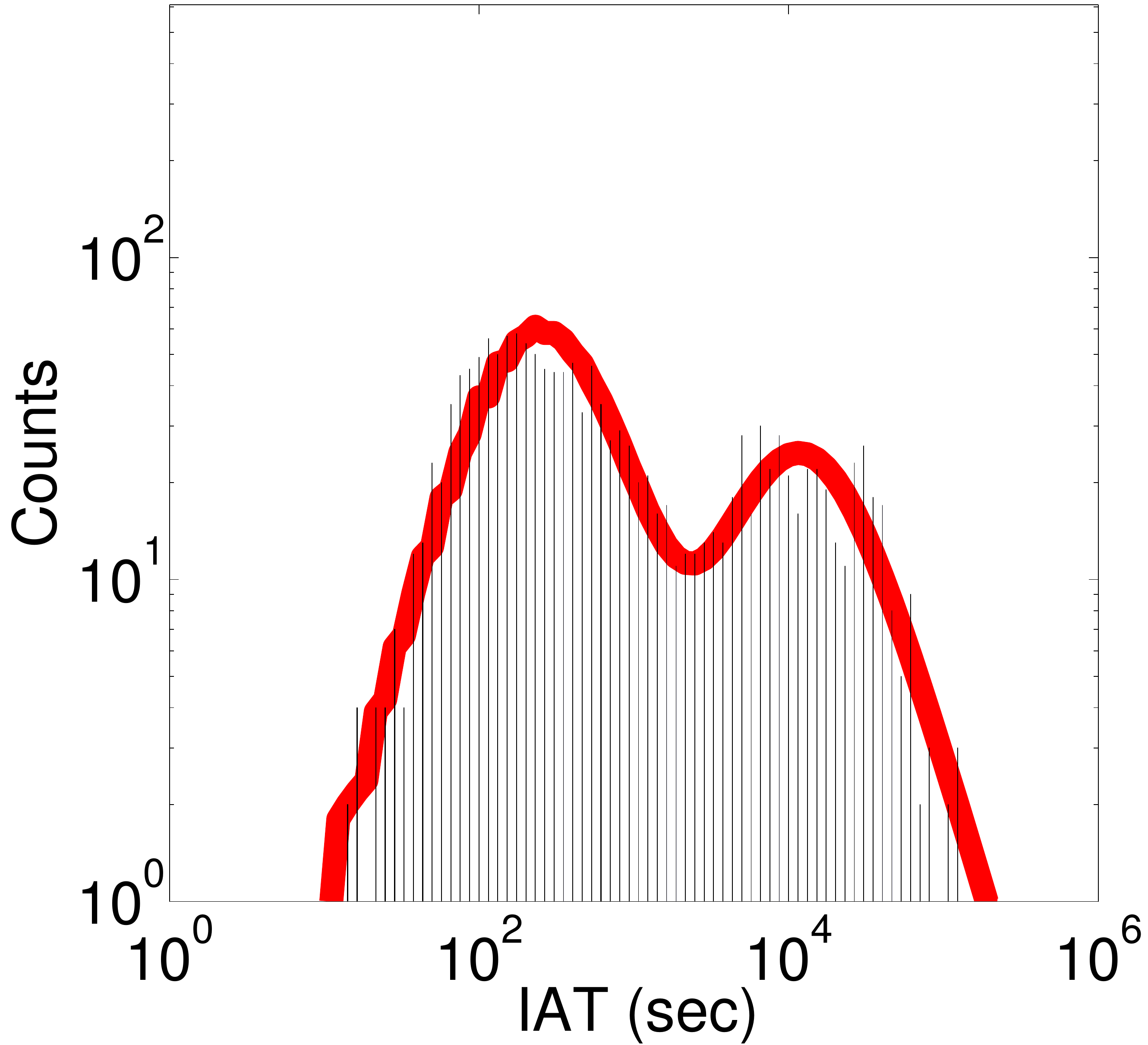}}
\subfigure{\includegraphics[width=0.32\columnwidth]{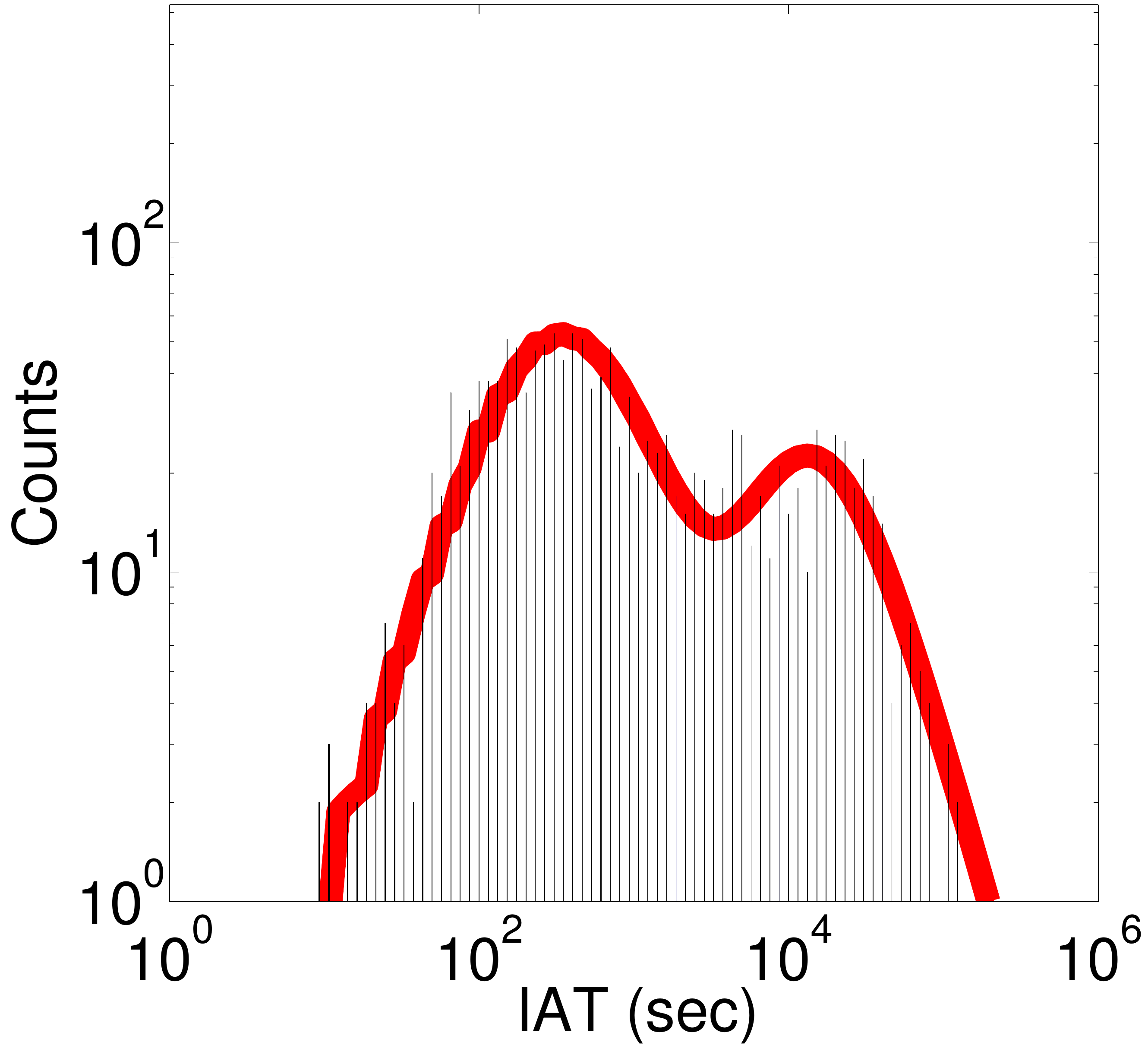}}
\subfigure{\includegraphics[width=0.32\columnwidth]{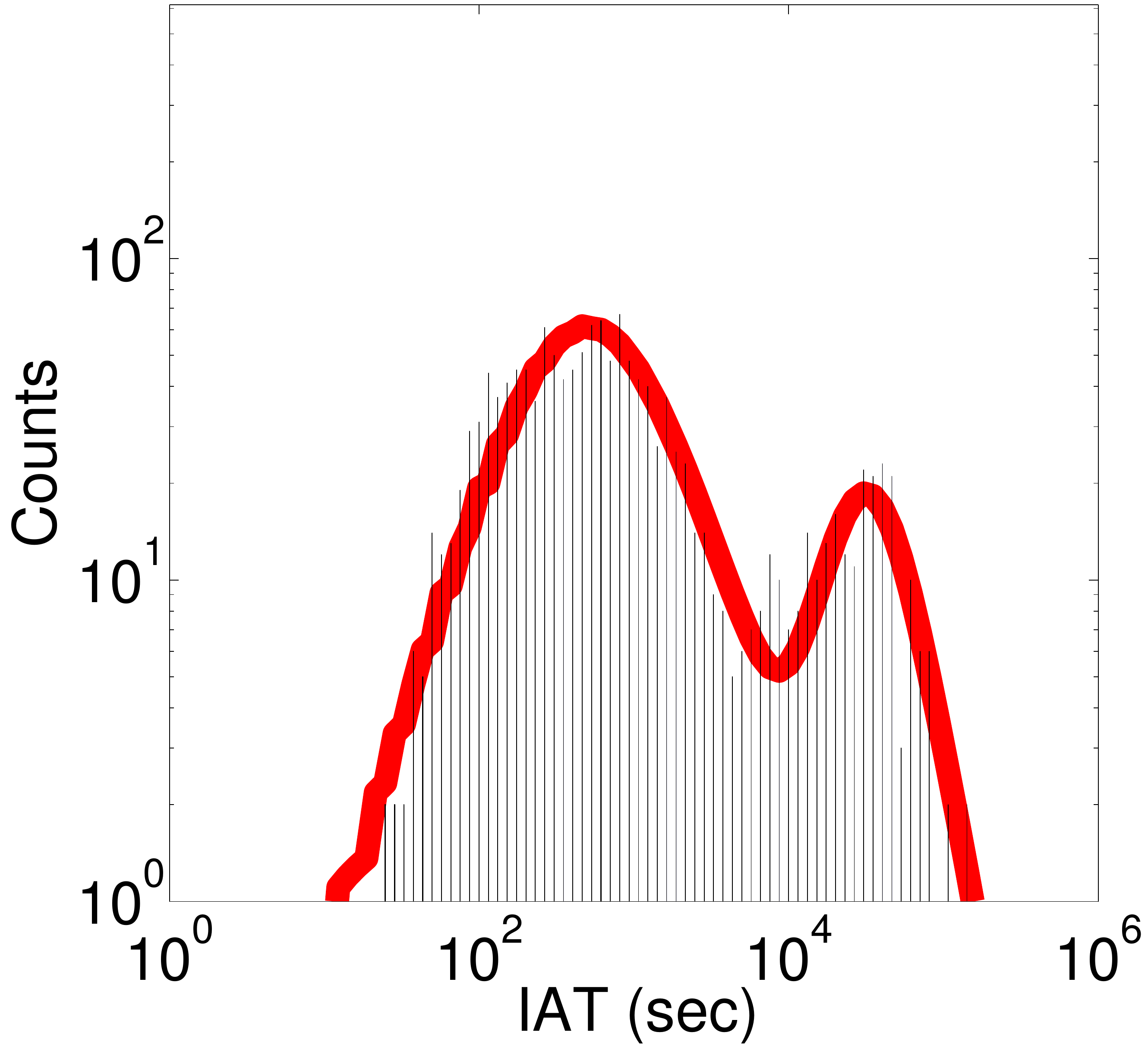}}
\subfigure{\includegraphics[width=0.32\columnwidth]{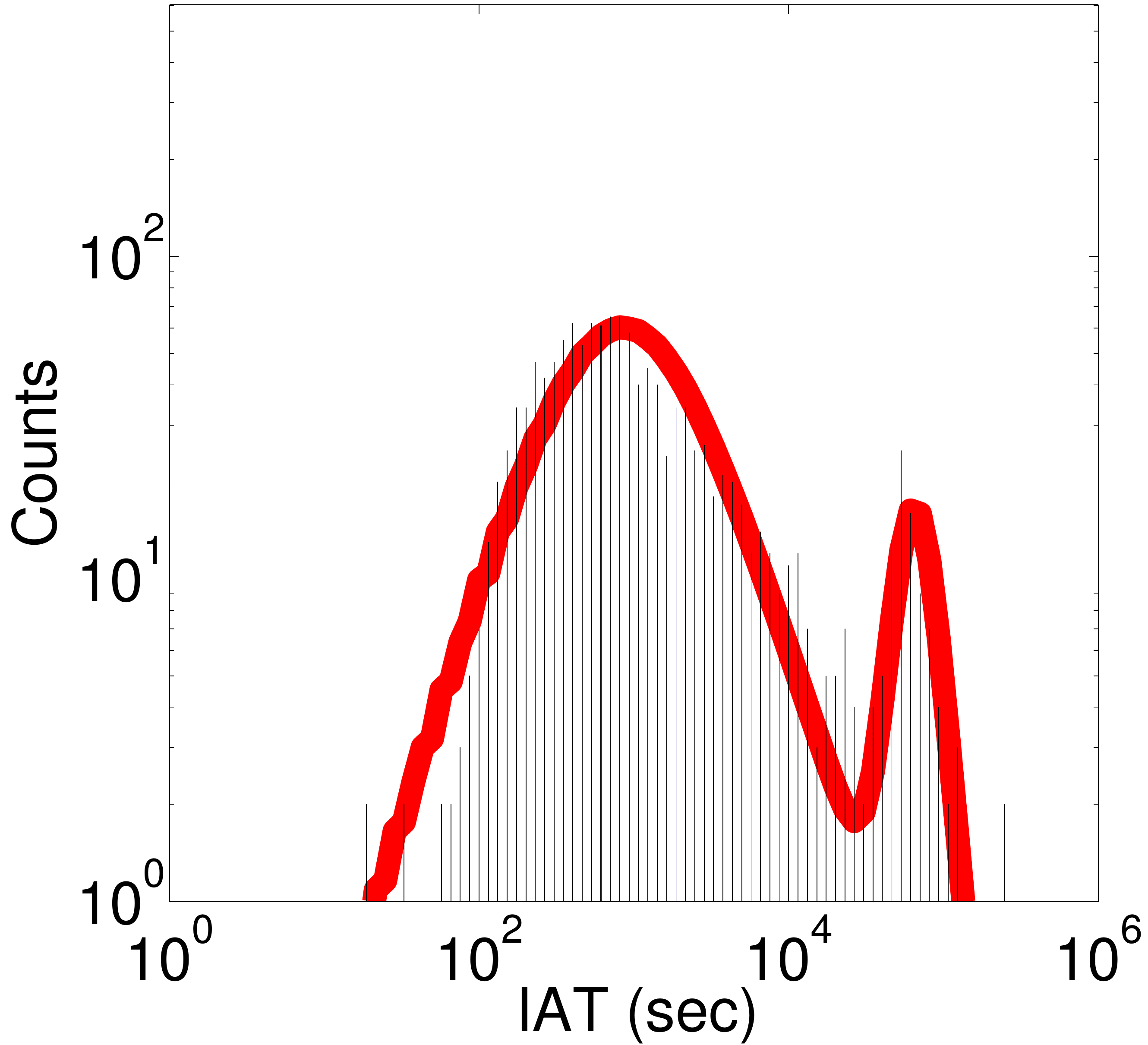}}
\subfigure{\includegraphics[width=0.32\columnwidth]{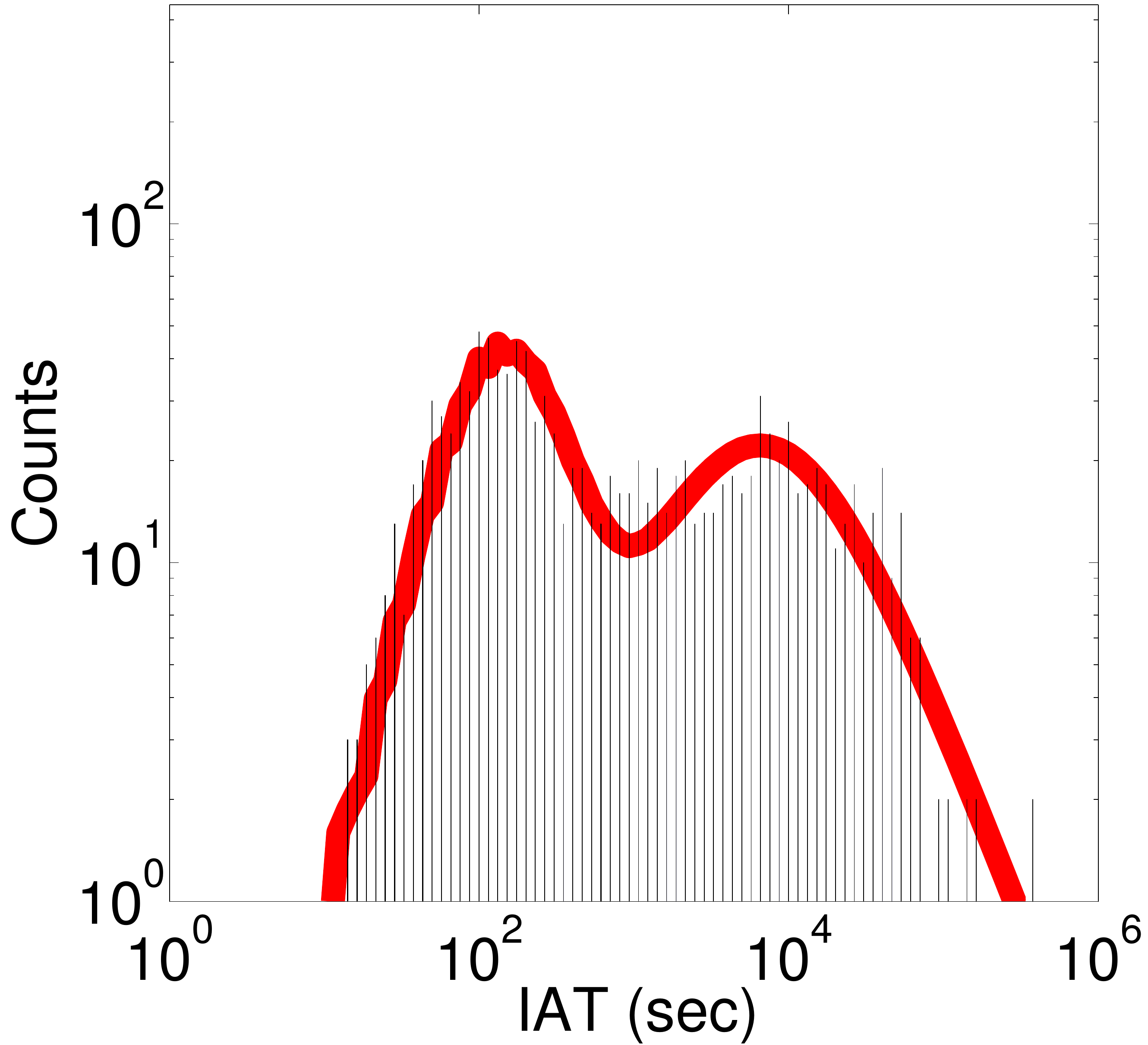}}
\subfigure{\includegraphics[width=0.32\columnwidth]{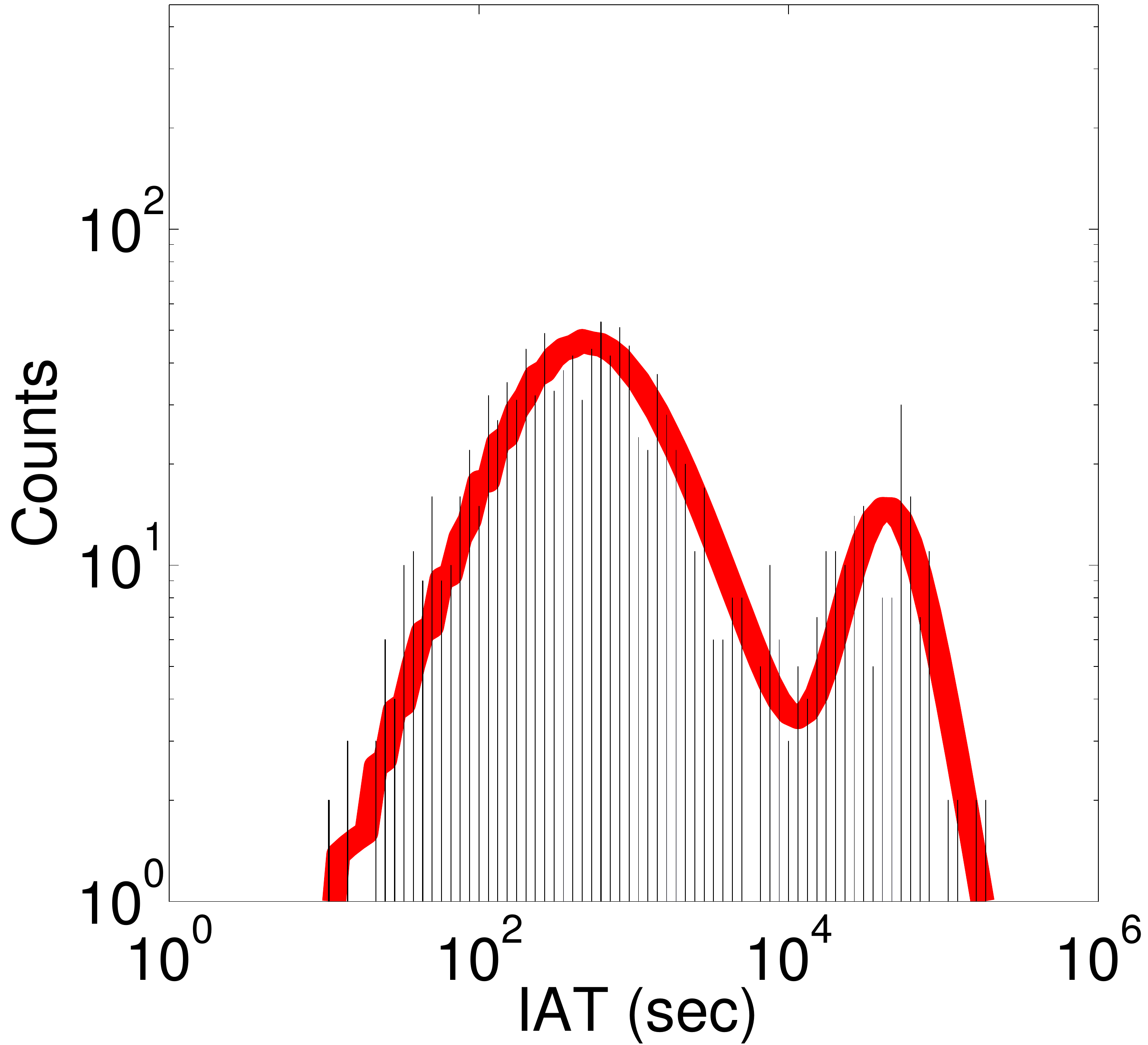}}
\subfigure{\includegraphics[width=0.32\columnwidth]{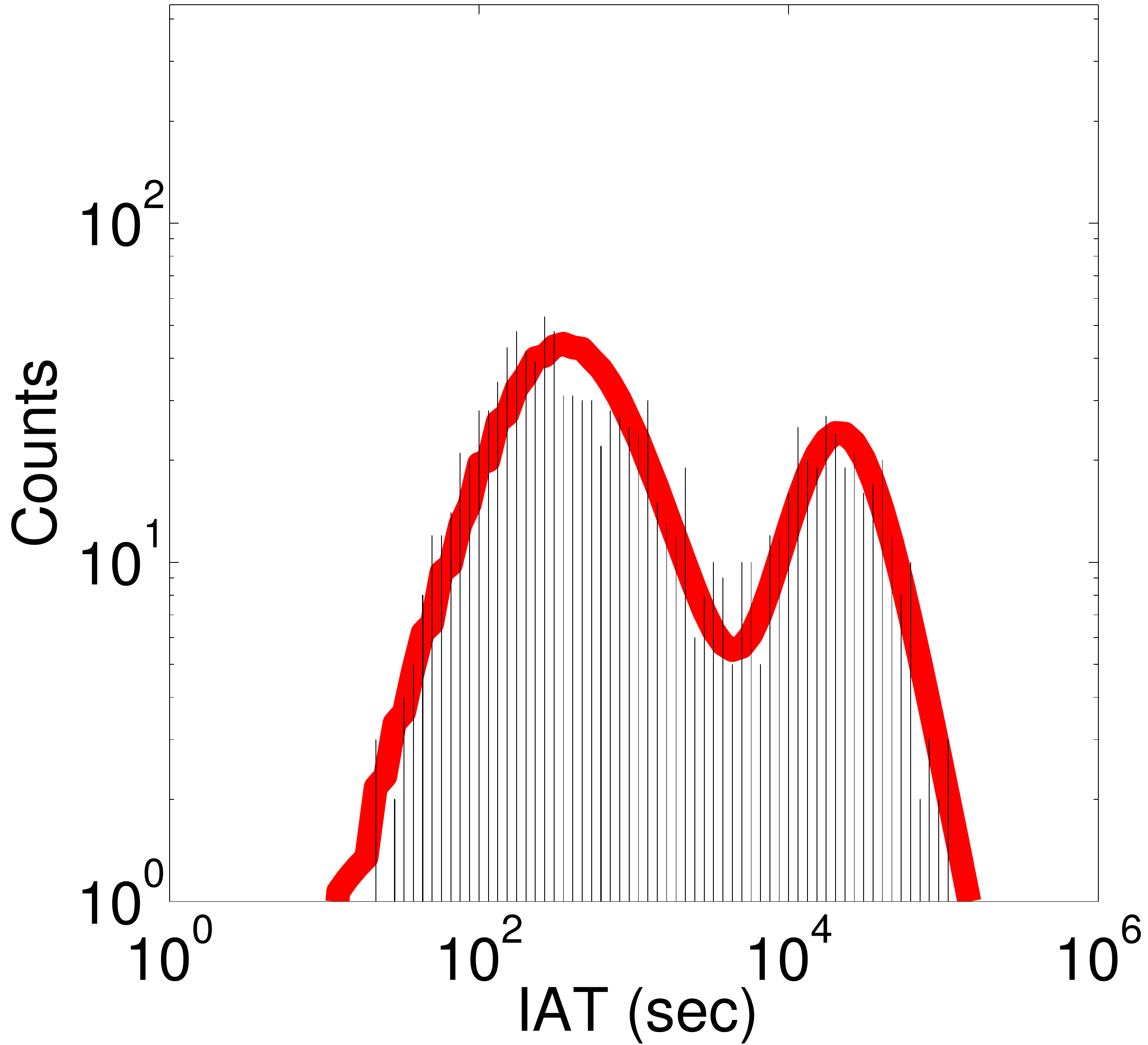}}
\subfigure{\includegraphics[width=0.32\columnwidth]{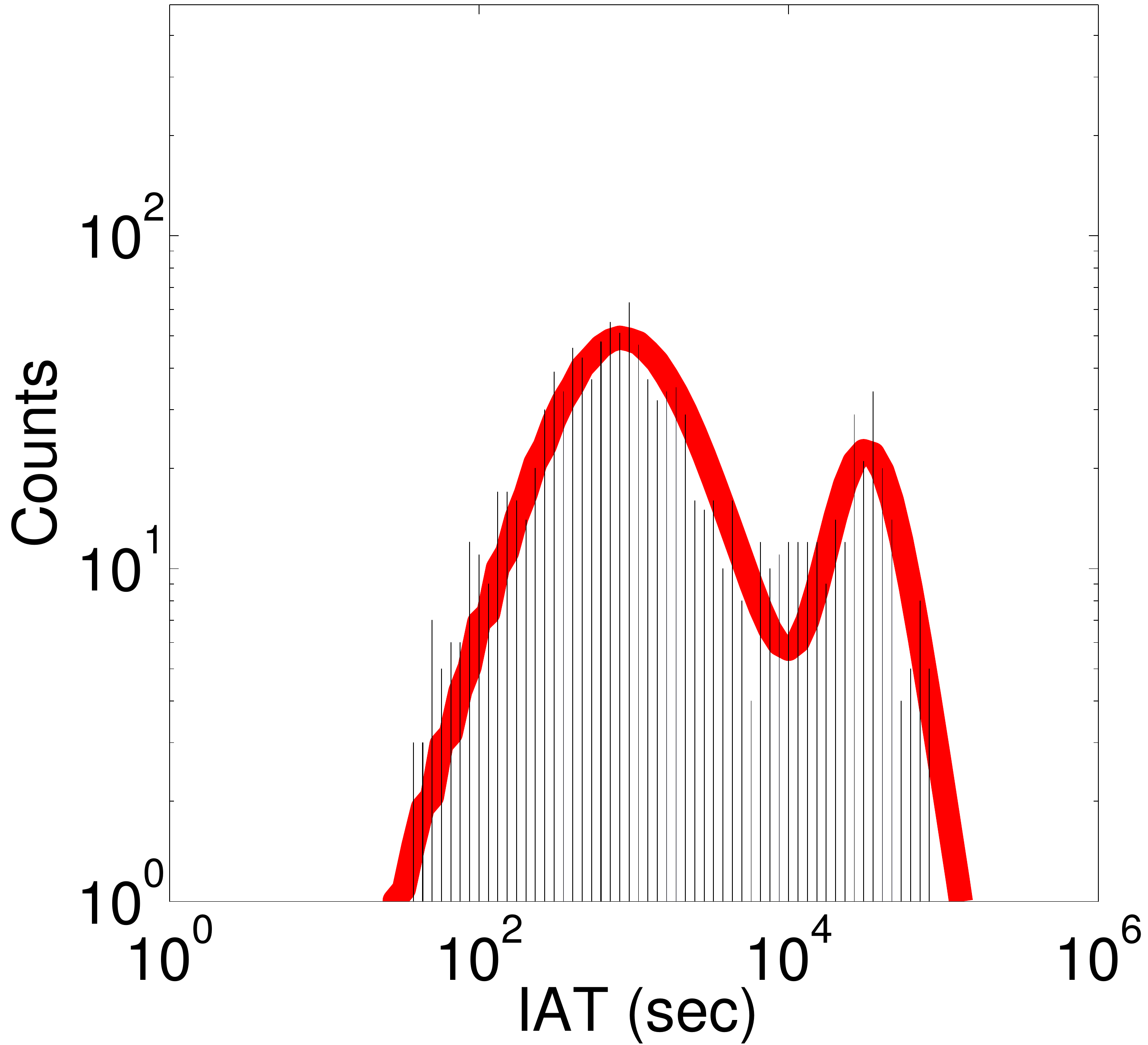}}
\subfigure{\includegraphics[width=0.32\columnwidth]{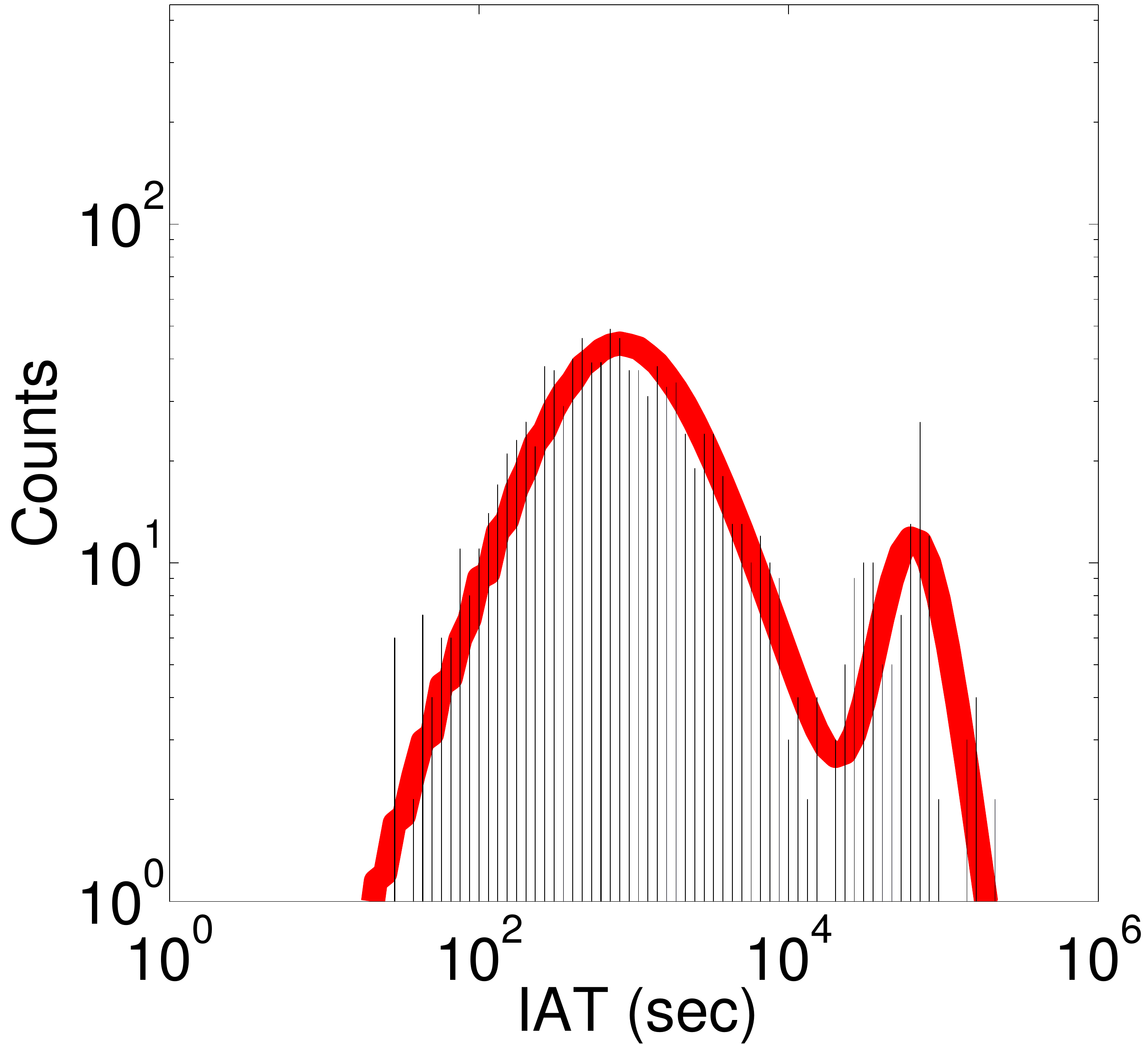}}
\subfigure{\includegraphics[width=0.32\columnwidth]{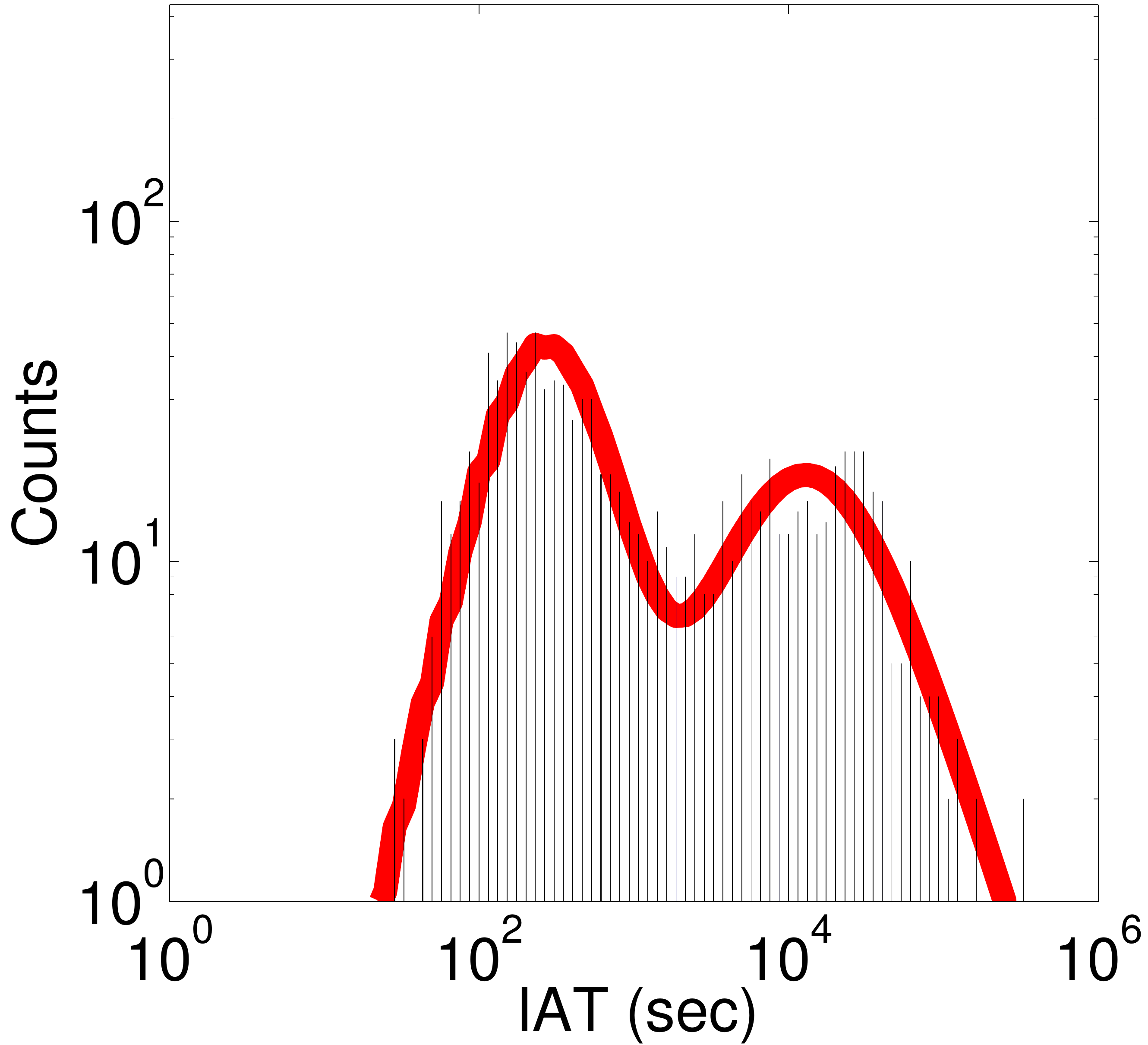}}
\caption{\textit{Consistency of the bi-modal behaviors}: in-session and take-off. Each sub-figure shows the marginal distribution of IATs (in logarithmic binning) from a user. The red curve is depicted by fitting a \mll distribution via expectation maximization (EM).}
\label{fig:Fitted_PDF}
\end{figure}

\begin{figure}[tb]
\centering
\subfigure{\includegraphics[width=0.30\columnwidth]{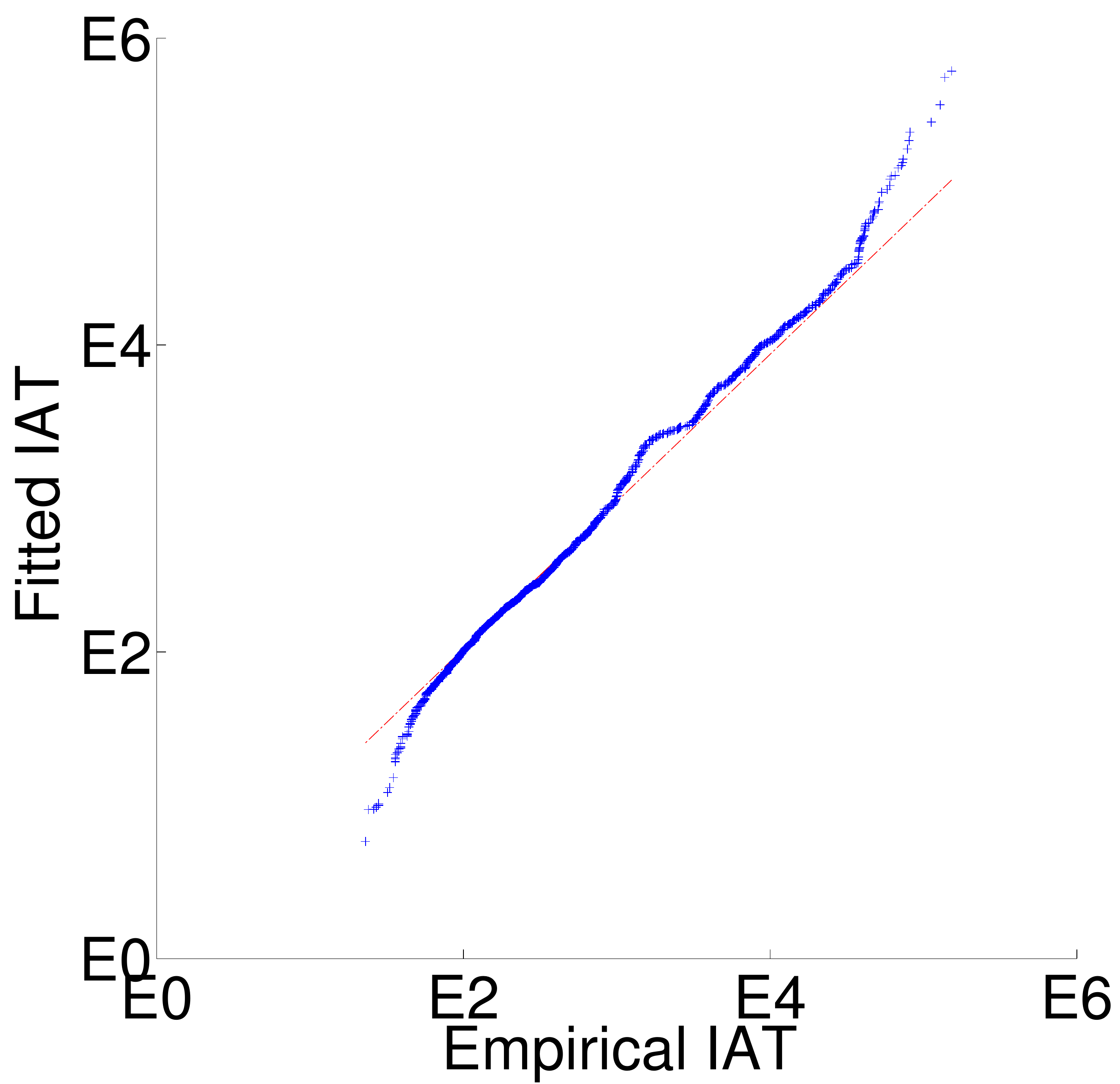}}
\subfigure{\includegraphics[width=0.30\columnwidth]{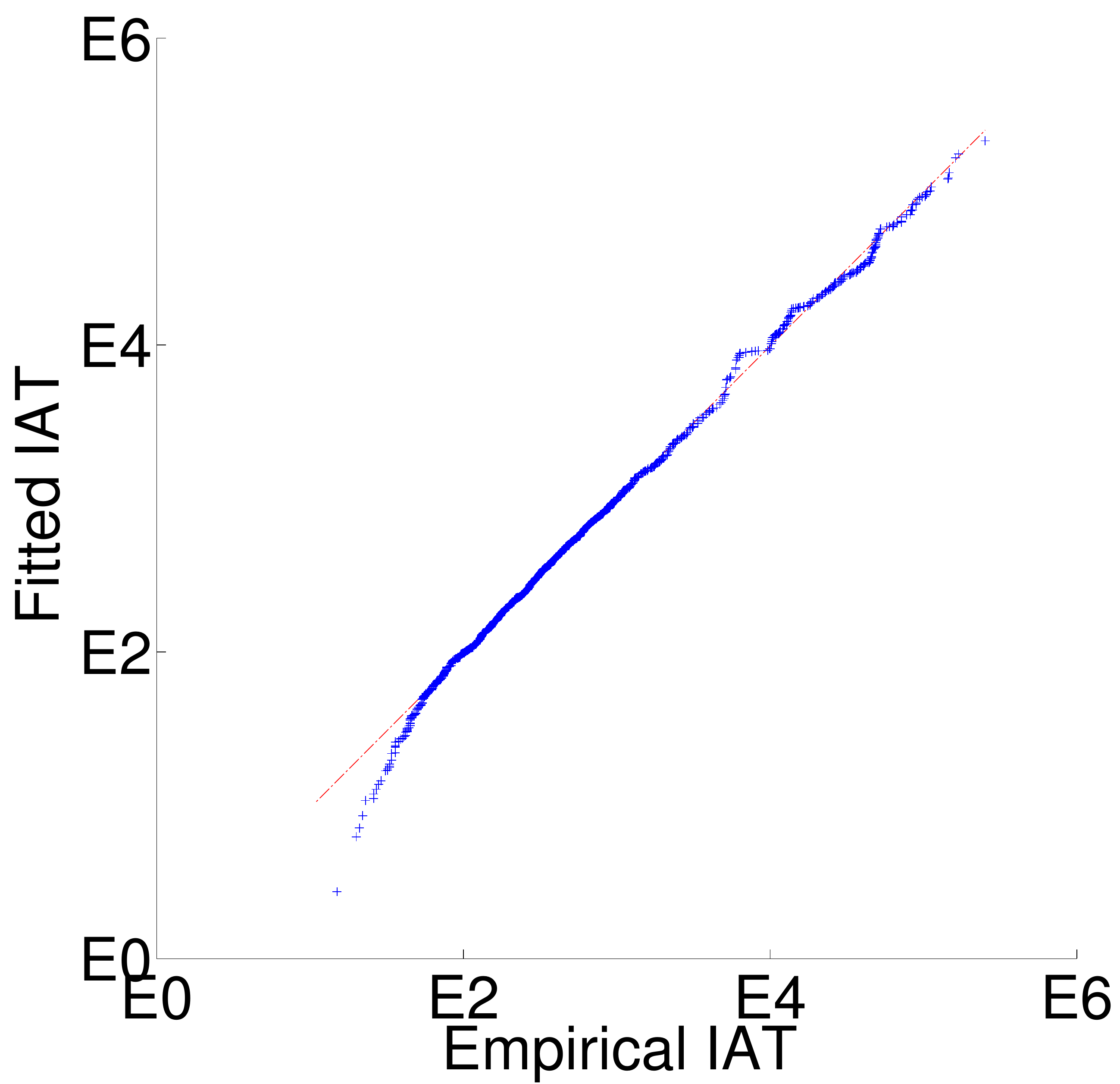}}
\subfigure{\includegraphics[width=0.30\columnwidth]{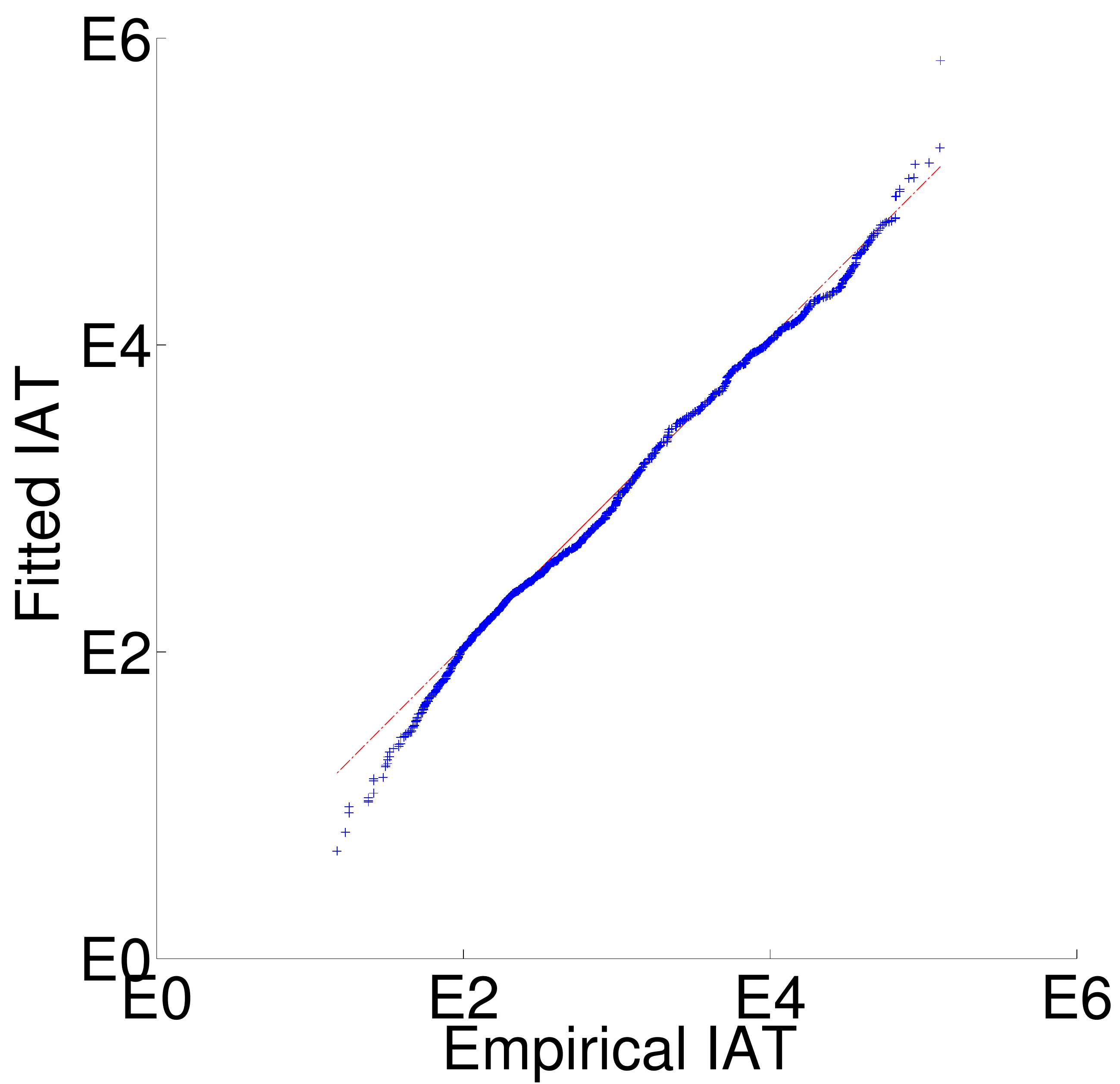}}
\subfigure{\includegraphics[width=0.30\columnwidth]{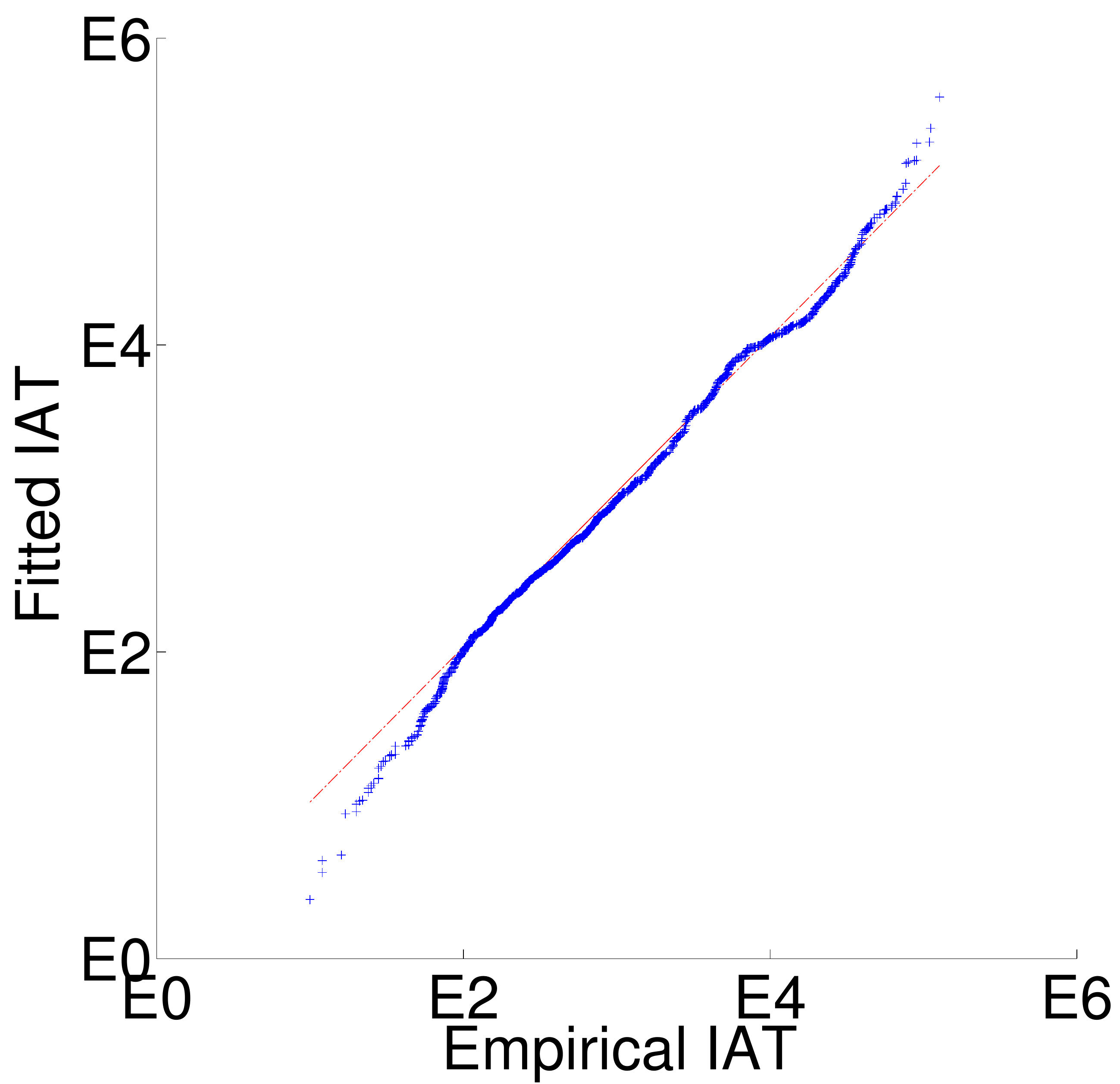}}
\subfigure{\includegraphics[width=0.30\columnwidth]{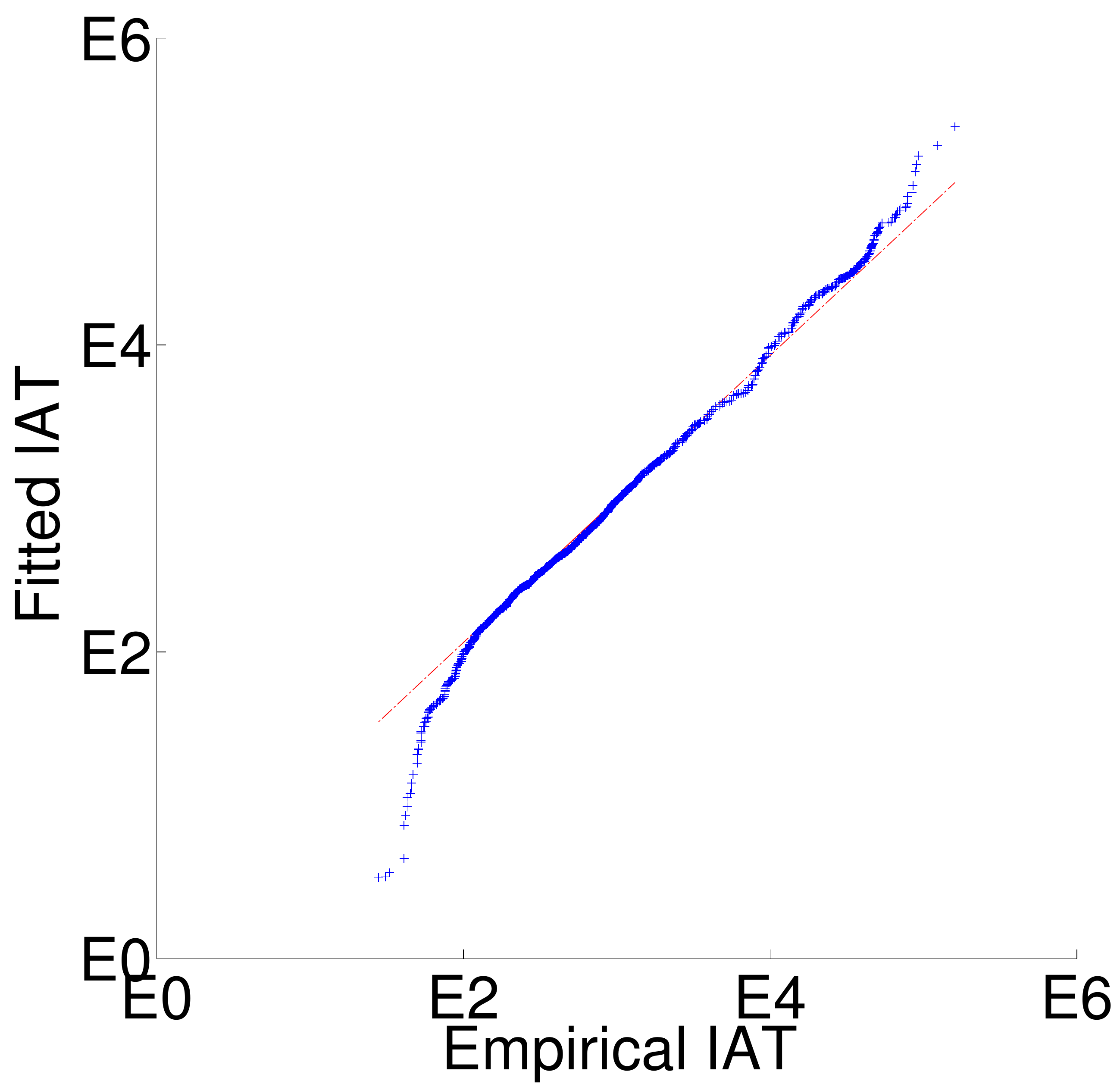}}
\subfigure{\includegraphics[width=0.30\columnwidth]{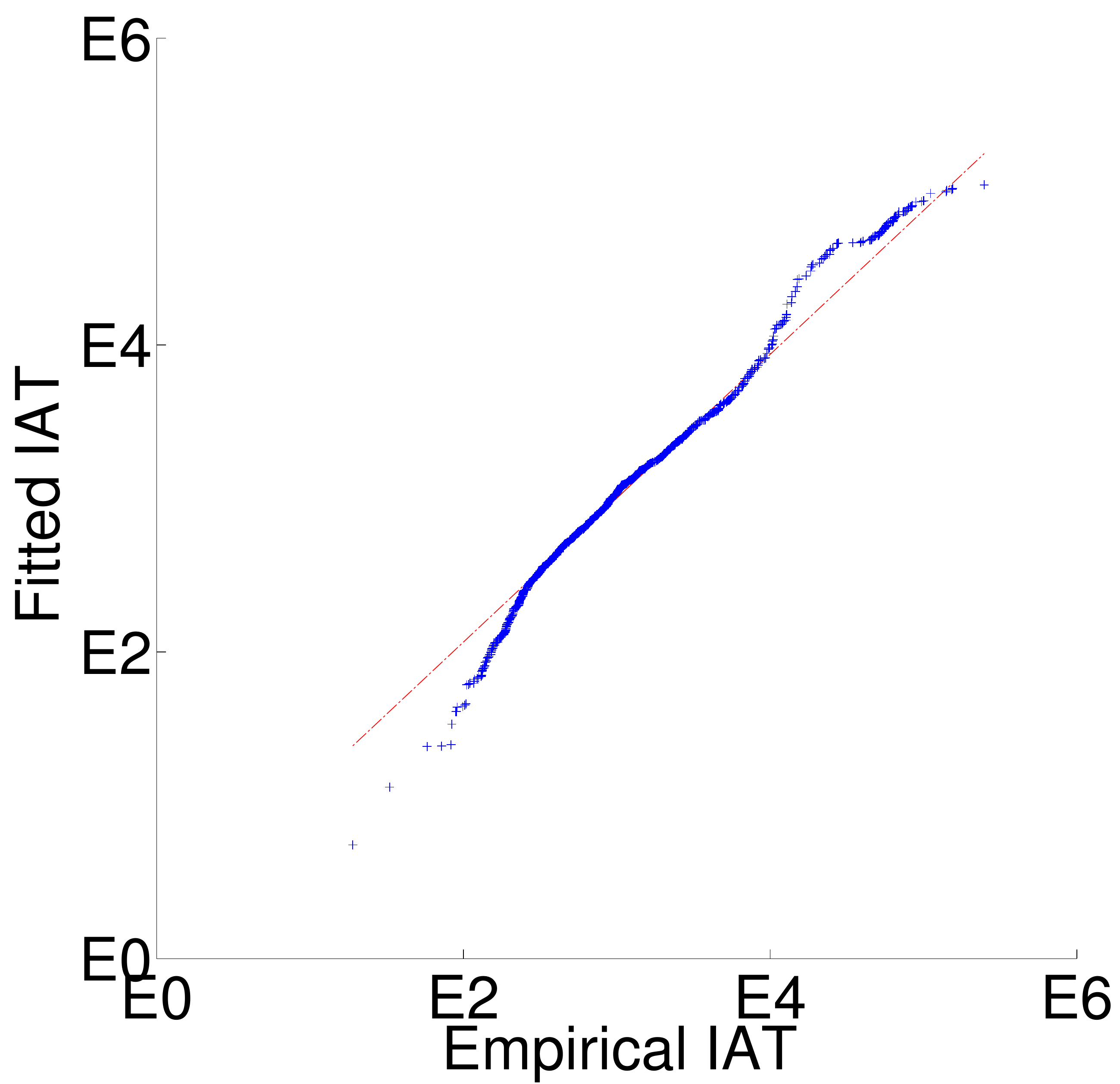}}
\subfigure{\includegraphics[width=0.30\columnwidth]{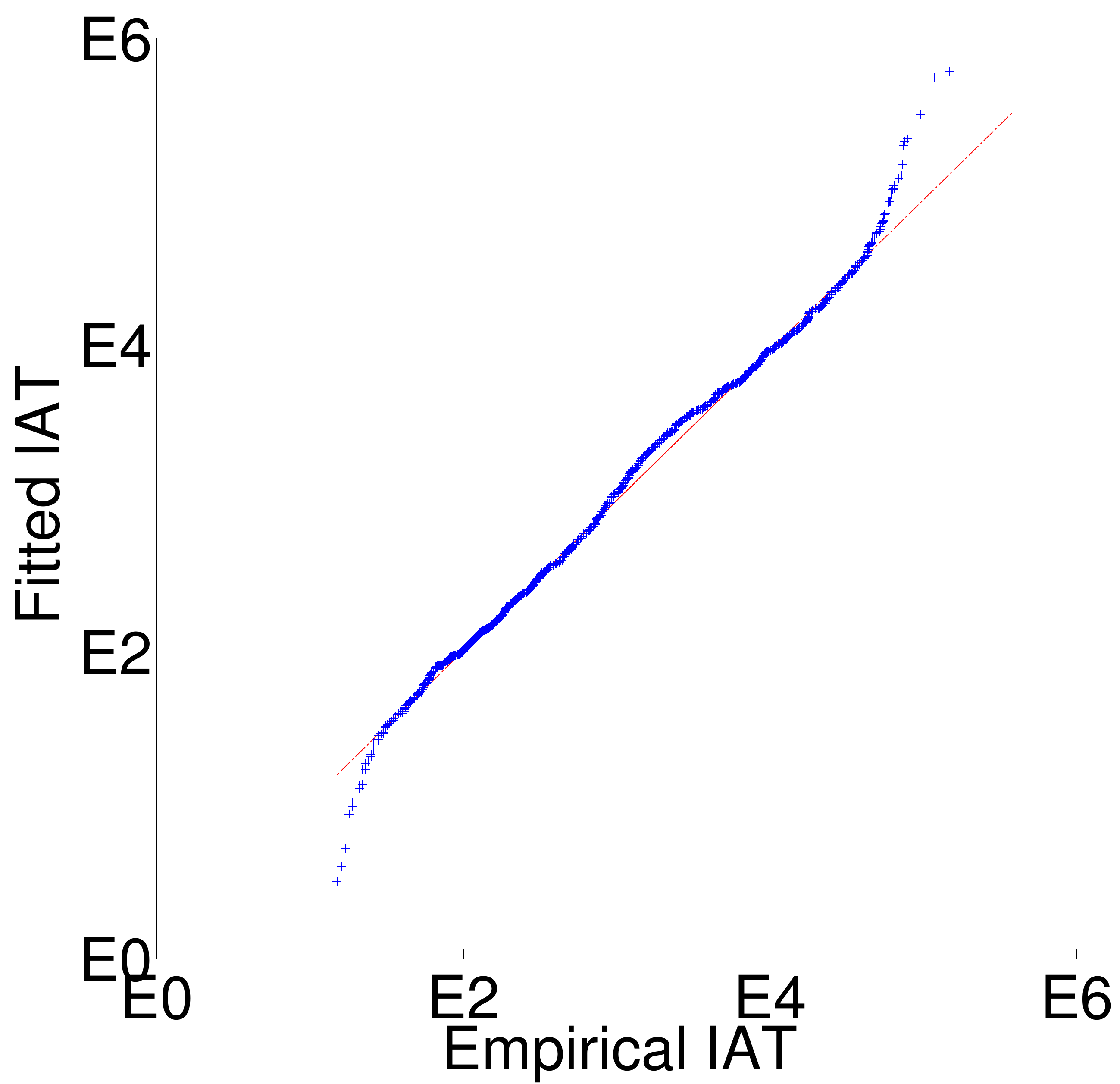}}
\subfigure{\includegraphics[width=0.30\columnwidth]{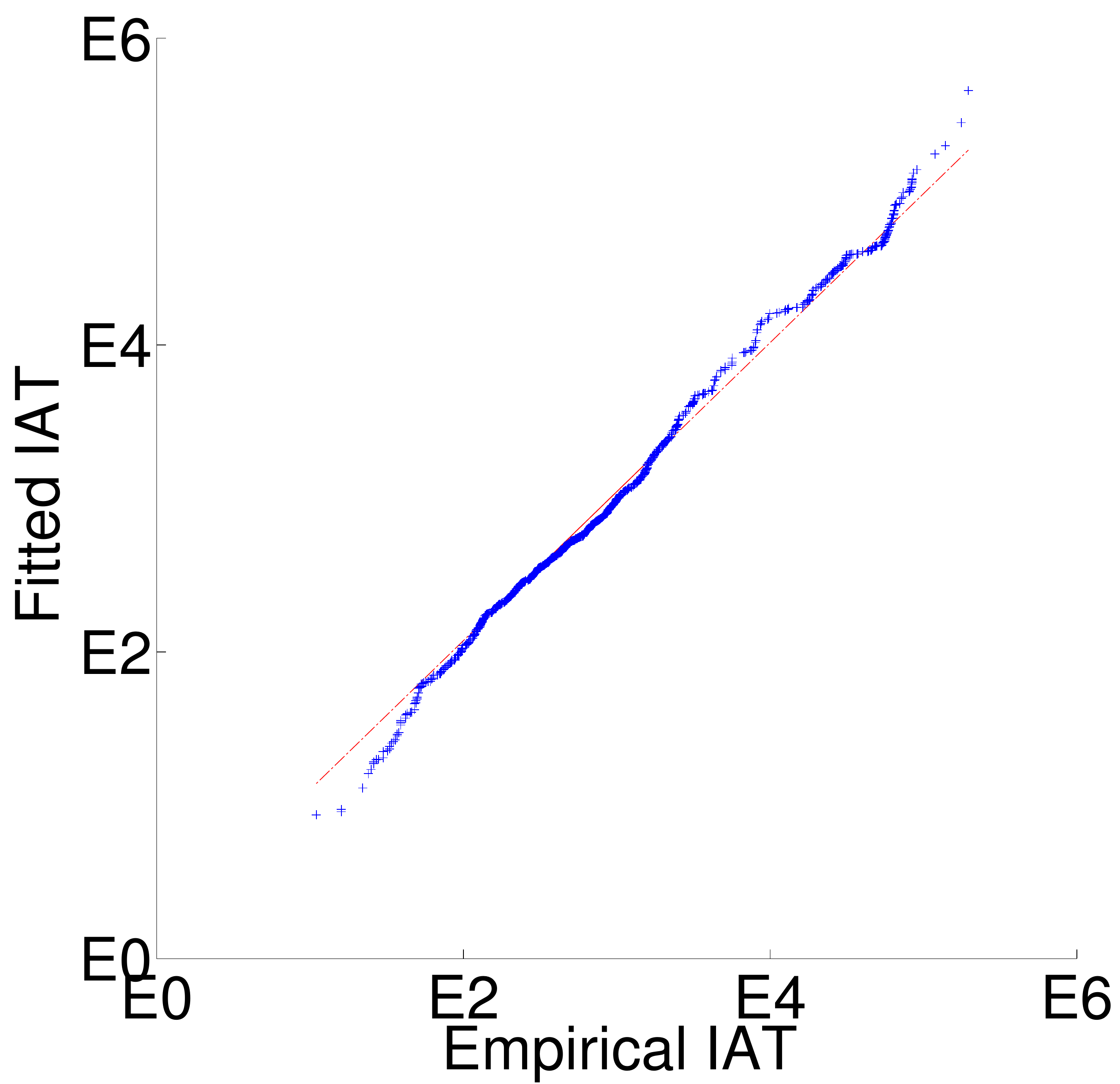}}
\subfigure{\includegraphics[width=0.30\columnwidth]{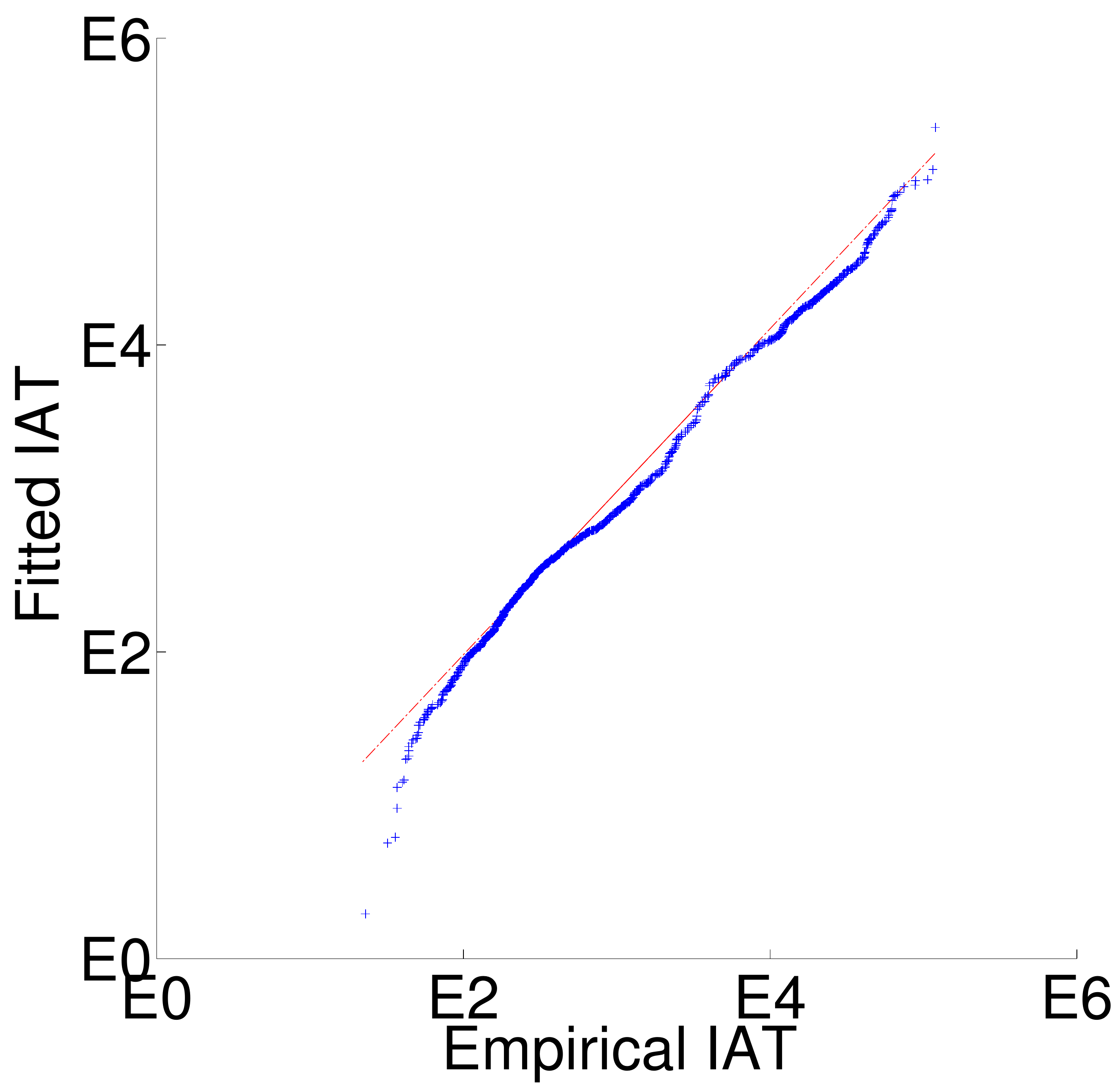}}
\subfigure{\includegraphics[width=0.30\columnwidth]{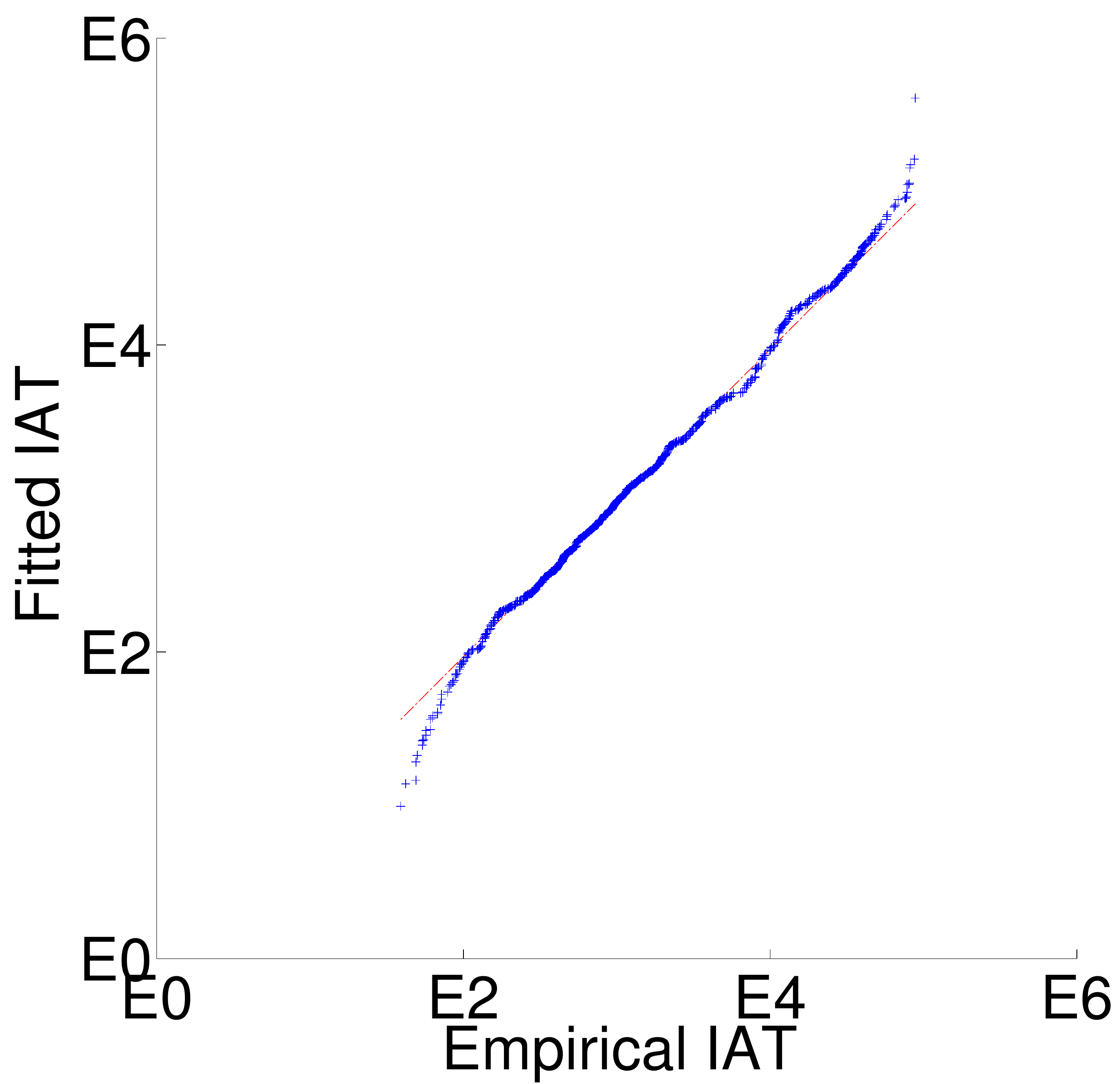}}
\subfigure{\includegraphics[width=0.30\columnwidth]{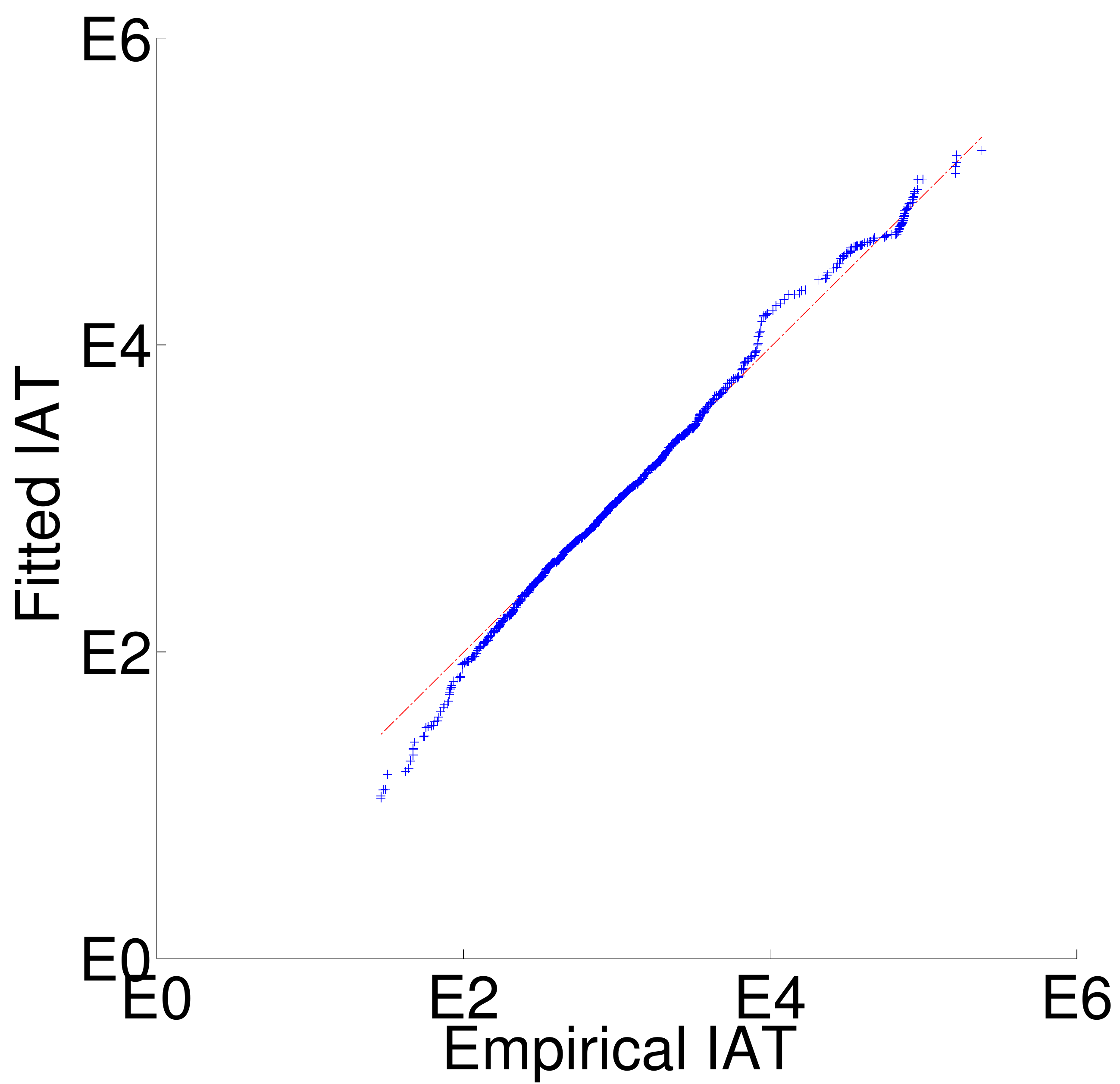}}
\subfigure{\includegraphics[width=0.30\columnwidth]{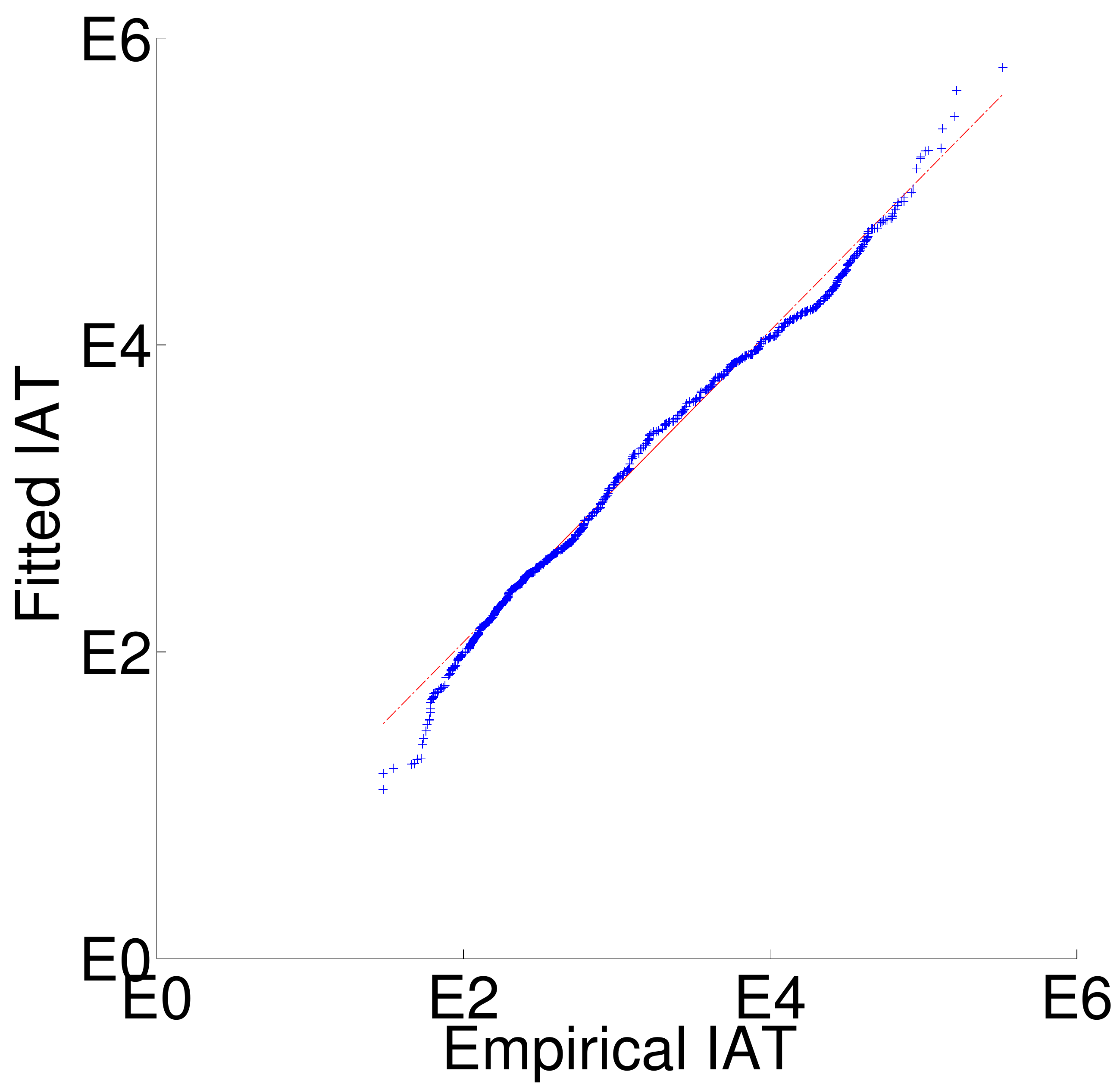}}
\caption{\textit{Validation by using Quantile-Quantile plot} (Q-Q plot). 45\tsc{$\circ$} line is ideal: all quantiles of the empirical data match the corresponding quantiles of the fitted samples. In each sub-figure, the majority of quantiles are matched very well by the proposed \mll distribution.}
\label{fig:Fitted_QQplot}
\end{figure}

The main idea of \mll is to use a mixture of two log-logistic (\LL) distributions to model the bi-modal pattern in Figure~\ref{fig:typical_behavior}(b). \LL is a skewed (in linear scale), power-law-like (heavy-tail) distribution, and there are two reasons for the choice of \LL: (a) it outperforms competitors (see Section~\ref{subsec: mll comparison}); (b) it has an intuitive explanation (the longer a person has waited, the longer (s)he will wait). \LL has been used successfully for modeling the IAT of the Internet communications of humans, such as posts on web blogs and comments on the Youtube\footnote{www.youtube.com}\cite{vaz2013self}. We remind its definition here:
\begin{definition}[Log-logistic distribution] 
Let $T$ be a non-negative continuous random variable and $T\sim \mathcal{LL}(t;\alpha,\beta)$. The CDF of a log-logistically distributed $T$ is given as:
\begin{equation}
F_{\mathcal{LL}}(t;\alpha,\beta) = \frac{1}{1+(t/\alpha)^{-\beta}}
\label{eq:CDF_LL}
\end{equation}
\noindent where $\alpha > 0$ is the median (or called scale parameter), and $\beta > 0$ is the shape parameter. The support $t \in [0,\infty)$.
The PDF of $T$ is given as:
\begin{equation}
f_{\mathcal{LL}}(t;\alpha,\beta) = \frac{(\beta/\alpha)(t/\alpha)^{\beta-1}}{[1+(t/\alpha)^{\beta}]^2}
\label{eq:PDF_LL}
\end{equation}
\end{definition}


With the knowledge of \LL, we present the definition of the proposed \mll distribution:
\begin{definition}[\mll distribution] Let $T$ be a non-negative random variable following \mll distribution. The probability density function (PDF) can be written as:
	\begin{eqnarray}
	f_{\mll}(t) &=& \theta\cdot f_\mathcal{LL}(t;\alpha_{\IN},\beta_{\IN})+ \nonumber\\ 
				& & (1-\theta)\cdot f_\mathcal{LL}(t;\alpha_{\OFF},\beta_{\OFF}) 
	\label{eq:mll}
	\end{eqnarray}
\noindent where $t\ge 0$, $\theta \in [0,1]$, $\alpha_{\IN}, \beta_{\IN}, \alpha_{\OFF}, \beta_{\OFF} > 0$.
\end{definition}

The proposed \mll distribution has the following properties:
\bit
	\item A mixture of two \LL (heavy-tail) distributions to qualitatively describe: in-session and take-off IAT.
	\item{Five parameters to characterize `Alice's search behavior: 
		\bit
			\item $\theta$ controls the proportion of in-session and take-off IAT.
			\item $\alpha_{\IN}$ represents the median of in-session IAT.
			\item $\beta_{\IN}$ is the ``concentration\footnote{The reciprocal of $\beta_{\IN}$ represents (approximately) the standard deviation of \LL.}'' of in-session IAT.
			\item $\alpha_{\OFF}$ represents the median of take-off IAT.
			\item $\beta_{\OFF}$ is the concentration of take-off IAT.
		\eit
	}
\eit

\mll distribution seems to model the marginal distribution of IAT very well, at least for `Alice' shown in Figure~\ref{fig:typical_behavior}(b), and also provides intuitive interpretations. But we still have the following questions:
\bit
	\item Is \mll sufficiently general and accurate to model and interpret other people's search behavior?
	\item Even so, does \LL outperform other famous distributions, say Exponential or Pareto (power-law)?
\eit

The answers to both questions are \textit{yes}, and the details are provided in the following two sections.

\subsection{Validation against empirical data}
\label{subsec: mll validation}

Figure~\ref{fig:Fitted_PDF} illustrates the empirical IAT from 12 most `prolific' users. Each sub-figure shows the marginal distribution of IATs (in logarithmic binning) from a user, and the red curve is depicted by fitting a \mll distribution via expectation maximization (EM). For brevity, we show only the top 12 most prolific users, but most of the remaining ones had similar behavior (see Figure 10(a), where the vast majority of users have very similar model parameters). Notice:
\bit
	\item The consistency of bi-modal behaviors. Not only `Alice' has the distinct pattern: in-session and take-off, but Bob and other users have this pattern as well.
	\item The generality of the proposed \mll. \mll is able to accurately model the marginal distribution of IAT from every user. (\mll also models other dataset; see Section \ref{subsec:Generality of mll} for details.)
\eit

\noindent Also from Figure~\ref{fig:Fitted_PDF}, we provide the following observation:
\begin{observation}[In-session and take-off]
The median in-session IAT is about five minutes, whereas the median of take-off IAT is approximately seven hours.
\label{ob:in-session and take-off}
\end{observation}
There are two types of IAT: in-session and take-off. The median of in-session IAT is about 5 minutes, which approximately represents the duration when a user is interested in the query results. On the other hand, the IAT of take-off queries is longer, ranging from tens of minutes (\eg, lunch break), hours (\eg, sleep time), to days (\eg, weekends). The median of take-off IAT is approximately seven hours, which corresponds to sleep time very well.

More validations are provided by Figure~\ref{fig:Fitted_QQplot}. For each user, Figure~\ref{fig:Fitted_QQplot} provides the Quantile-Quantile plot (Q-Q plot) between the empirical IAT and the samples drawn from the fitted \mll distribution. In each sub-figure, X axis represents the IAT from a user and Y axis are the samples randomly drawn from the fitted \mll distribution. 45\tsc{$\circ$} line is ideal, meaning that the empirical data and the fitted samples follow the same distribution). As it can be seen, in each sub-figure the majority of quantiles are matched very well by the proposed \mll distribution.

By now we have strong evidences supporting the goodness of fit for \mll; we still need to answer the question: why not using a mixture of other well-known ``named'' distributions, say Exponential or Pareto (power-law)?

\subsection{Why not other well-known distributions?}
\label{subsec: mll comparison}

We compare the goodness of fit among the following three candidates:
\bit
	\item A mixture of two Exponential distributions.
	\item A mixture of two Pareto distributions.
	\item The proposed \mll distribution.
\eit
by using the following criteria:	
\bit
	\item P value reported by two-sample Kolmogorov-Smirnov (K-S) test.
	\item Data log-likelihood.
	\item Bayesian information criterion (BIC).
\eit

It turns out \mll outperforms other candidates in all three criteria. Note that for each user, the candidate models are fitted by the training set (randomly drawn from her/his IAT), whereas the P-value and log-likelihood are reported by using the testing set (data not in the training set).

\begin{figure}[tb]
\centering
    \includegraphics[width=0.6\linewidth]{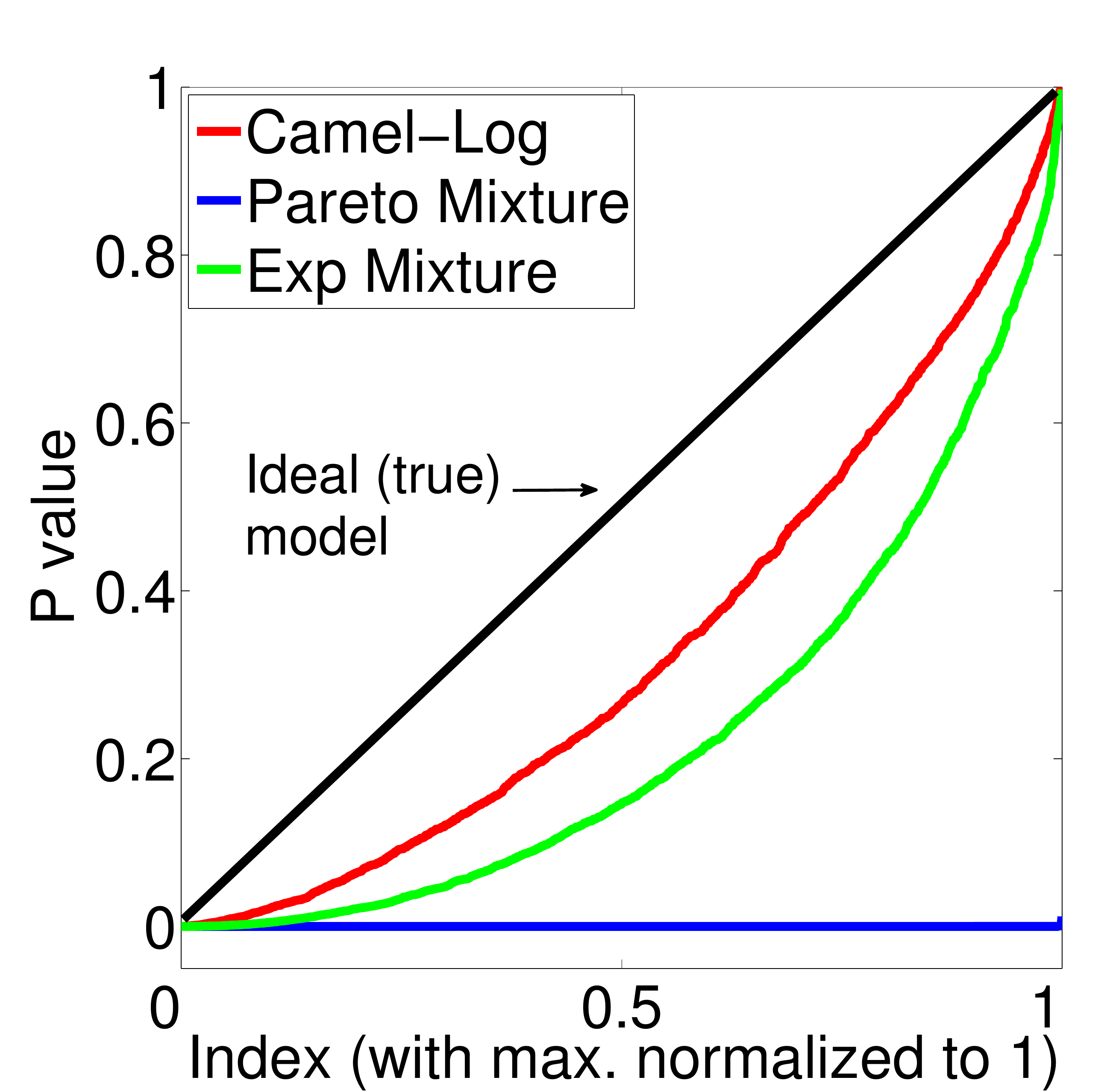}
\caption{\textit{\mll wins with respect to K-S tests}. Sorted P-values are reported from K-S tests on: the proposed \mll (in red), Pareto mixture (in blue) and Exponential mixture (in green). The 45\tsc{$\circ$} straight line represents the ideal (true) model: p-value follows the uniform(0,1) distribution. The proposed \mll is the closest to the true model.}
\label{fig:pvalue}
\end{figure}

Figure~\ref{fig:pvalue} provides the p-value reported by K-S test on each user, with the null hypothesis ($H_0$): the user's IAT follows the fitted candidate distribution. If $H_0$ is true, the p-value will follow a uniform(0,1) distribution, depicted by the 45\tsc{$\circ$} straight line. From Figure~\ref{fig:pvalue}, the proposed \mll is the candidate closest to the true model; exponential mixture fits well but not as close, whereas Pareto mixture does not fit at all (with constantly low p-values).

\begin{table}[tb]
	\centering
	\caption{Evaluation with log-likelihood and BIC: \%-of users that \mll explains better (higher is better)}
	\begin{tabular}{|c||c|c|}
	\hline
	\multicolumn{3}{|c|}{Log-likelihood (of the testing set)} \\ \hline
	Compared against:& Exponential mix. & Pareto mix. \\ \hline	
	\mll & 78\% & $>$ 99\% \\ \hline\hline
	\multicolumn{3}{|c|}{Bayesian information criterion (BIC)} \\ \hline
	Compared against:& Exponential mix. & Pareto mix. \\ \hline	
	\mll & 66\% & $>$ 99\% \\ \hline
	\end{tabular}	
	\label{tbl:LL_BIC}
\end{table}

We also provide log-likelihoods to show \mll better explains users' behaviors. Table~\ref{tbl:LL_BIC} presents \%-of users that \mll explains better (achieves higher likelihood), compared to other candidates. The proposed \mll achieves a higher log-likelihood on 78\% of the users (compared to Exponential mixture), and more than 99\% of the users (compared to Pareto mixture). 

Furthermore, since each candidate model uses different number of parameters: \mll (five), Exponential mixture (three), and Pareto mixture (three), we also evaluate the BIC that strongly\footnote{Compared to Akaike information criterion (AIC).} penalizes using more parameters and therefore prefers a parsimonious model. Table~\ref{tbl:LL_BIC} presents the BIC scores: the proposed \mll achieves a lower BIC\footnote{Given any two estimated models, the model with the lower value of BIC is the one to be preferred.} on 66\% of the users (compared to Exponential mixture), and more than 99\% of the users (compared to Pareto mixture).

From the evaluation of p-value, log-likelihood and BIC among three candidate models, we summarize:
\bit
	\item Exponential mixture fits well, and the proposed \mll fits even better.
	\item Compared to other two candidates, even \mll using two more parameters, it is the preferred model by BIC for the majority cases.
	\item Pareto mixture is out of the winner circle.
\eit

Both qualitative (Section~\ref{subsec: mll validation}) and quantitative (this section) evidences are favorably supporting the goodness-of-fit of \mll. Now we ask: how general \mll is? Does \mll model other Internet-based, human behaviors? The answer is \textit{yes}: \mll models the IAT between posts on Reddit\footnote{http://www.reddit.com/} very well.

\subsection{Generality of \mll}
\label{subsec:Generality of mll}
\input{Generality}

%% file: Generality.tex
\begin{figure}[tb]
\centering
\subfigure{\includegraphics[width=0.15\textwidth]{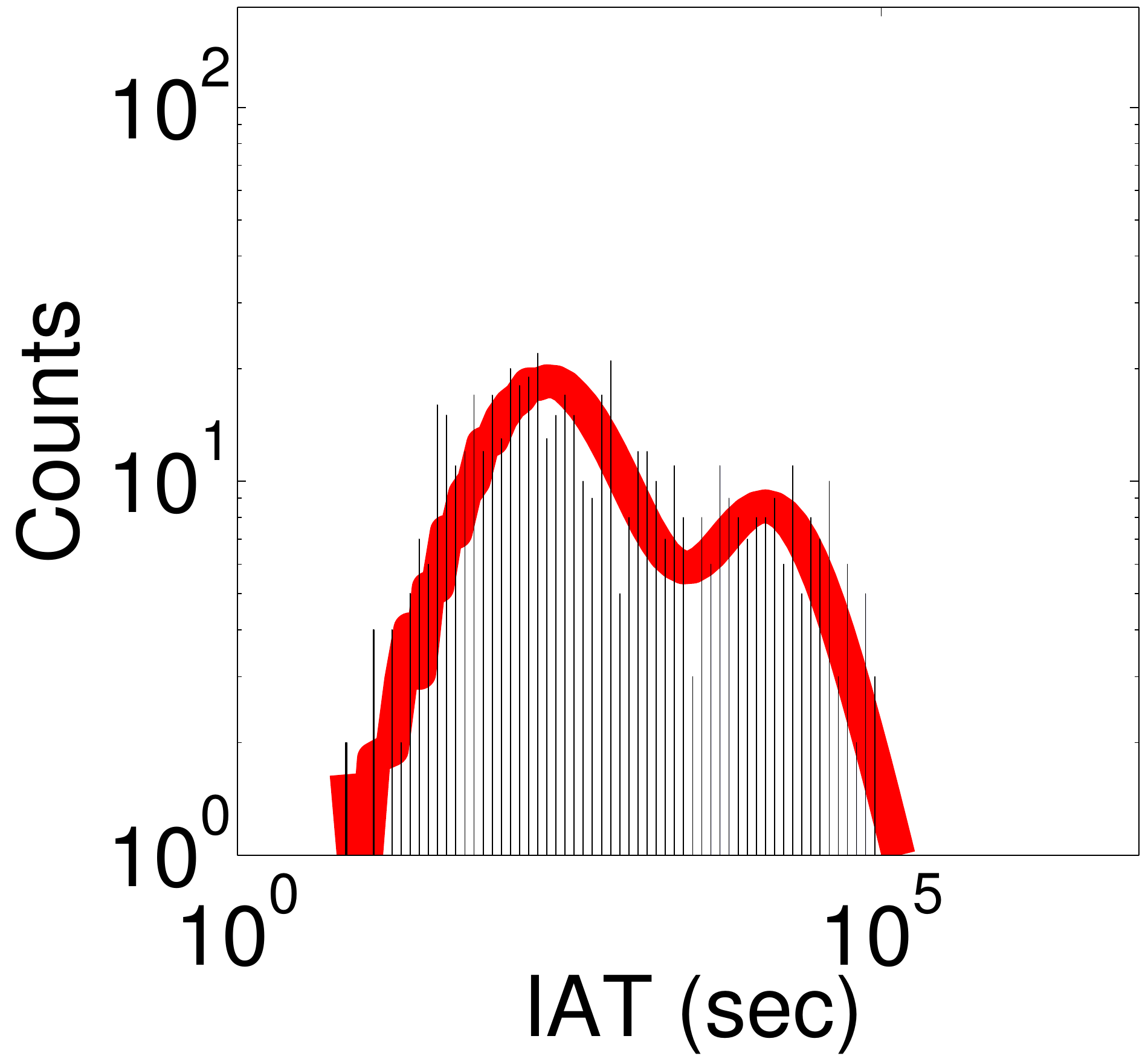}}
\subfigure{\includegraphics[width=0.15\textwidth]{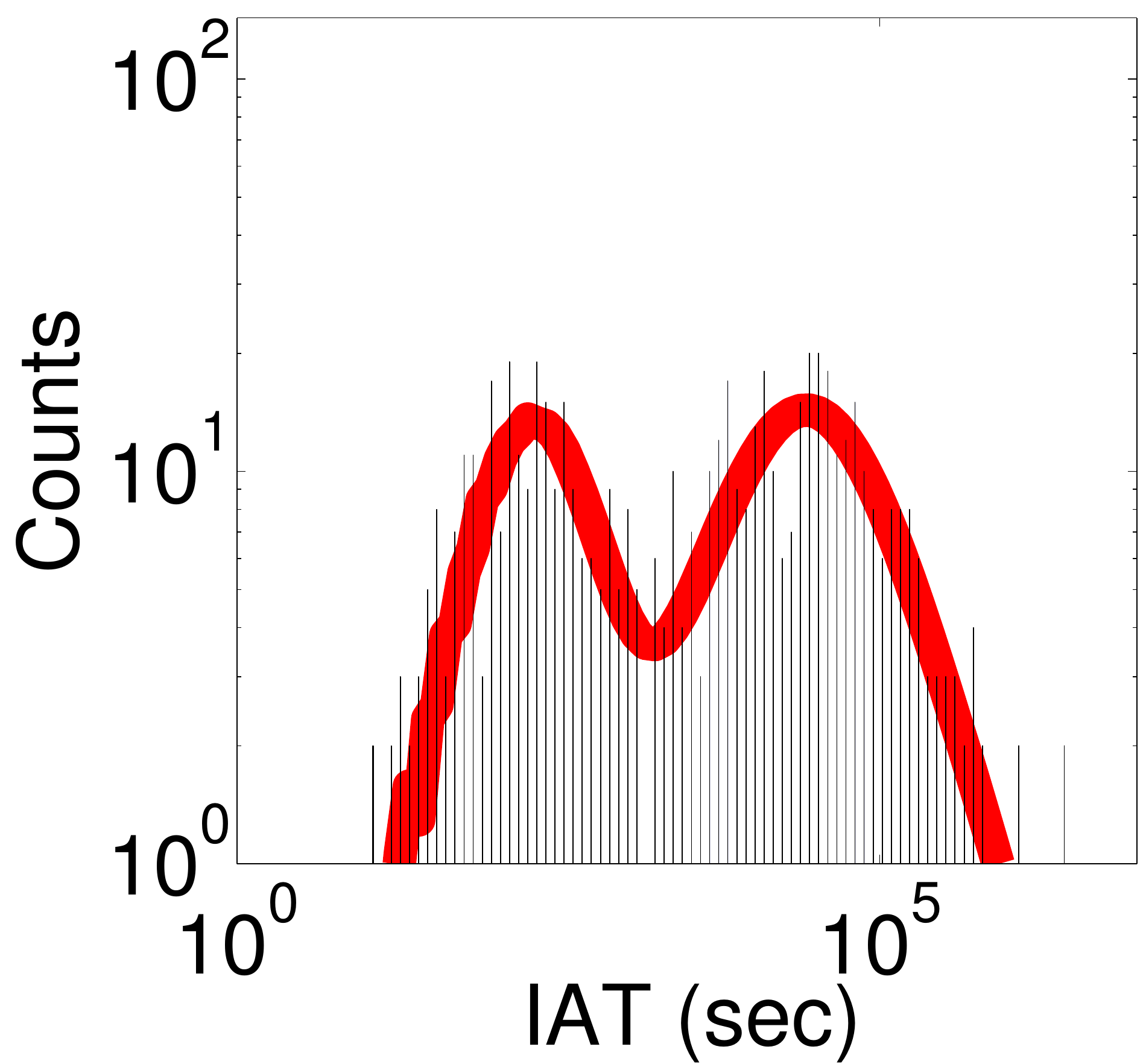}}
\subfigure{\includegraphics[width=0.15\textwidth]{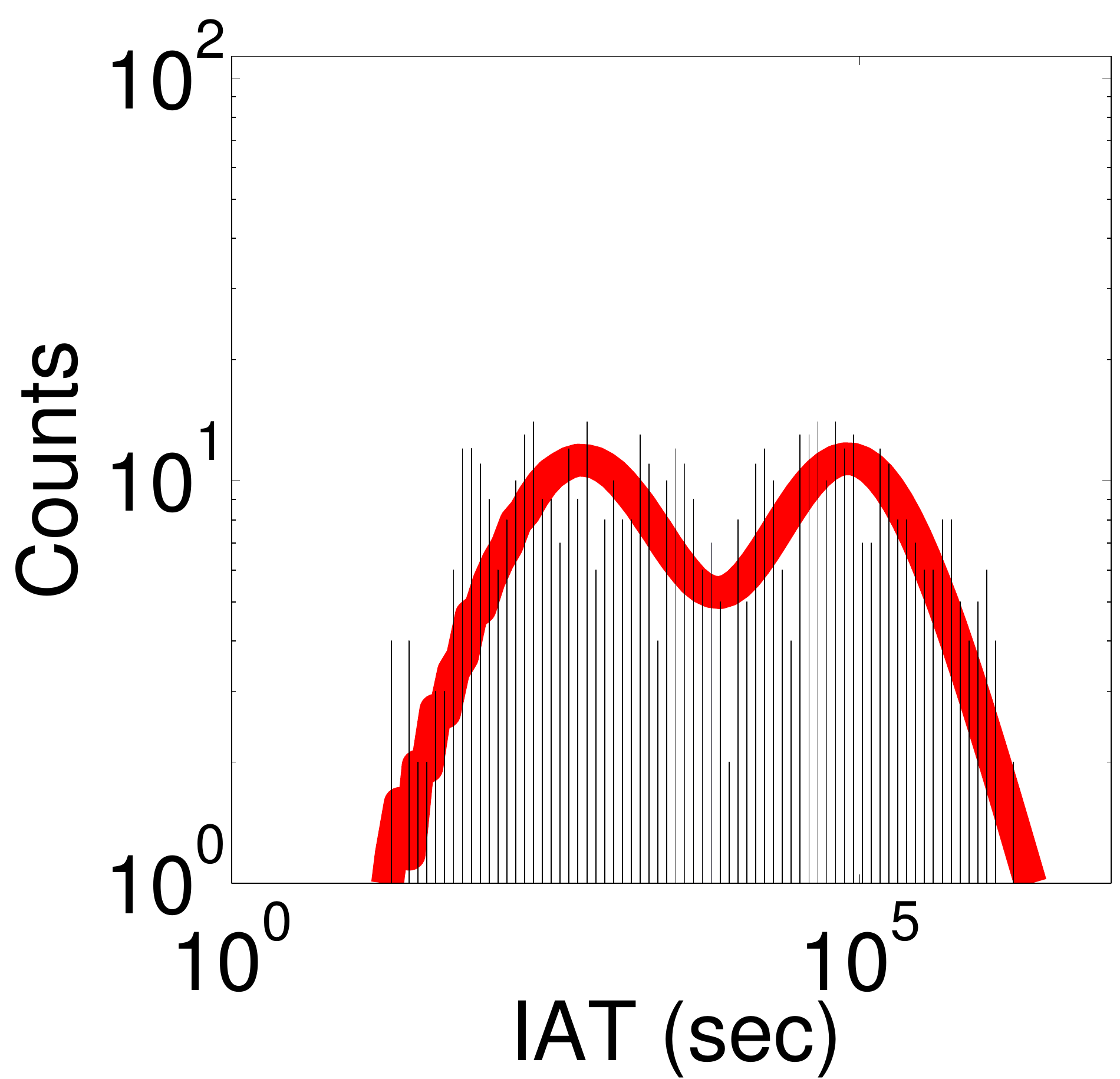}}
\subfigure{\includegraphics[width=0.15\textwidth]{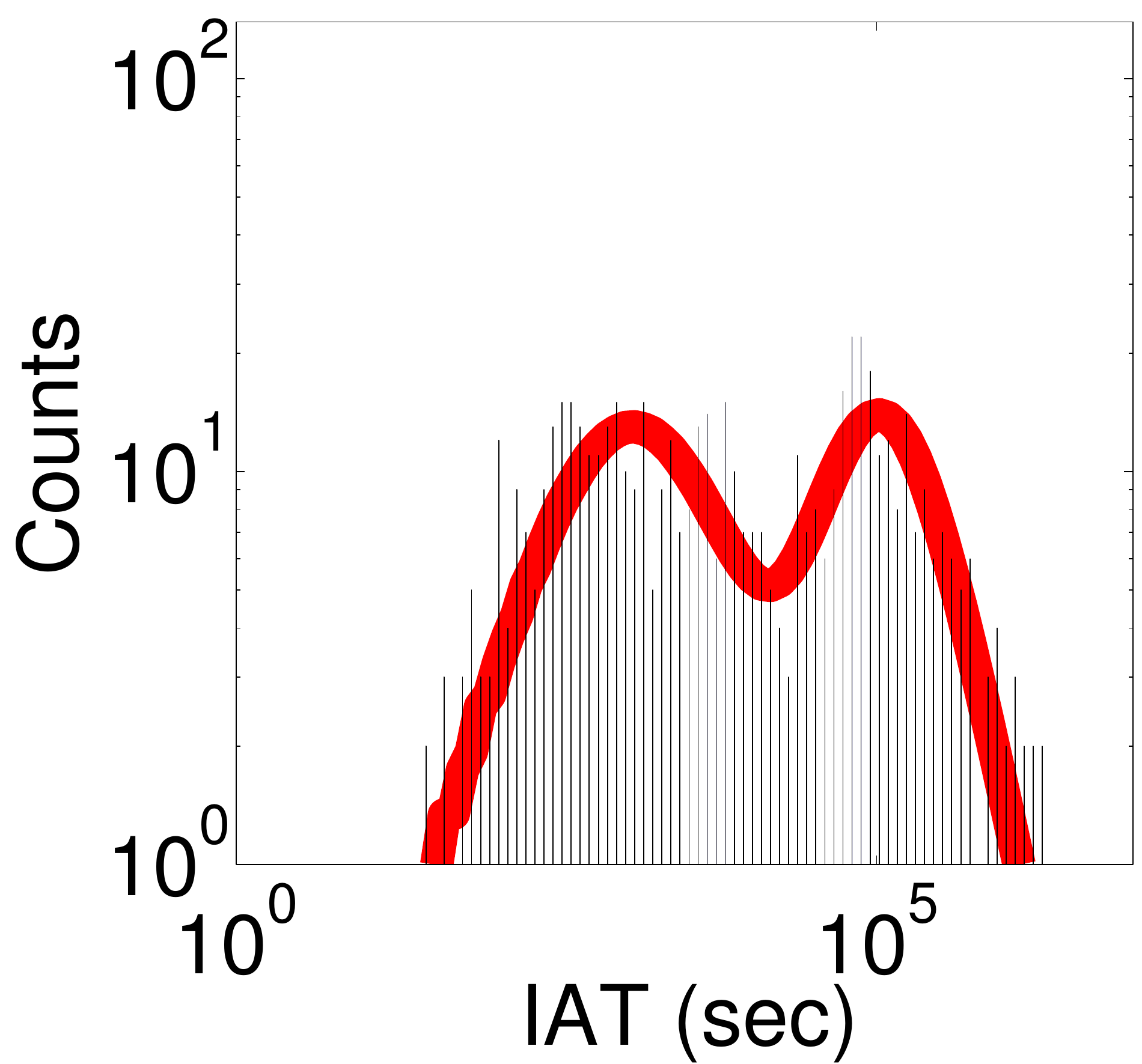}}
\subfigure{\includegraphics[width=0.15\textwidth]{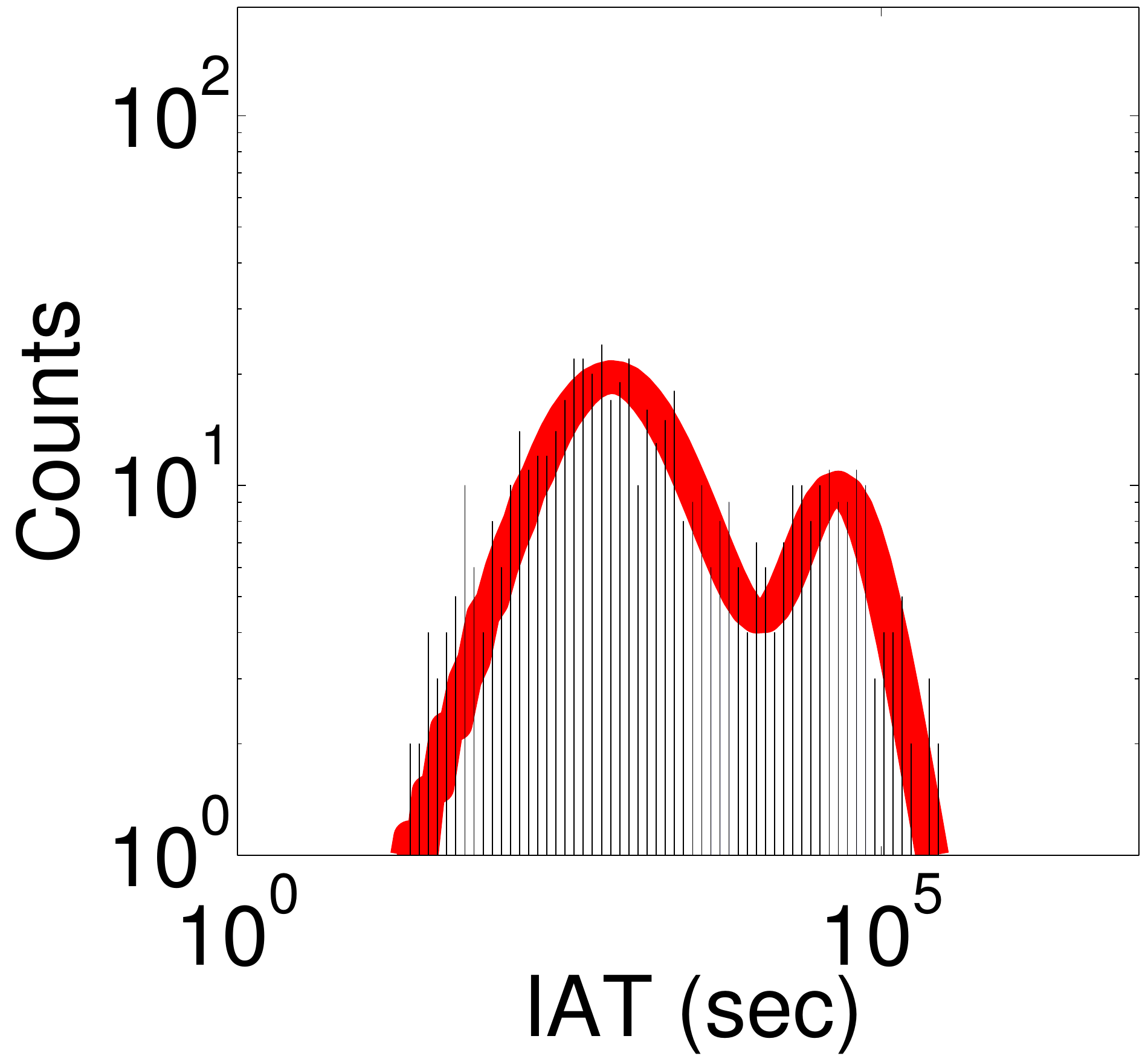}}
\subfigure{\includegraphics[width=0.15\textwidth]{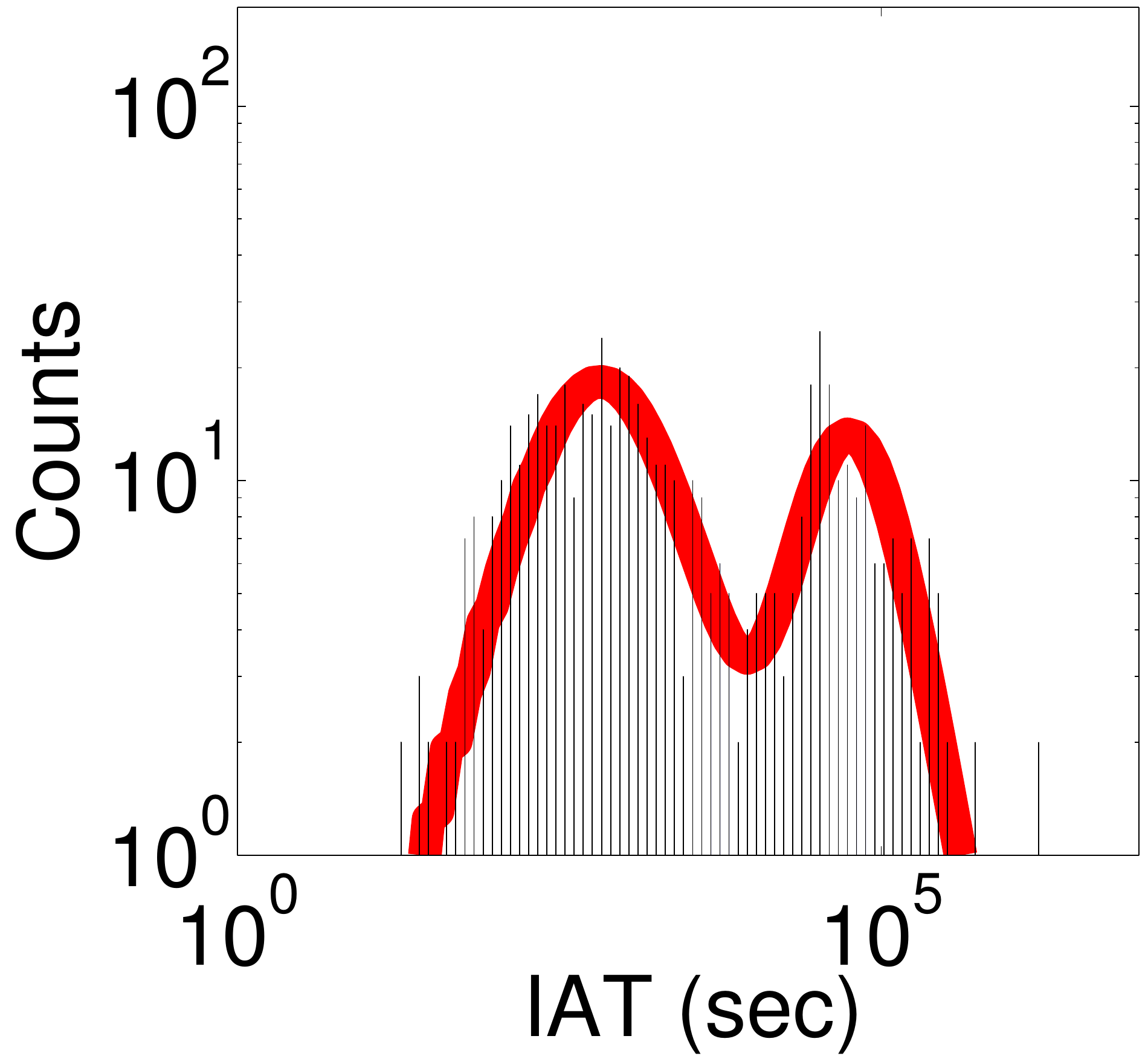}}
\subfigure{\includegraphics[width=0.15\textwidth]{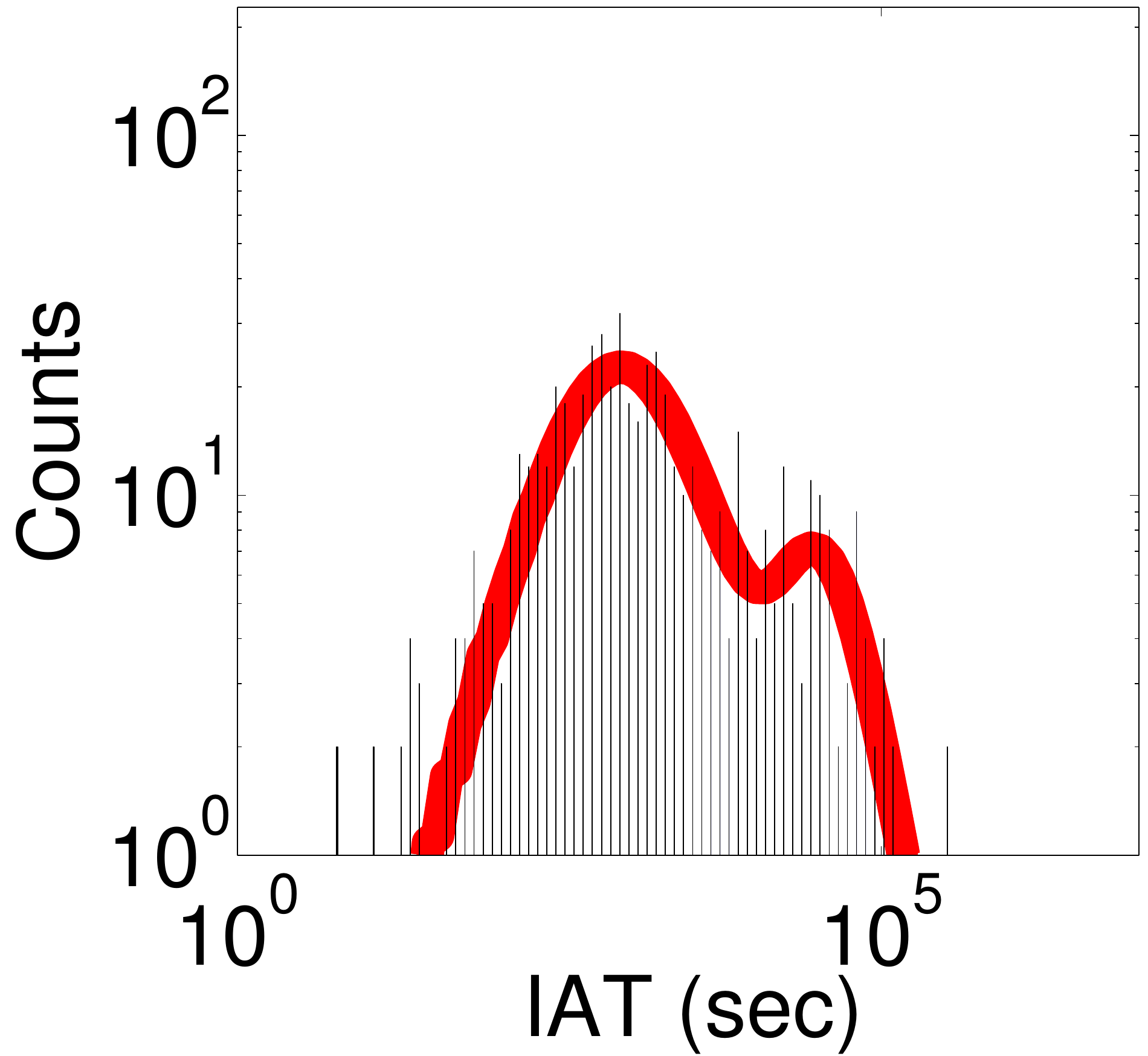}}
\subfigure{\includegraphics[width=0.15\textwidth]{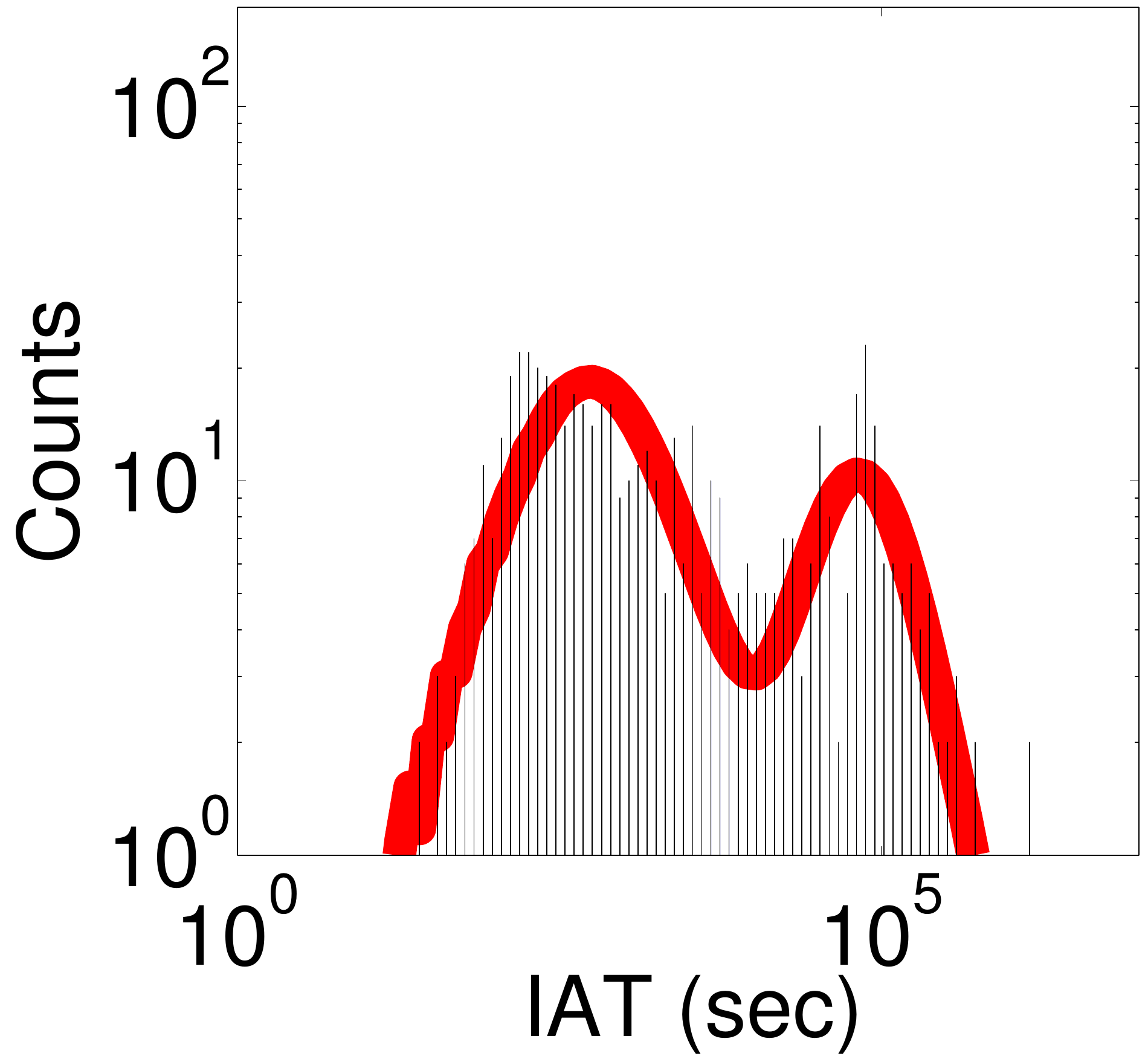}}
\subfigure{\includegraphics[width=0.15\textwidth]{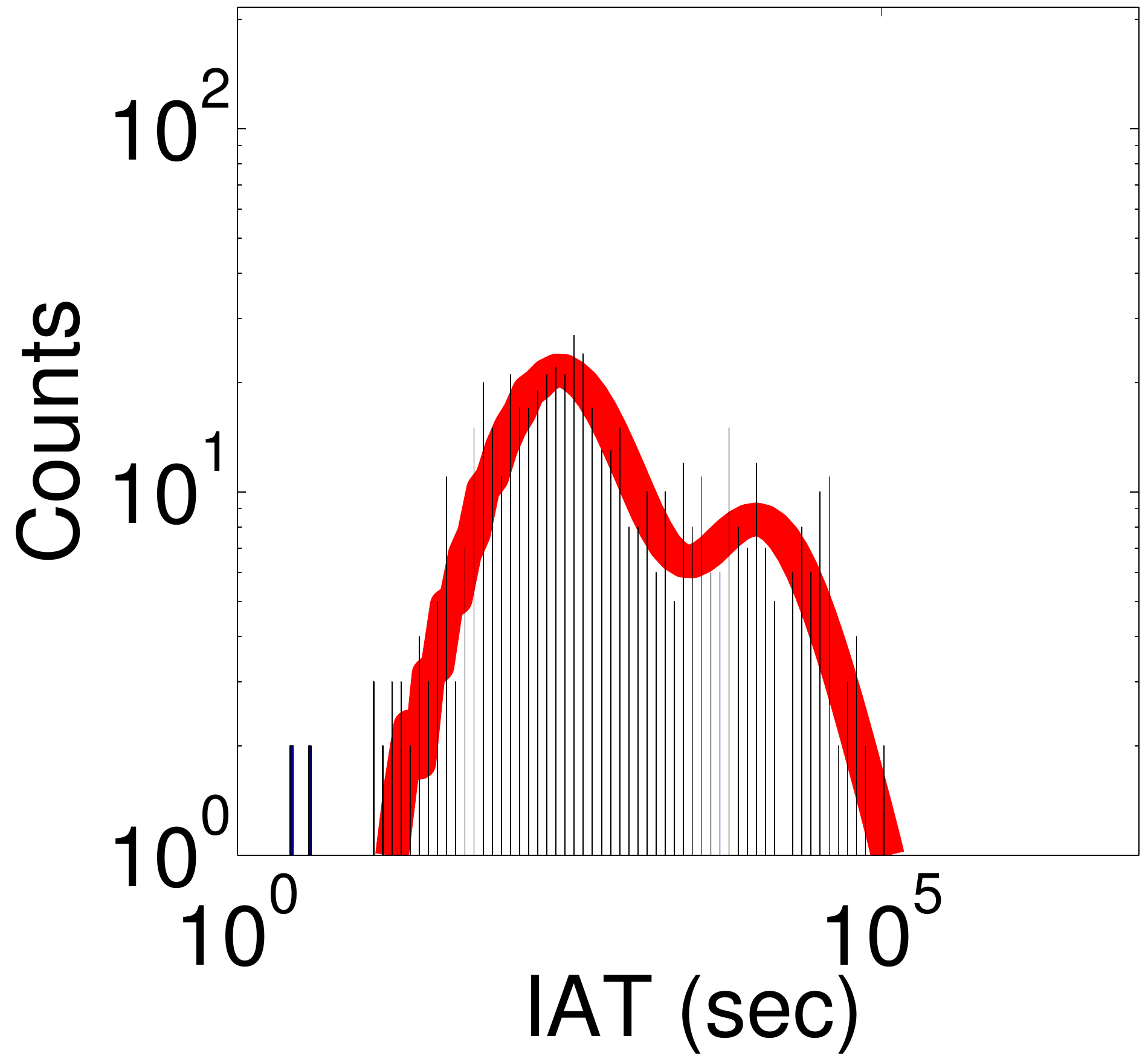}}
\subfigure{\includegraphics[width=0.15\textwidth]{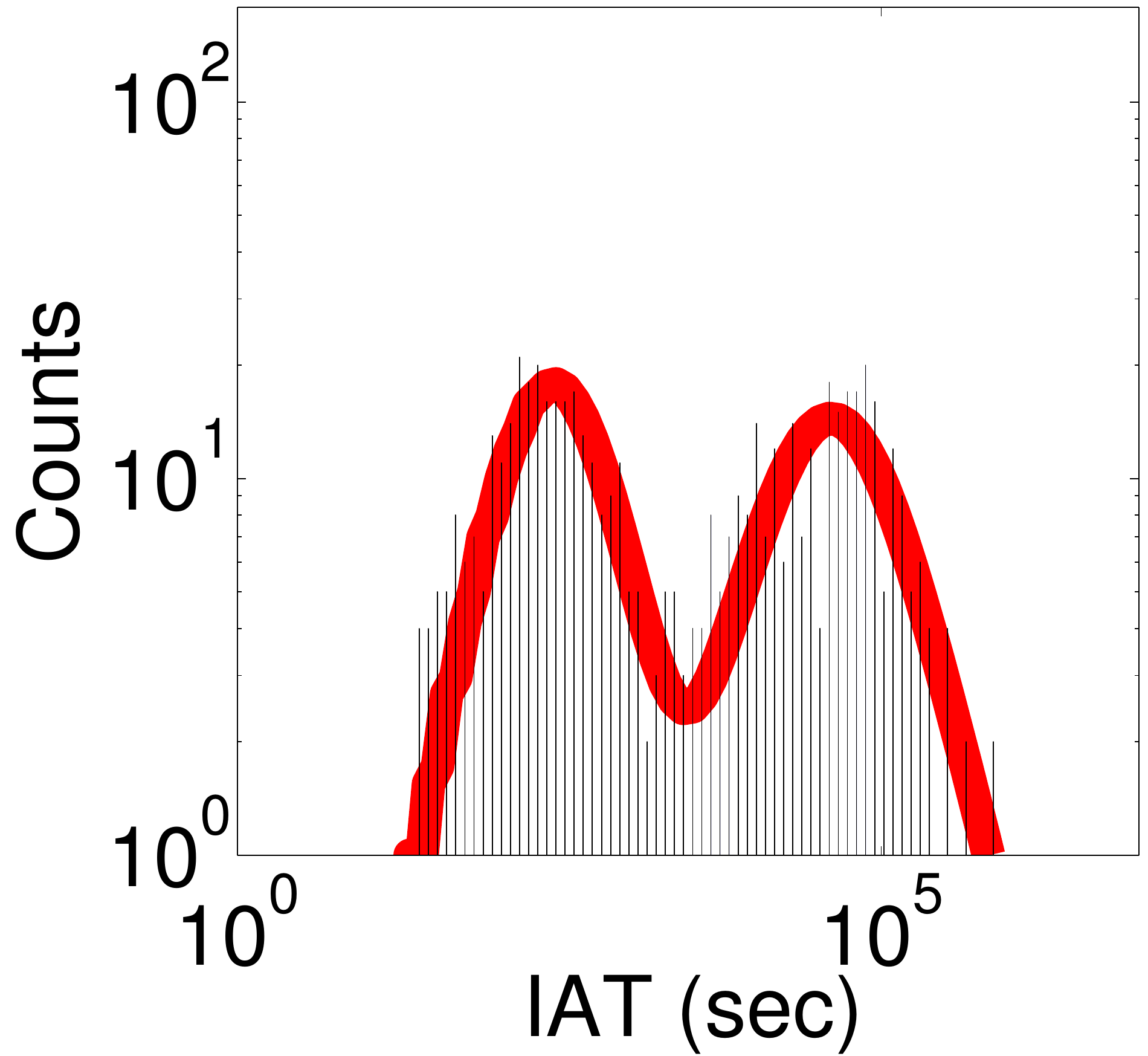}}
\subfigure{\includegraphics[width=0.15\textwidth]{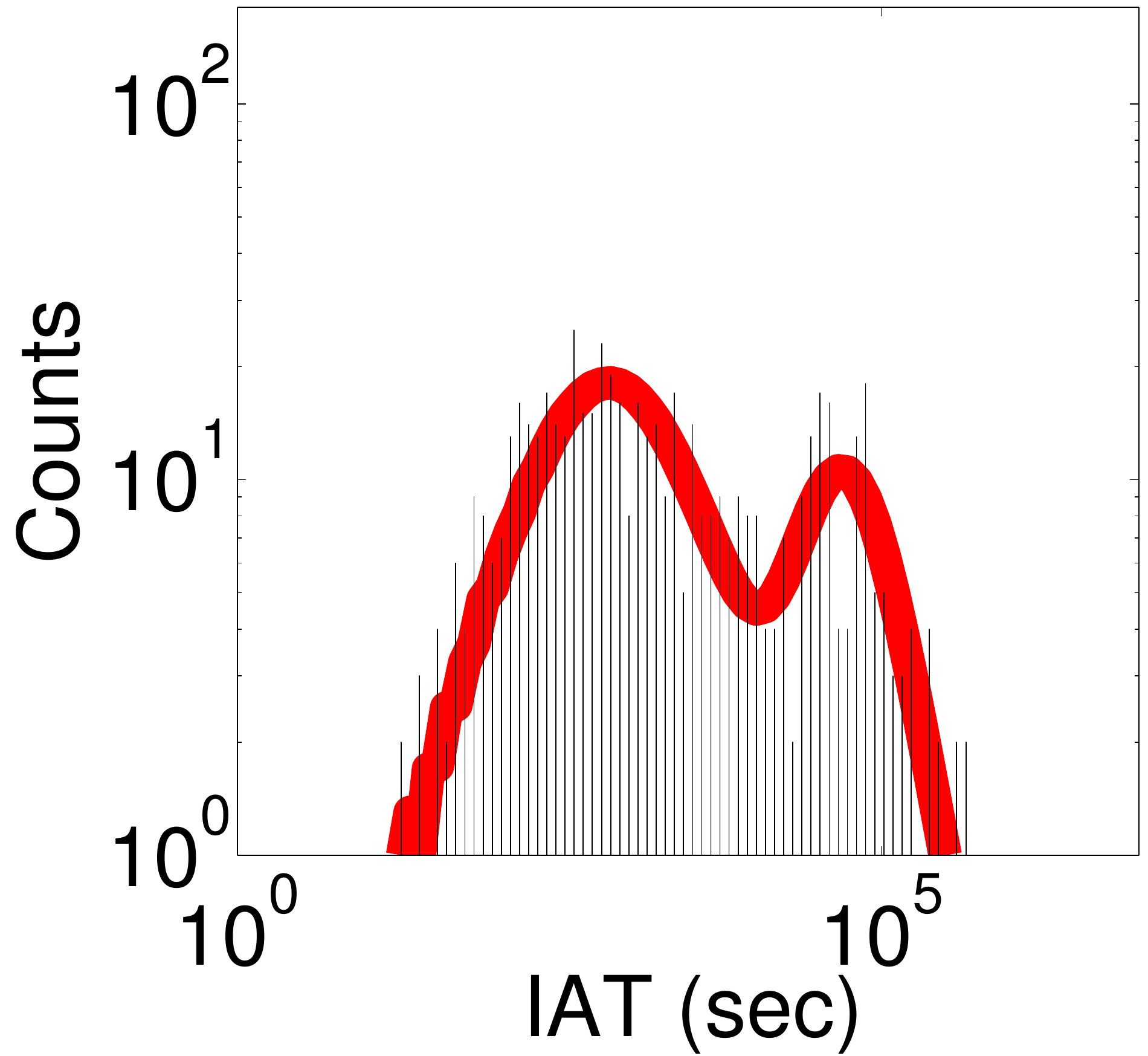}}
\subfigure{\includegraphics[width=0.15\textwidth]{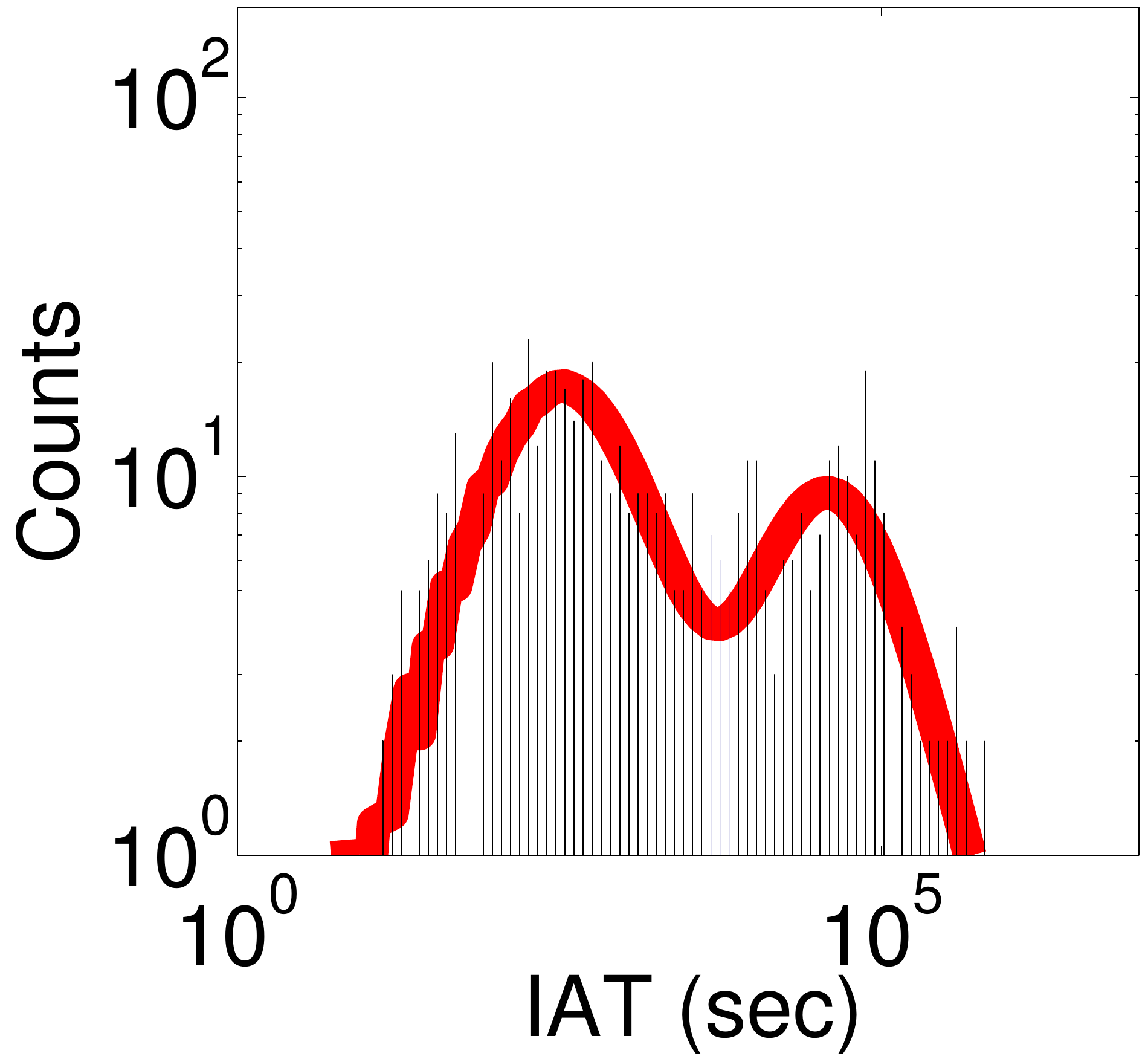}}
\caption{\textit{\mll fits the Reddit dataset (marginal PDF)}. Each sub-figure shows the marginal distribution of IATs and the proposed \mll fitting results (in red). Notice that \mll fits well. Further notice the consistency of the bi-modal (in-session, take-off) behaviors.}
\label{fig:PDF mll fits Reddit dataset}
\end{figure}

\begin{figure}[tb]
\centering
\subfigure{\includegraphics[width=0.15\textwidth]{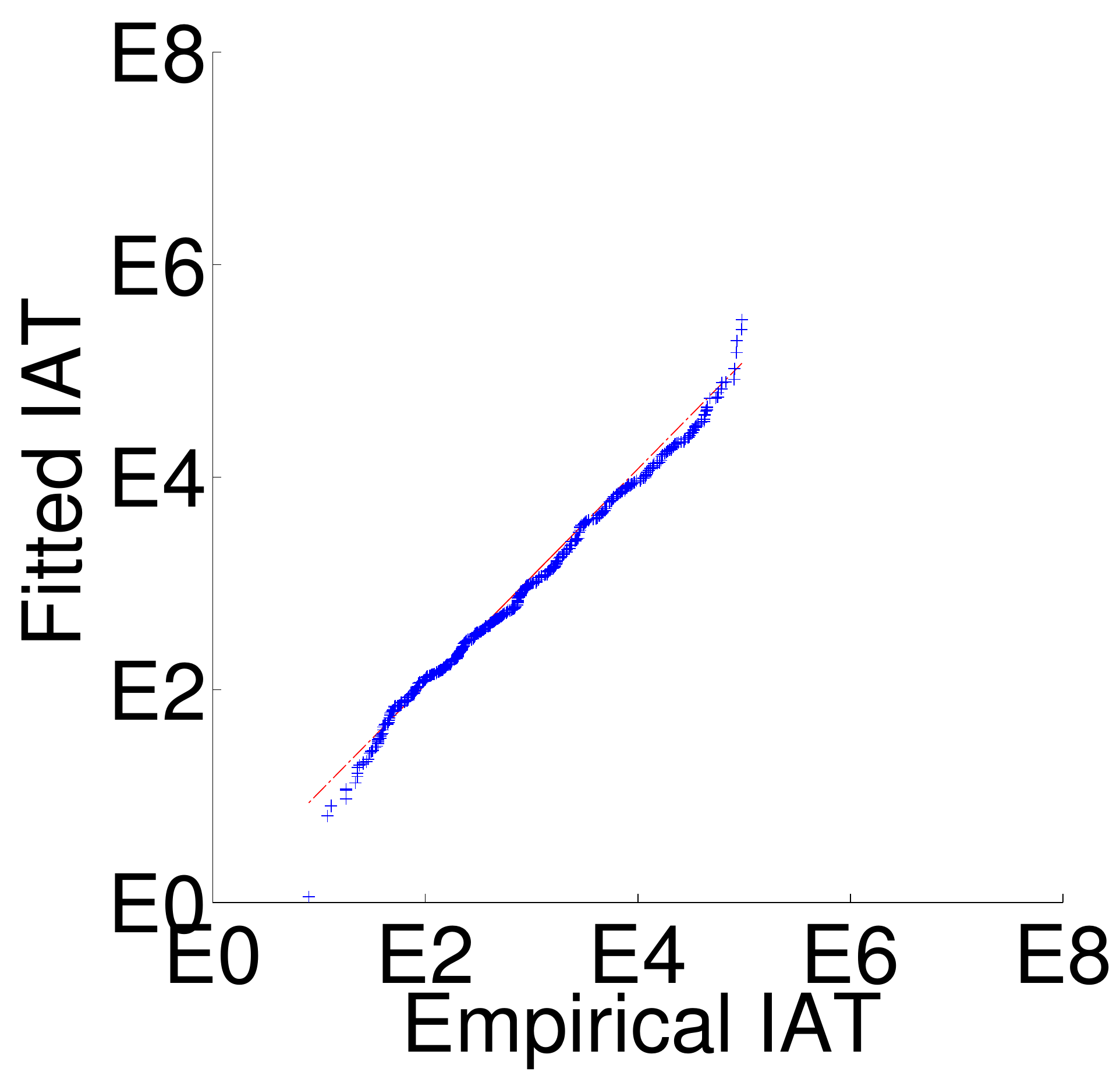}}
\subfigure{\includegraphics[width=0.15\textwidth]{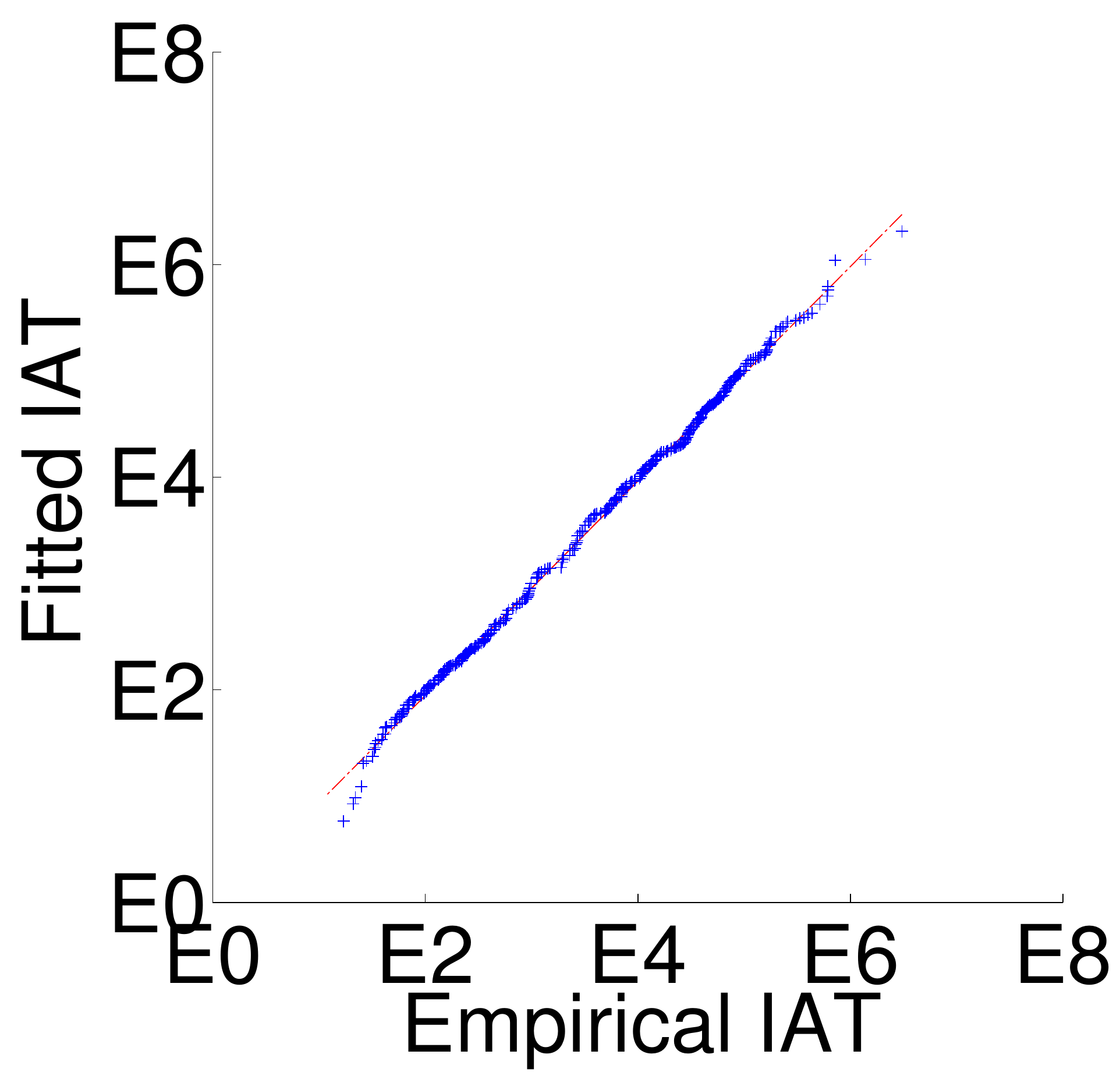}}
\subfigure{\includegraphics[width=0.15\textwidth]{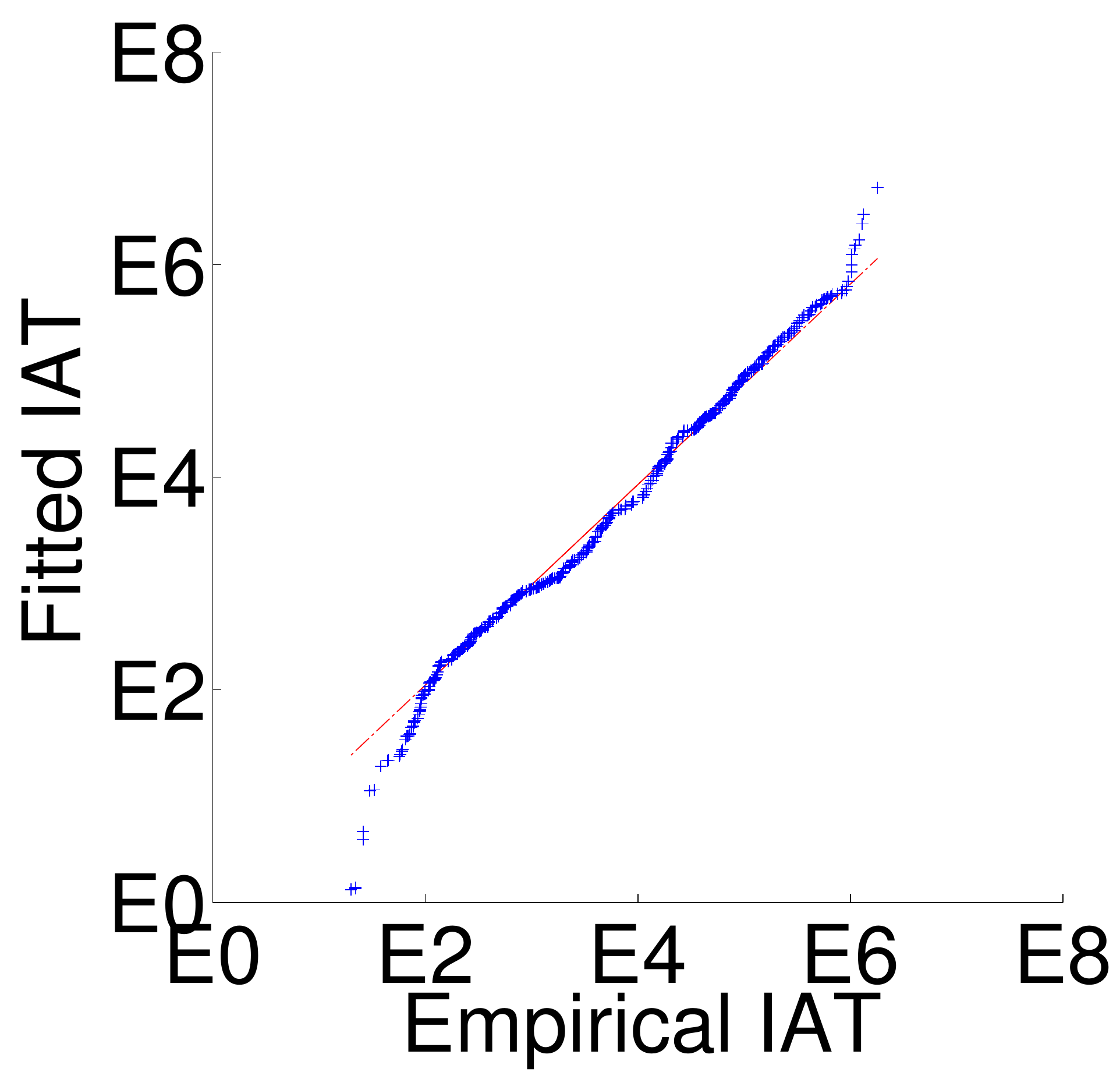}}
\subfigure{\includegraphics[width=0.15\textwidth]{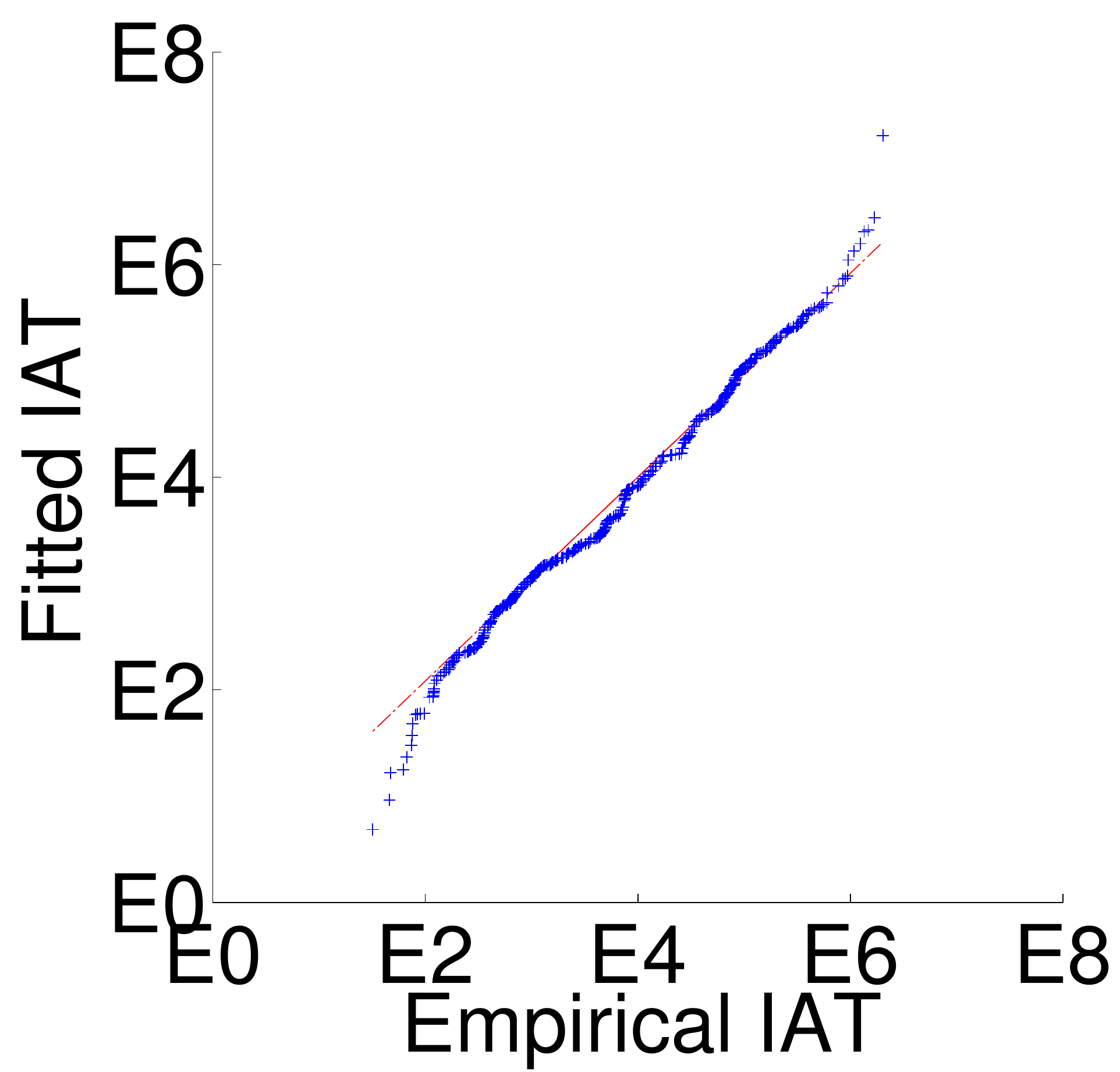}}
\subfigure{\includegraphics[width=0.15\textwidth]{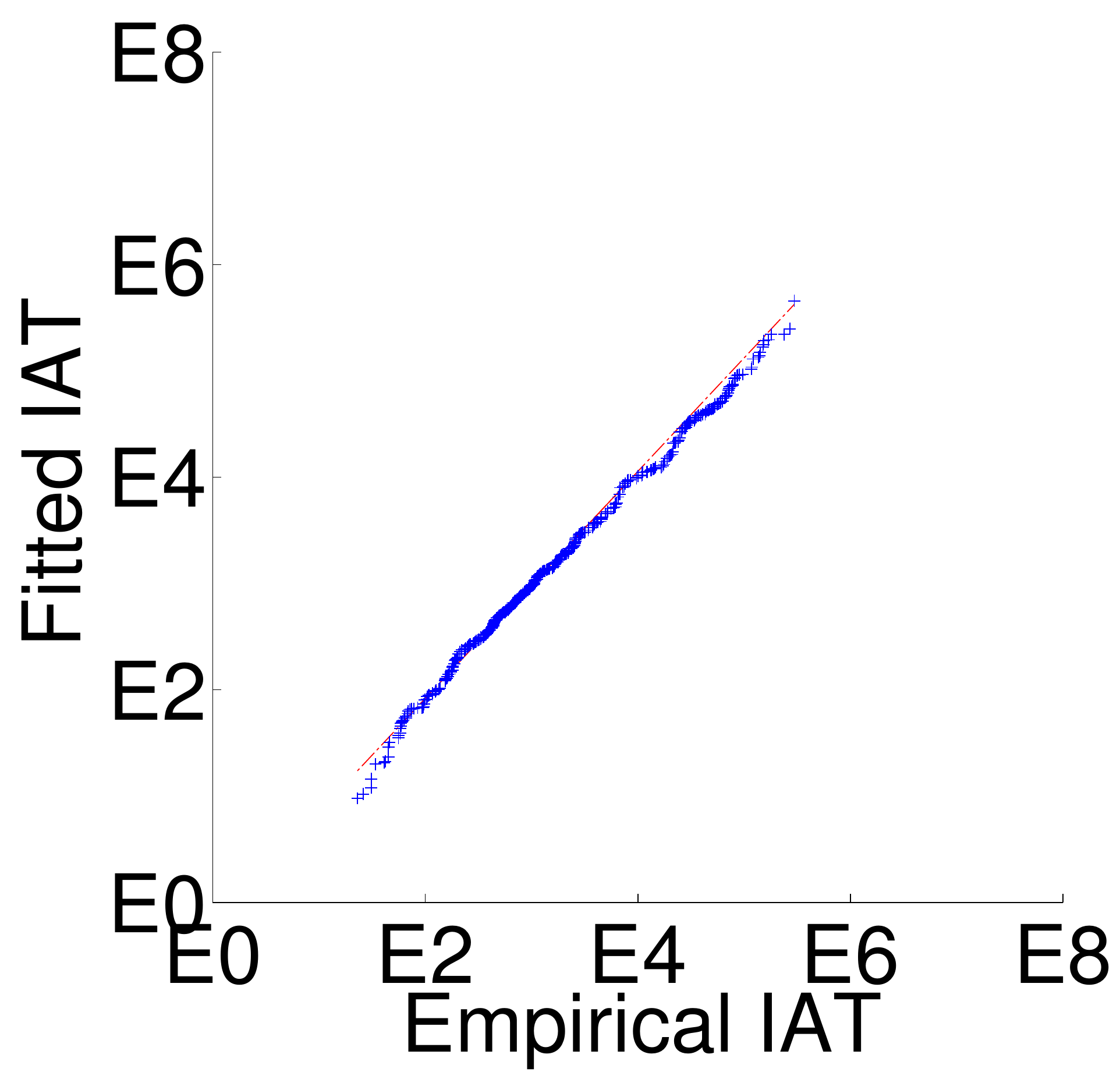}}
\subfigure{\includegraphics[width=0.15\textwidth]{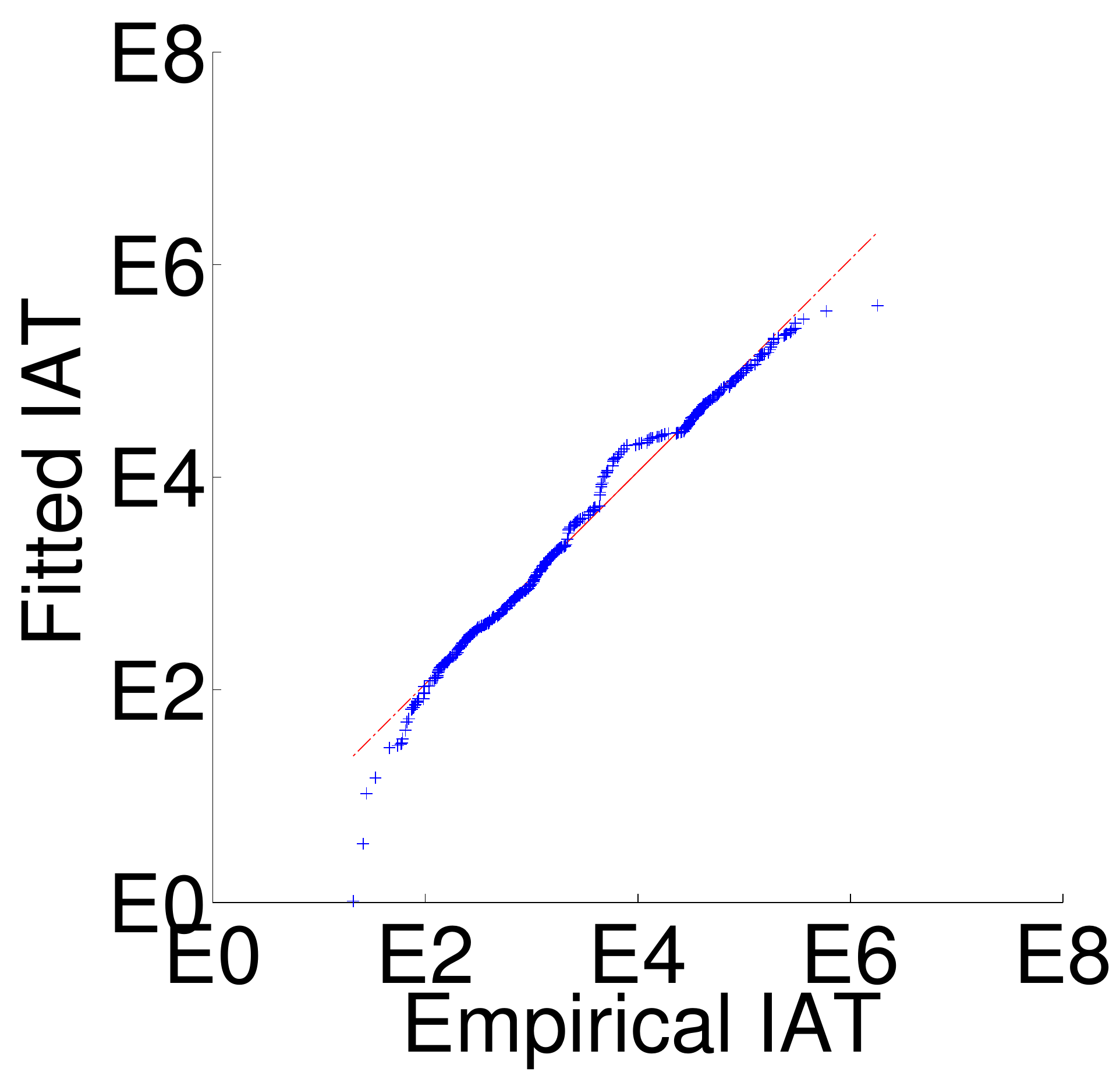}}
\subfigure{\includegraphics[width=0.15\textwidth]{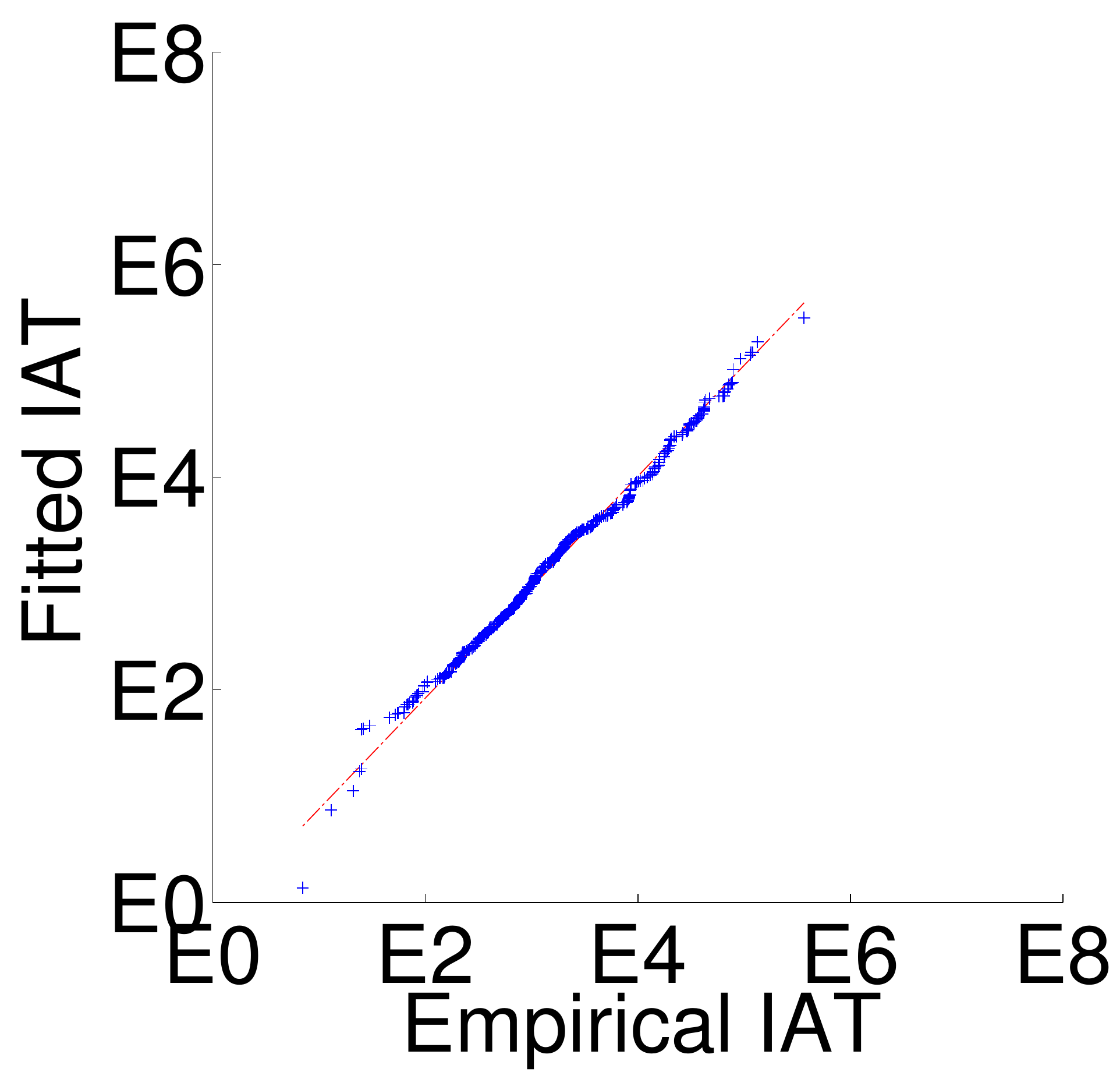}}
\subfigure{\includegraphics[width=0.15\textwidth]{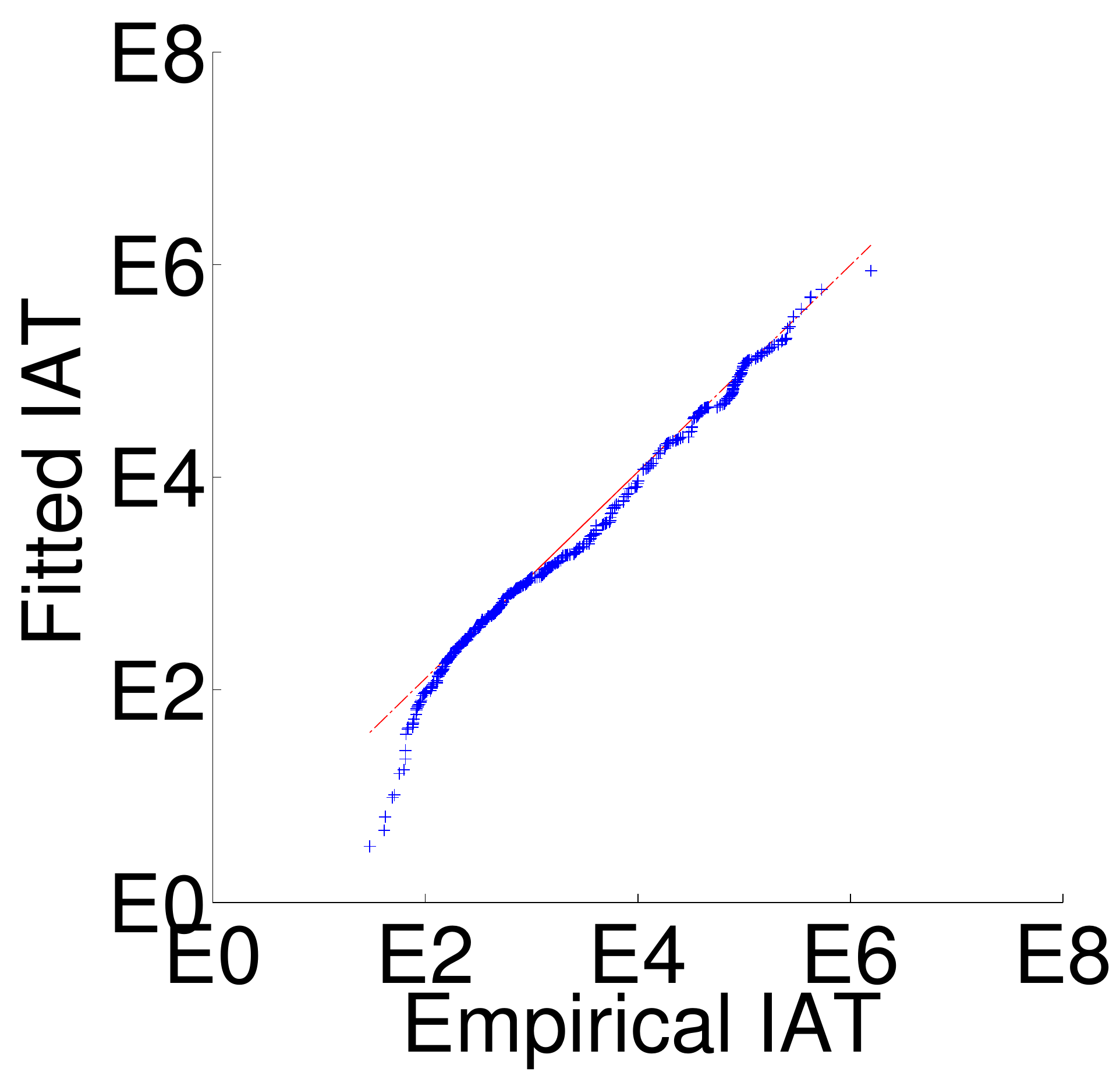}}
\subfigure{\includegraphics[width=0.15\textwidth]{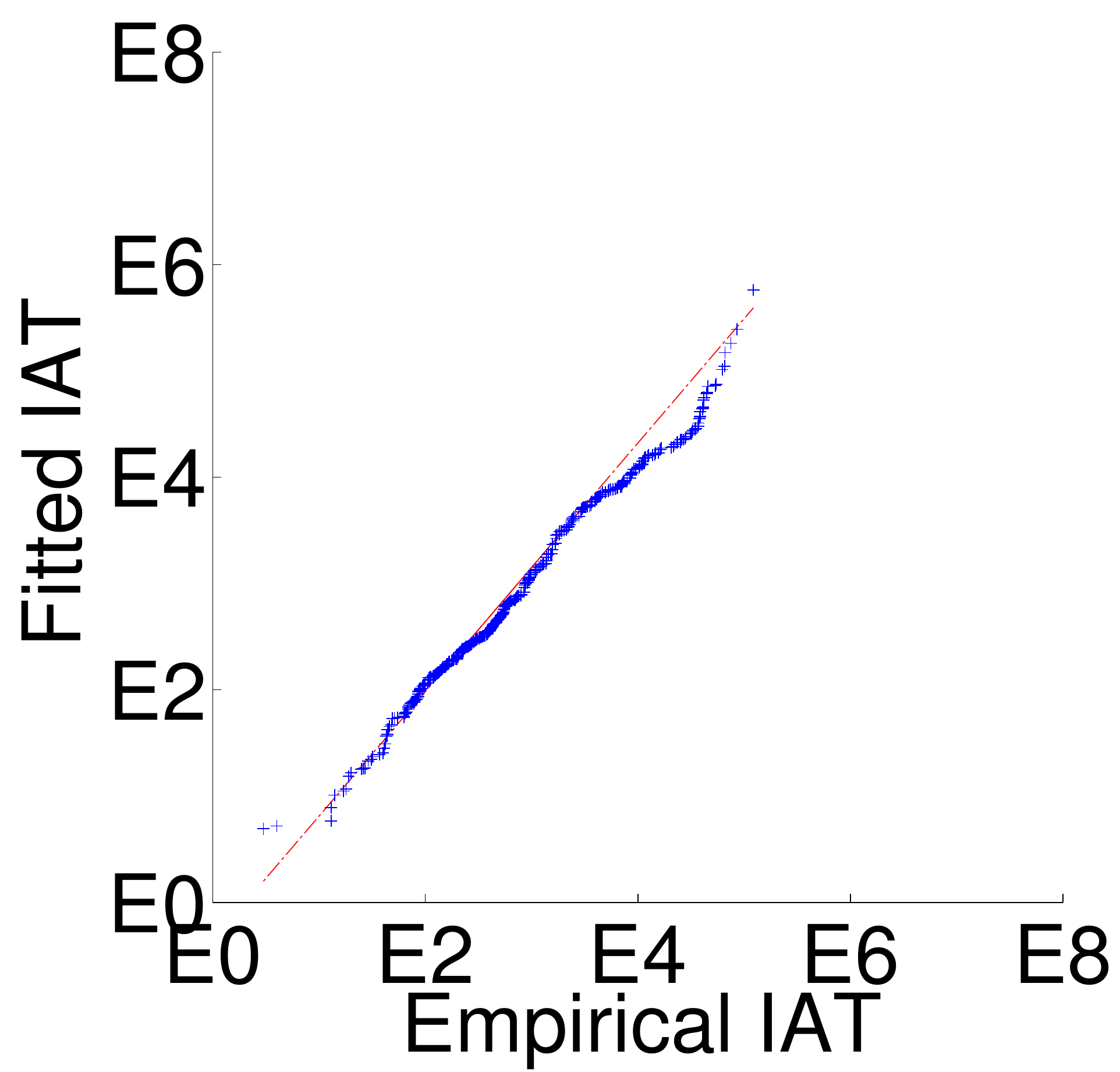}}
\subfigure{\includegraphics[width=0.15\textwidth]{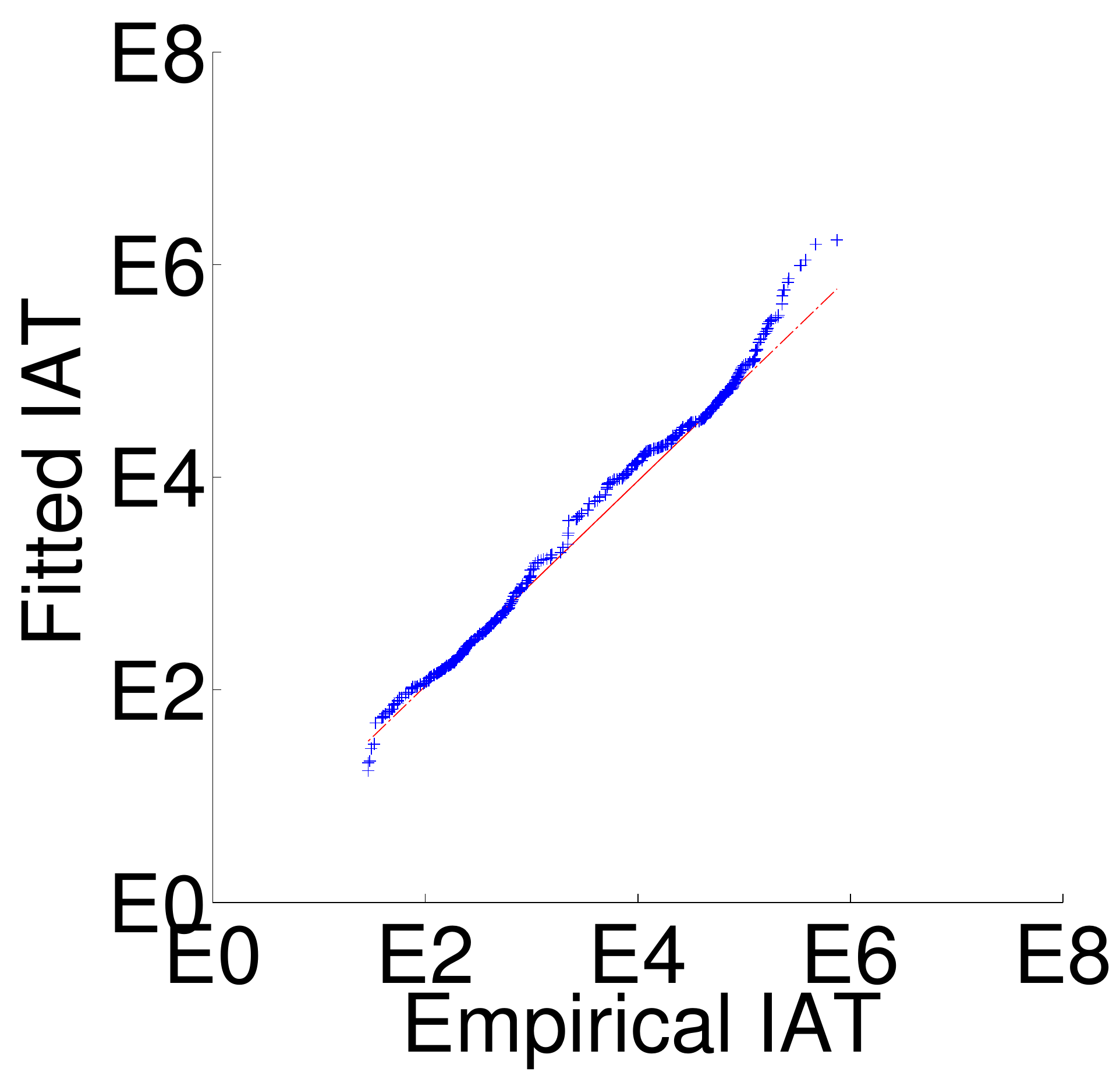}}
\subfigure{\includegraphics[width=0.15\textwidth]{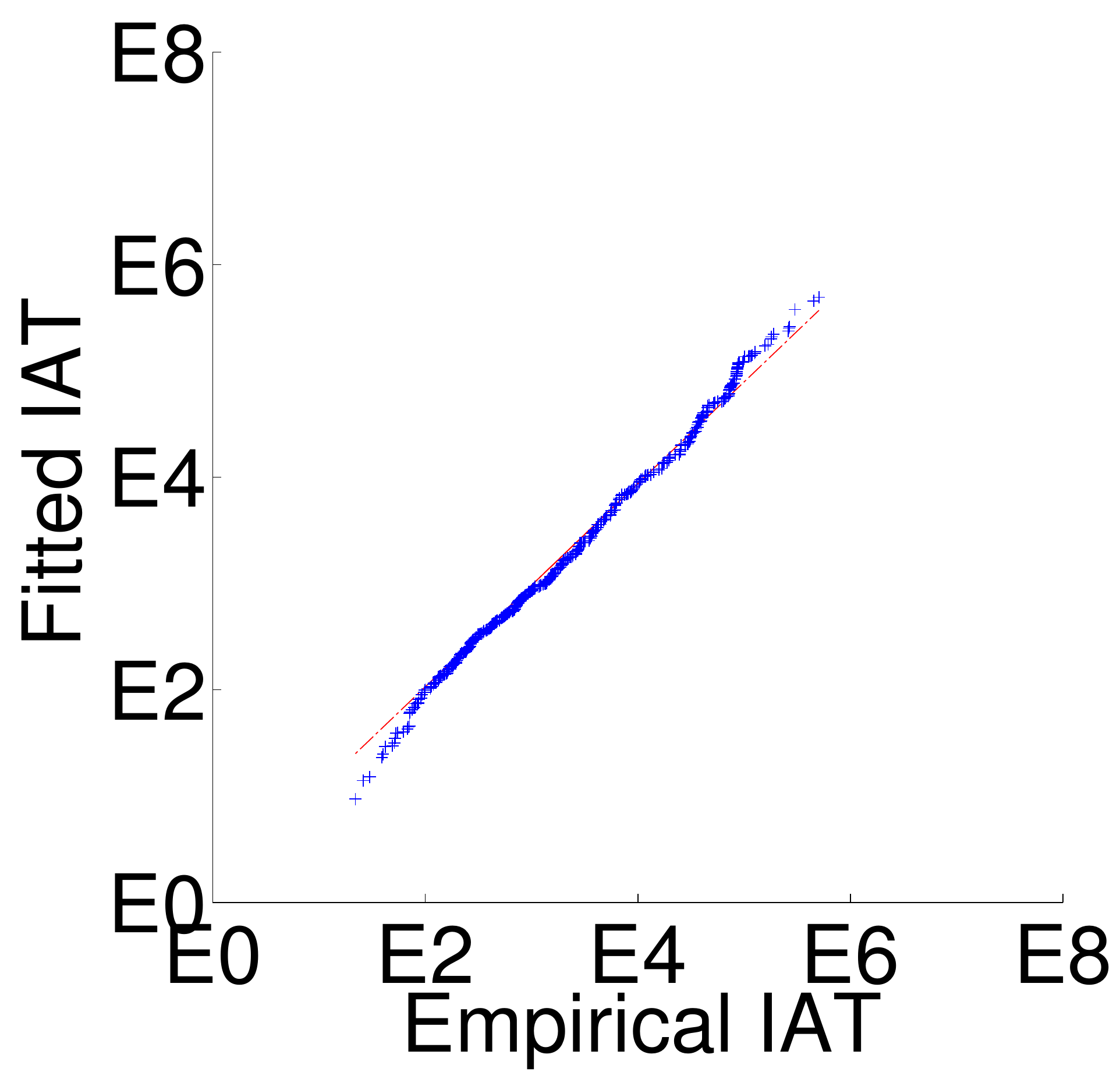}}
\subfigure{\includegraphics[width=0.15\textwidth]{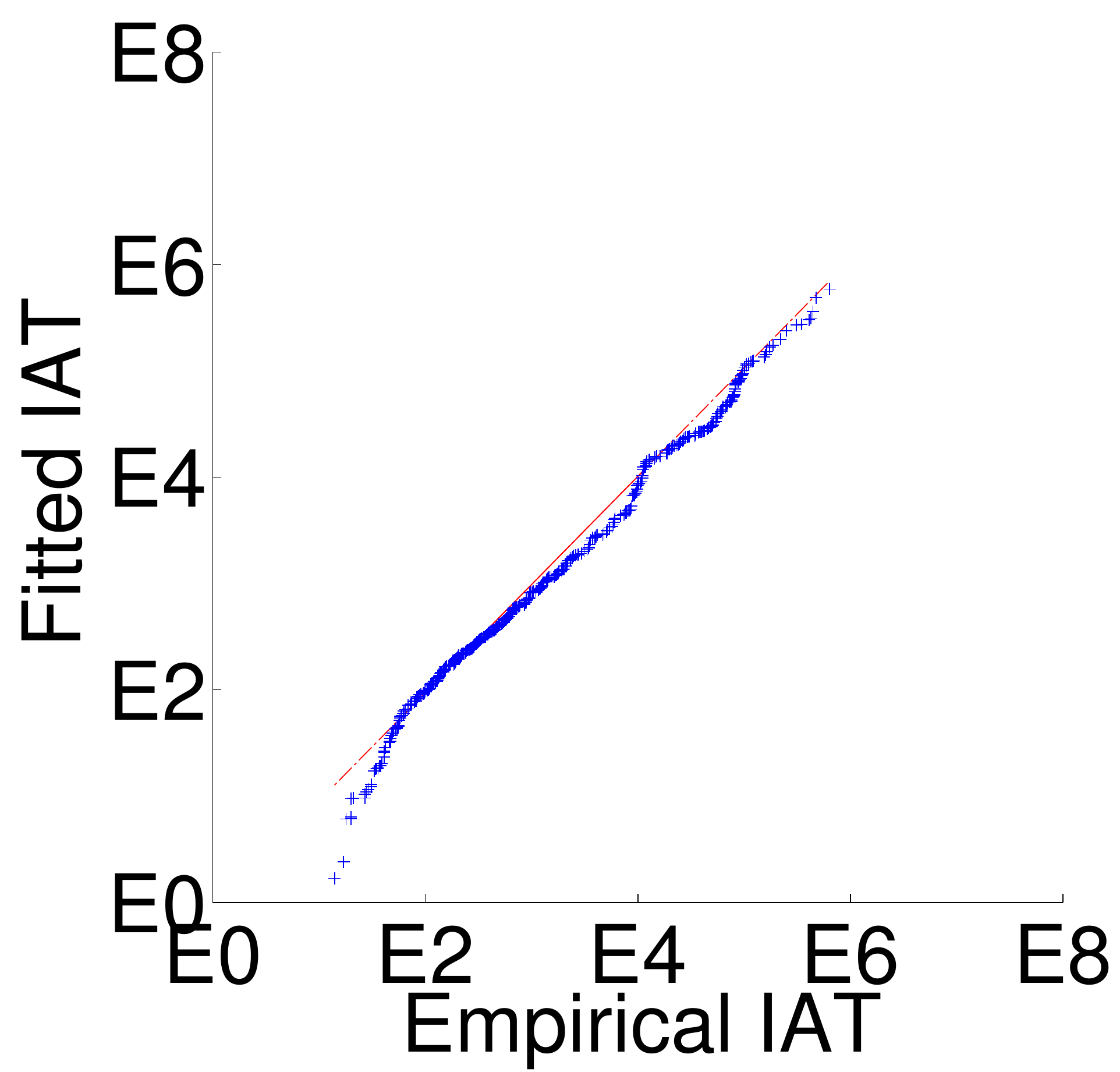}}
\caption{\textit{\mll fits the Reddit dataset (Q-Q plot)}. Each sub-figure shows the Q-Q plot (ideal: 45$^\circ$ line) between the real data and the samples randomly drown from the fitted \mll. Notice that the majority of quantiles match very well.}
\label{fig:QQplot mll fits Reddit dataset}
\end{figure}
Starting immediately, we evaluate the proposed \mll on modeling the IAT from the Reddit dataset\footnote{The dataset contains 16,927 unique users; for each user, we collect the timestamp of 500 his/her posts.}. Figure~\ref{fig:PDF mll fits Reddit dataset} shows 12 typical users behaviors and the \mll fits. Notice that (a) the \mll fits the marginal distribution well, and (b) the consistency of the bi-modal (in-session, take-off) behaviors. Here, the median of in-session IAT is is approximately nine minutes, whereas the median of take-off IAT is around 10 hours. Recall in the Observation~\ref{ob:in-session and take-off} (for web queries), the median of in-session IAT is about five minutes, whereas the median of take-off IAT is approximately seven hours. This makes sense, since compared to web queries, (a) each post/comment on Reddit requires few more minutes to compose (longer in-session IAT); (b) people post on Reddit less frequently (longer take-off IAT).

Figure~\ref{fig:QQplot mll fits Reddit dataset} also shows that \mll fits the Reddit dataset well by Q-Q plot. Notice that the majority of quantiles match very well. Therefore, the generality of the proposed \mll is demonstrated: \mll fits and explains multiple datasets (both Google queries and Reddit posts).

Since \mll characterizes each user's search behavior by five parameters, we ask: how to use these parameters, specifically the ratio ($\Ratio$) and the log-median ($\Mean$), to detect anomalies as Figure~\ref{fig:typical_behavior}(c) shows?

%% file: cop.tex
\pdfoutput=1

Are there regularities, in the parameters of all the users? It turns out that yes, some of the parameters are correlated. The two that show a stronger correlation are the ratio $\Ratio$ ($\triangleq \frac{\theta}{1-\theta}$) and the log-median $\Mean$ ($\triangleq\log(\alpha_{IN})$). Thus, our goal is to model the joint distribution.

Jumping ahead, given that both their marginals follow \LL (see Section~\ref{subsec:Marginal distribution of R and M}), how should we combine them, to reach a joint distribution that models Figure~\ref{fig:typical_behavior}(c)? The main idea is to use a powerful statistical tool, Copulas (see Section~\ref{subsec:A crash introduction to Copulas}). For convenience, the final CDF of the proposed \cop (details in Section~\ref{subsec:Proposed cop}) is provided here:
\begin{eqnarray}
 & &F_{\cop}(r,m;\Ken, \alpha_{\Ratio}, \beta_{\Ratio}, \alpha_{\Mean}, \beta_{\Mean}) \nonumber \\
 &=& e^{-([\log(1+(r/\alpha_{\Ratio})^{-\beta_{\Ratio}})]^\Ken+[\log(1+(m/\alpha_{\Mean})^{-\beta_{\Mean}})]^\Ken)^{1/\Ken}} \nonumber 
\end{eqnarray}

\subsection{Marginal distribution of $\Ratio$ and $\Mean$}
\label{subsec:Marginal distribution of R and M}

\begin{figure*}[tb]
\centering
\subfigure[Marginal distribution of R (lin. scale)]{
    \includegraphics[width=0.28\textwidth]{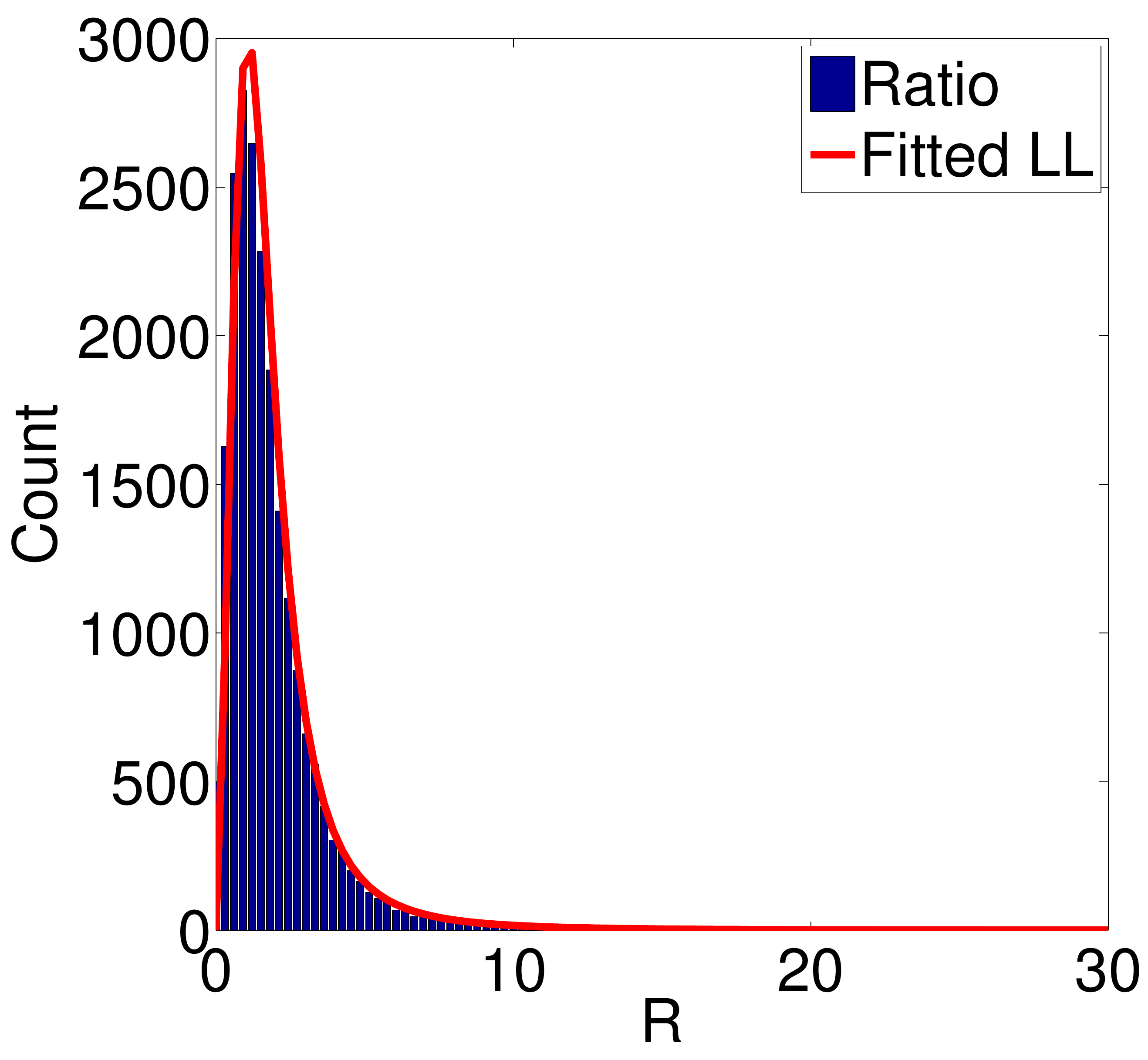}
}
\subfigure[Q-Q plot of R (log scale)]{
    \includegraphics[width=0.28\textwidth]{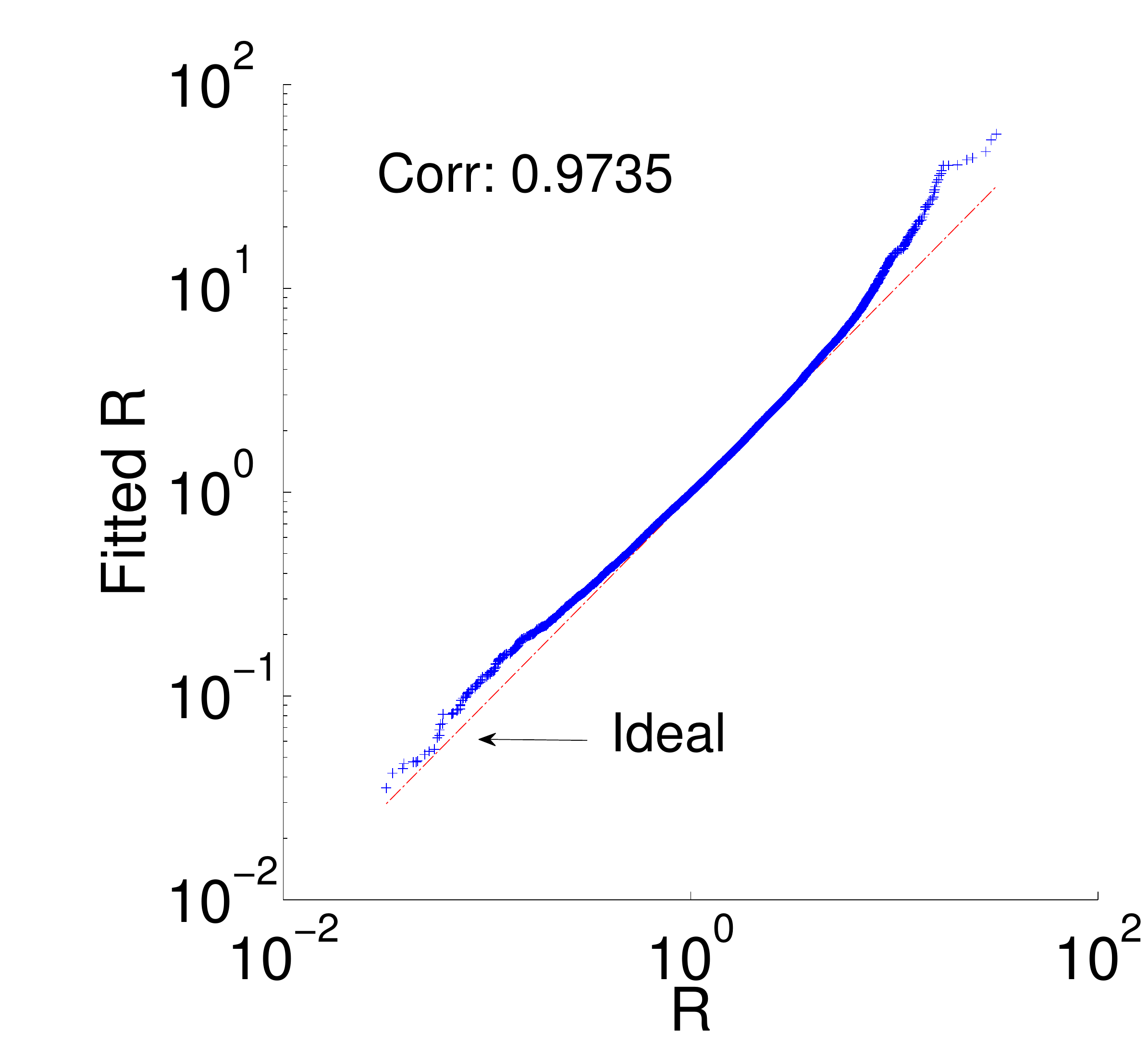}
}
\subfigure[Odds Ratio of R]{
    \includegraphics[width=0.28\textwidth]{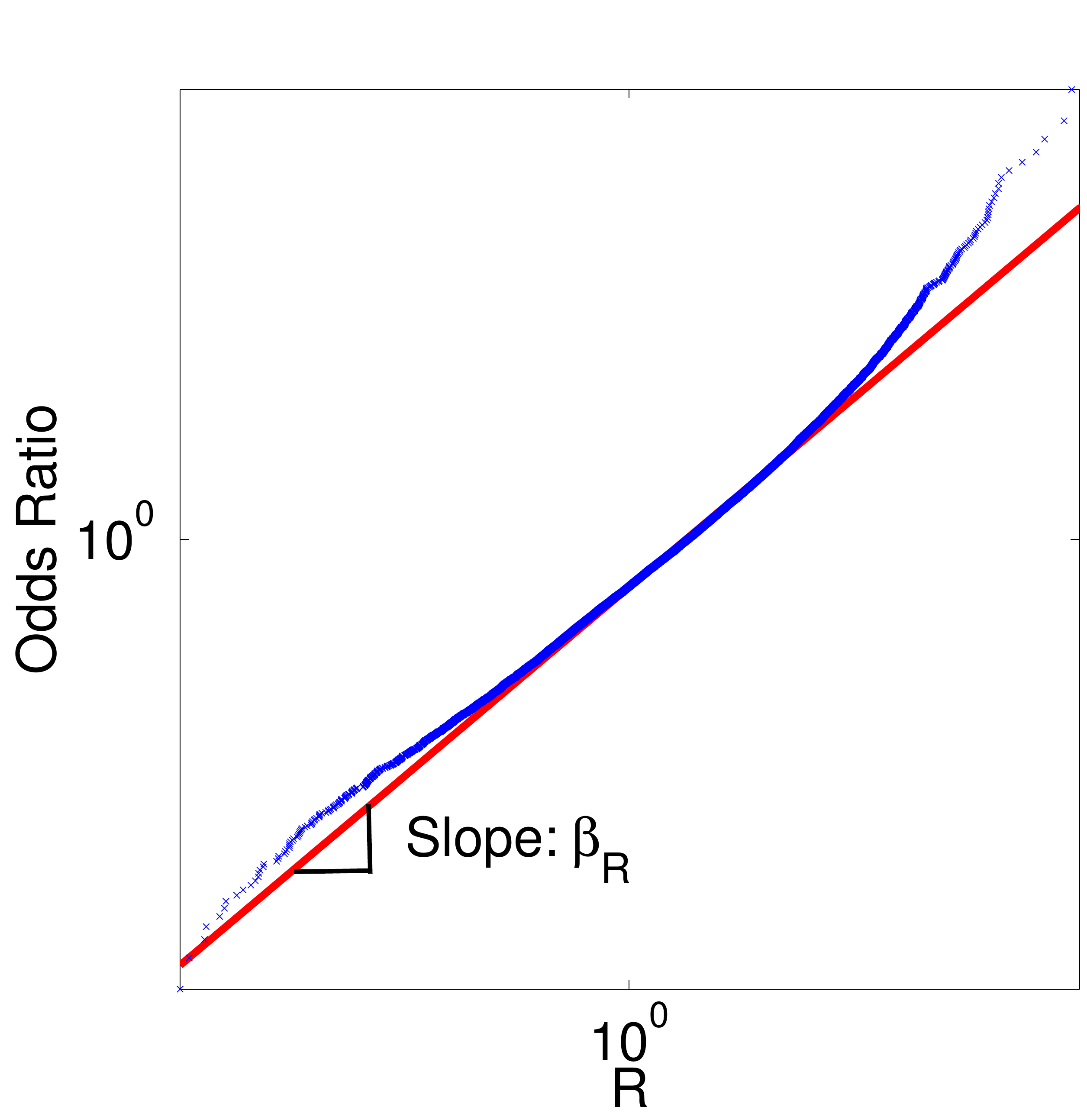}
}
\subfigure[Marginal distribution of M (lin. scale)]{
    \includegraphics[width=0.28\textwidth]{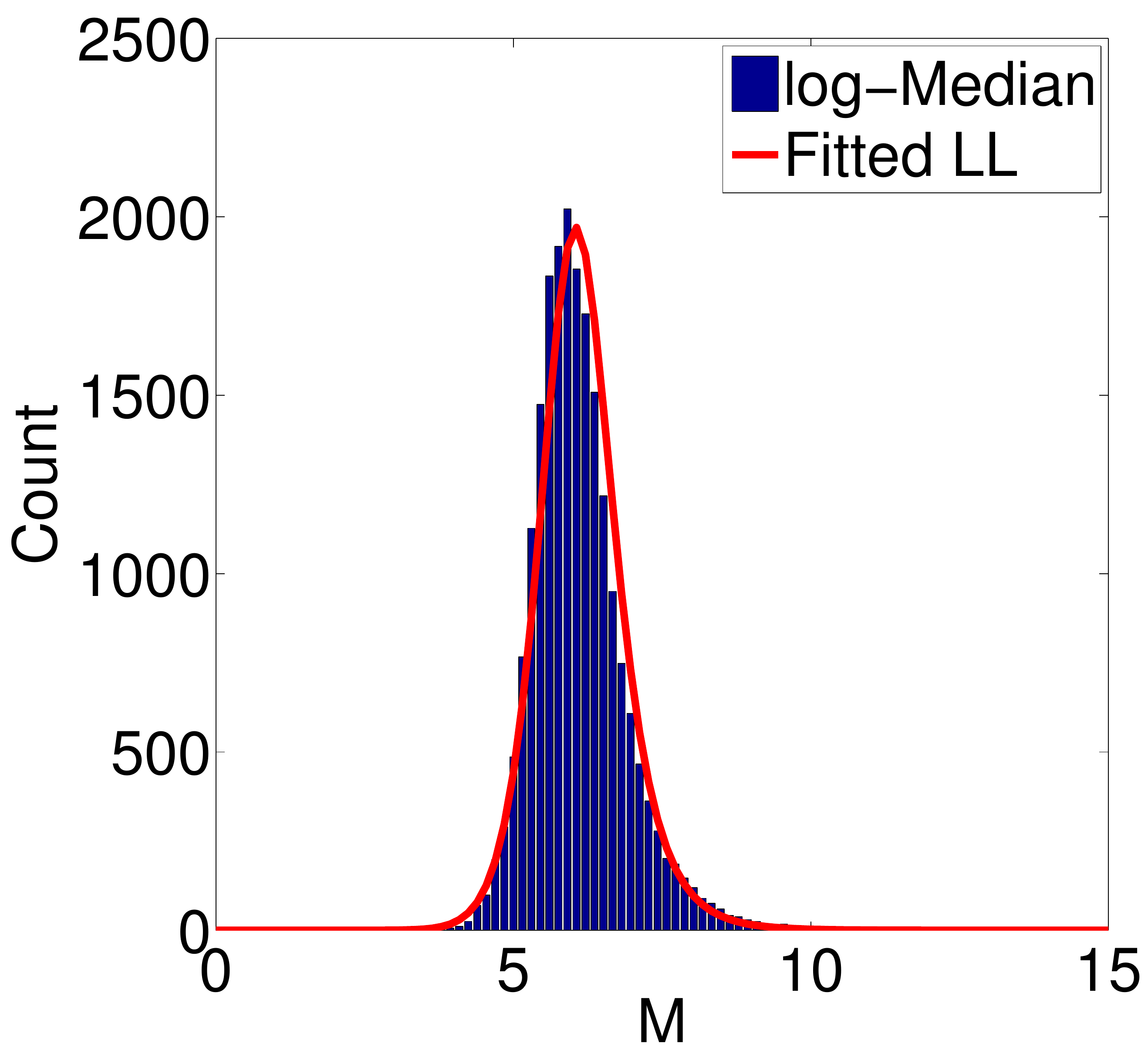}
}
\subfigure[Q-Q plot of M (log scale)]{
    \includegraphics[width=0.28\textwidth]{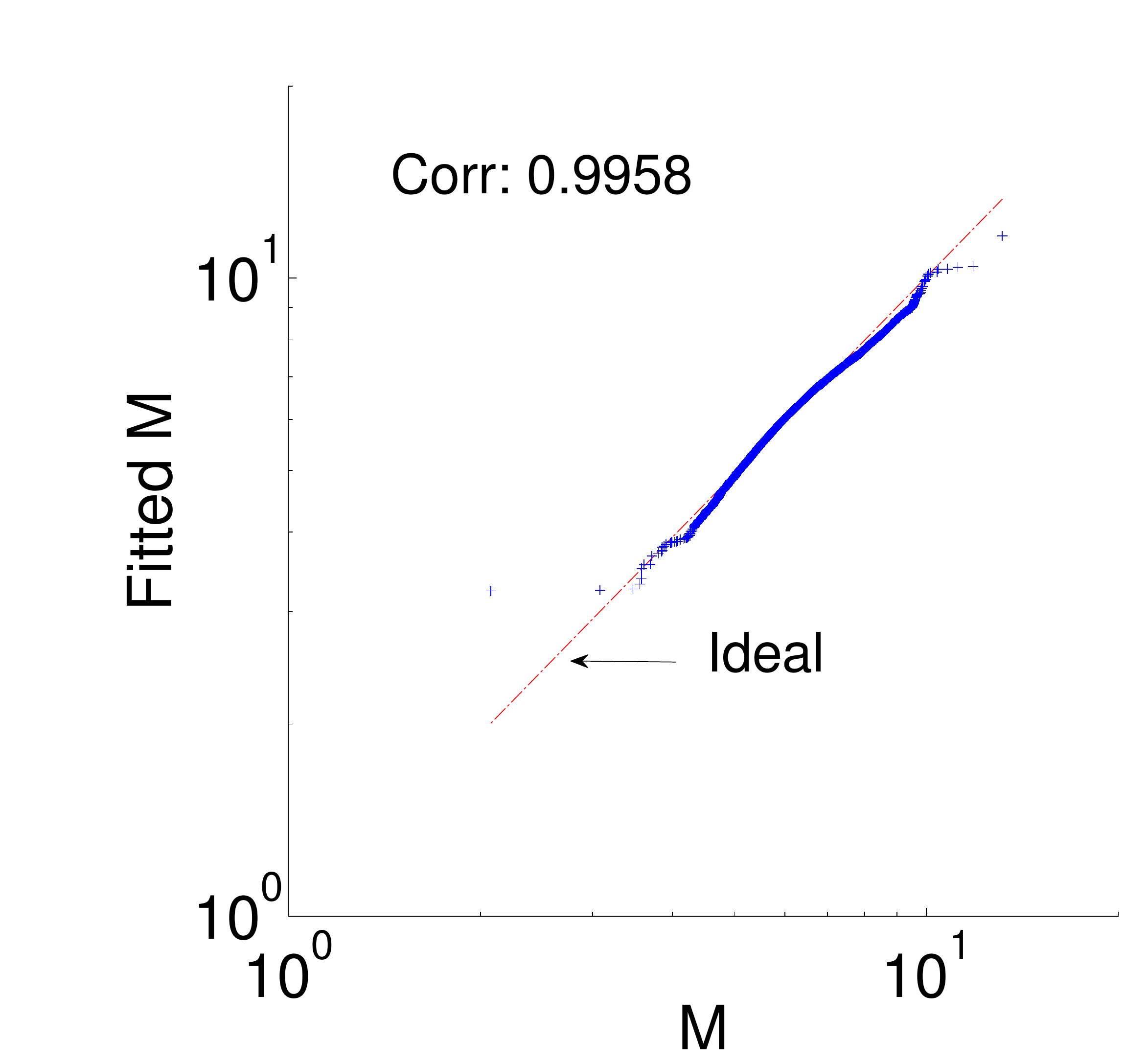}
}
\subfigure[Odds Ratio of M]{
    \includegraphics[width=0.28\textwidth]{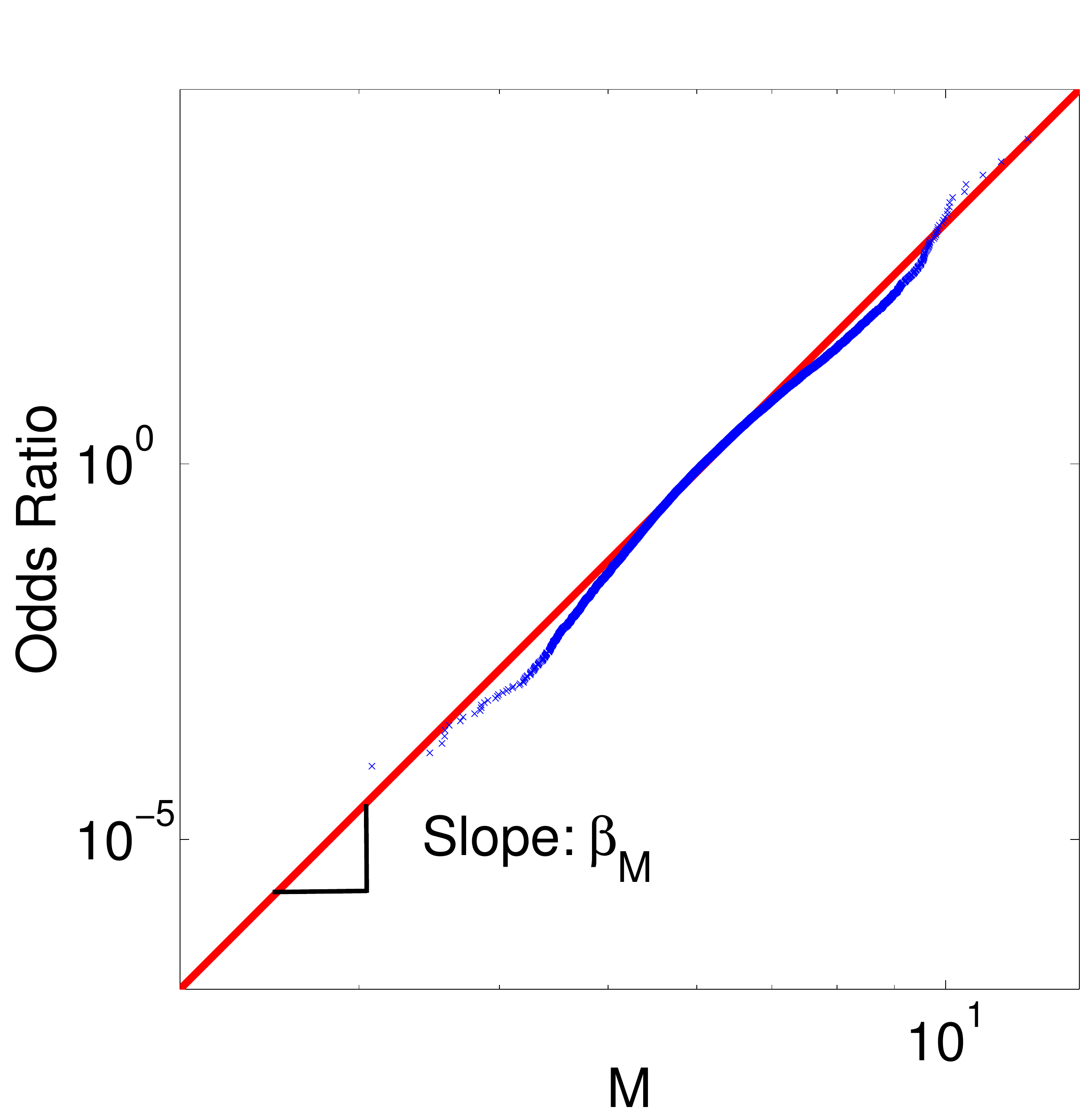}
}
\caption{\textit{Marginal distributions follow \LL distributions}: (a) Marginal distribution of $\Ratio$ and the \LL fitting. (b) Q-Q plot between empirical $\Ratio$ and fitted \LL. (c) Odds Ratio (OR) between empirical $\Ratio$ and fitted \LL. (d)(e)(f) provide the corresponding plots for $\Mean$. In (c), the OR of $\Ratio$ seems to entirely follow the linear line, which serves as another evidence that its marginal distribution follows a \LL. The same statement also holds for (d). K-S tests are conducted for both $\Ratio$ and $\Mean$; under the 95\% confidence level, we retain the null hypothesis: the empirical data follows the fitted \LL.}
\label{fig:Validation of marginal distribution of parameter}
\end{figure*}

With the parameters extracted by \mll (specifically, $\theta$ and $\alpha_{\IN}$ for each user), we define two random variables that are particularly useful for anomaly detection:
\bit
	\item \textit{Ratio}: $\Ratio \triangleq \theta/(1-\theta)$ that represents approximately how many ``query and click''s happening within a search session (in-session) v.s. take-off.
	\item \textit{Log-median}: $\Mean \triangleq \log(\alpha_{\IN})$ represents the median of in-session IAT in log scale.	
\eit

Intuitively, $\Ratio$ and $\Mean$ represent an aggregate behavior, in terms of a statistical distribution of parameters (specifically, $\theta$ and $\alpha_{\IN}$) used to characterize each user. Figure~\ref{fig:Validation of marginal distribution of parameter} illustrates the marginal distribution of $\Ratio$ in (a) and $\Mean$ in (d), respectively. Note that all the \LL fittings are done by using Maximum Likelihood Estimate (MLE).

To better examine the distribution behavior both in the head and tail, we propose to use the Odds Ratio (OR) function.
\begin{lemma}[Odds Ratio] 
\label{OR_func}
In logarithmic scale, $OR(t)$ has a linear behavior, with a slope $\beta$ and an intercept ($-\beta\log\alpha$), if $T$ follows Log-logistic distribution.
 From the definition of OR function, we have:
	\begin{align}
		Odds Ratio(t) &= OR(t) = \frac{F_T(t)}{1-F_T(t)} = \bigg(\frac{t}{\alpha}\bigg)^\beta \\ \nonumber
		&\Rightarrow \log OR(t) = \beta\log(t) - \beta\log\alpha \hspace{12pt} \blacksquare	
	\end{align}
\end{lemma}

Figure~\ref{fig:Validation of marginal distribution of parameter}(c)(f) show the OR of $\Ratio$ and $\Mean$, respectively. For both random variables, their ORs seem to entirely follow the linear line, which serves as another evidence that their marginal distributions follow \LL. K-S tests are also conducted for both $\Ratio$ and $\Mean$; under 95\% confidence level, we retain the null hypothesis: $\Ratio$ (and $\Mean$) follows the fitted \LL.

\begin{observation}[Common user behavior]
The mode of the ratio $\Ratio$ is approximately three, which suggests a common user behavior: ``click-click-click$-$taken off$-$then click (new session).''
\end{observation}

The marginals of $\Ratio$ and $\Mean$ follow \LL, but how about their two-dimensional joint distribution ($F_{\Ratio,\Mean}$)? Can we use a multivariate normal (MVN) distribution to describe them?

\subsection{Why not multivariate normal (MVN)?}

Modeling multivariate distribution is a rather challenging task. One popular method is to use a multivariate normal (MVN) distribution. However, we provide four reasons against the use of MVN in modeling the joint distribution of $\Ratio$ and $\Mean$:
\bit
	\item Marginals are not Normal. As shown in Section~\ref{subsec:Marginal distribution of R and M}, the marginals of $\Ratio$ and $\Mean$ follow \LL, as opposed to MVN's marginals being normally distributed.
	\item Contour of covariance is not an ellipsoid. As shown in Figure~\ref{fig:typical_behavior}(c) and later in Fig~\ref{fig:Gumbel_with_diff_Tau}(d), the contour of $\Ratio$ and $\Mean$ do not follow MVN's ellipsoid contour.
	\item MVN models negative values. The support of MVN includes negative values whereas both $\Ratio$ and $\Mean$ are non-negative. 
	\item Low log-likelihood. The log-likelihood of MVN is an order magnitude lower than the log-likelihood achieved by proposed \cop distribution.
\eit

We ask: is there any other candidate that models a multivariate distribution, with marginals following \LL? The short answer is \textit{yes}: the proposed \cop by using Gumbel Copula.

\subsection{A crash introduction to Copulas}
\label{subsec:A crash introduction to Copulas}

In statistics, Copulas are widely-used to model a multivariate, joint distribution considering the dependency structures between random variables (\eg, $\Ratio$ and $\Mean$). The main concept of Copulas is to associate univariate marginals (\eg, $F_{\Ratio}, F_{\Mean}$) with their full multivariate distribution. Here, we remind the mathematical definition of copula as below:
\begin{definition}[Copula]
A copula $\Copula (u,v)$ is a dependence function defined as:
\begin{equation}
	\Copula : [0,1]\times[0,1]\rightarrow [0,1]
\label{eq:copula}
\end{equation}
\noindent Given two random variables $\Ratio$, $\Mean$ and their marginal CDFs $F_{\Ratio}$, $F_{\Mean}$, a copula $\Copula (u,v)$ generates a joint CDF that captures the correlation between $\Ratio$ and $\Mean$: $F_{\Ratio,\Mean}(\ratio,\mean) = \Copula(F_{\Ratio}(\ratio),F_{\Mean}(\mean))$.
\end{definition}
In theory, Copulas can capture any type of dependency between variables: positive, negative, or independence. The existence of such Copula is guranteed by Sklar's Theorem\footnote{The details of Sklar's theorem can be found in \cite{schweizer2011probabilistic}.}. 

One type of Copulas is very popular in modeling joint distribution of random variables with heavy tails: \textit{Gumbel Copula}. We remind the definition of Gumbel Copula as below: 
\begin{definition}[Gumbel Copula] A Gumbel Copula is defined as:
	\begin{equation}
	\Copula (u,v) = e^{ -[\phi(u)^\Ken + \phi(v)^\Ken]^{1/\Ken} }	
	\end{equation}
\noindent where $\Ken \geq 1$ and $\phi(\cdot) = -\log(\cdot)$. 
\end{definition}
\noindent Notice that $\Copula(u,v) = u\cdot v$ when $\Ken = 1$, indicating that $u, v$ are independent.

With this tool, we are ready to proceed to the proposed \cop.

\subsection{Proposed \cop}
\label{subsec:Proposed cop}

\begin{figure}[tb]
\centering
\subfigure[$\Ken = 1$]{
    \includegraphics[width=0.45\columnwidth]{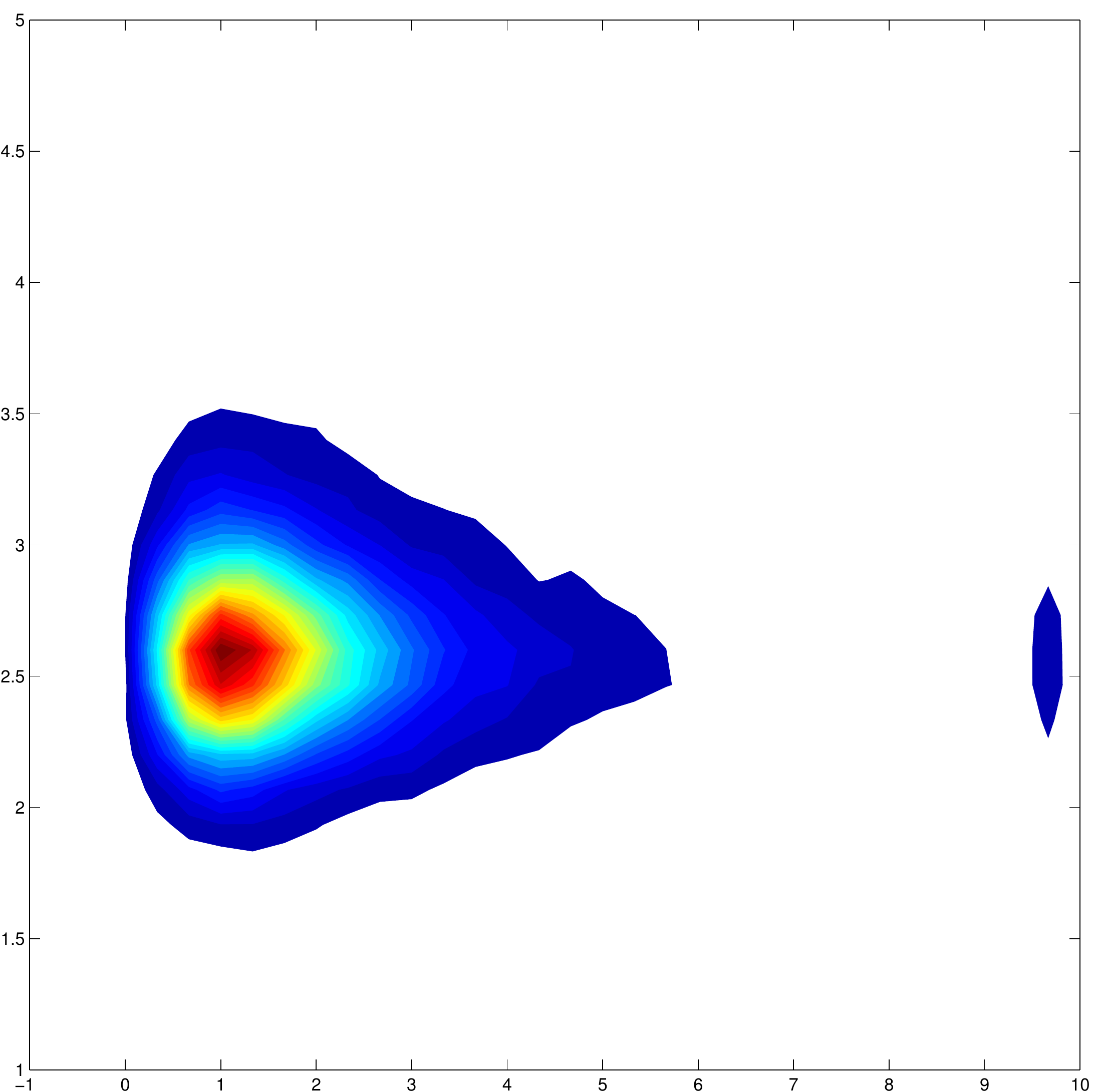}
}
\subfigure[$\Ken = 1.12$]{
    \includegraphics[width=0.45\columnwidth]{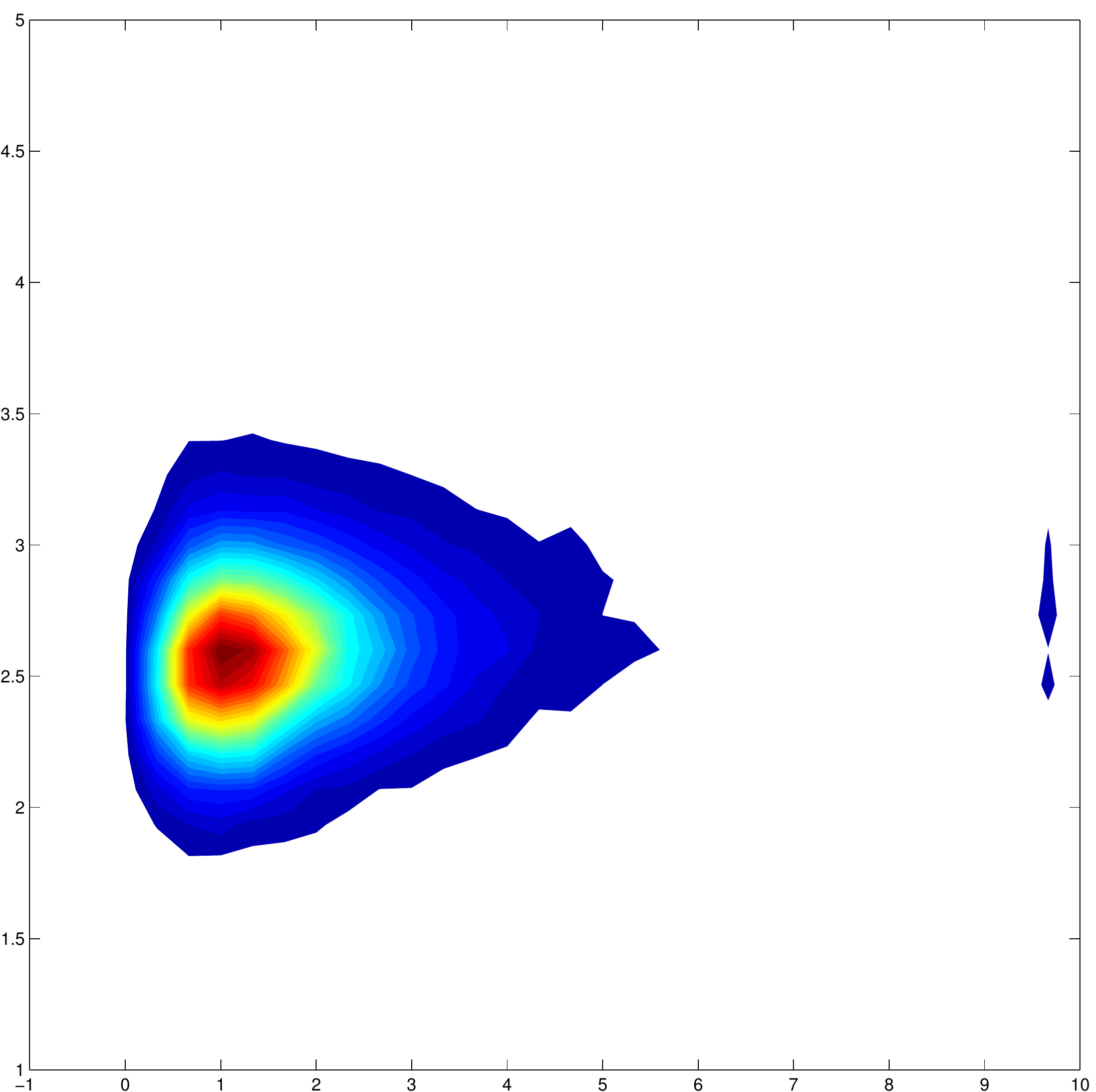}
}
\subfigure[$\Ken = 1.3$]{
    \includegraphics[width=0.45\columnwidth]{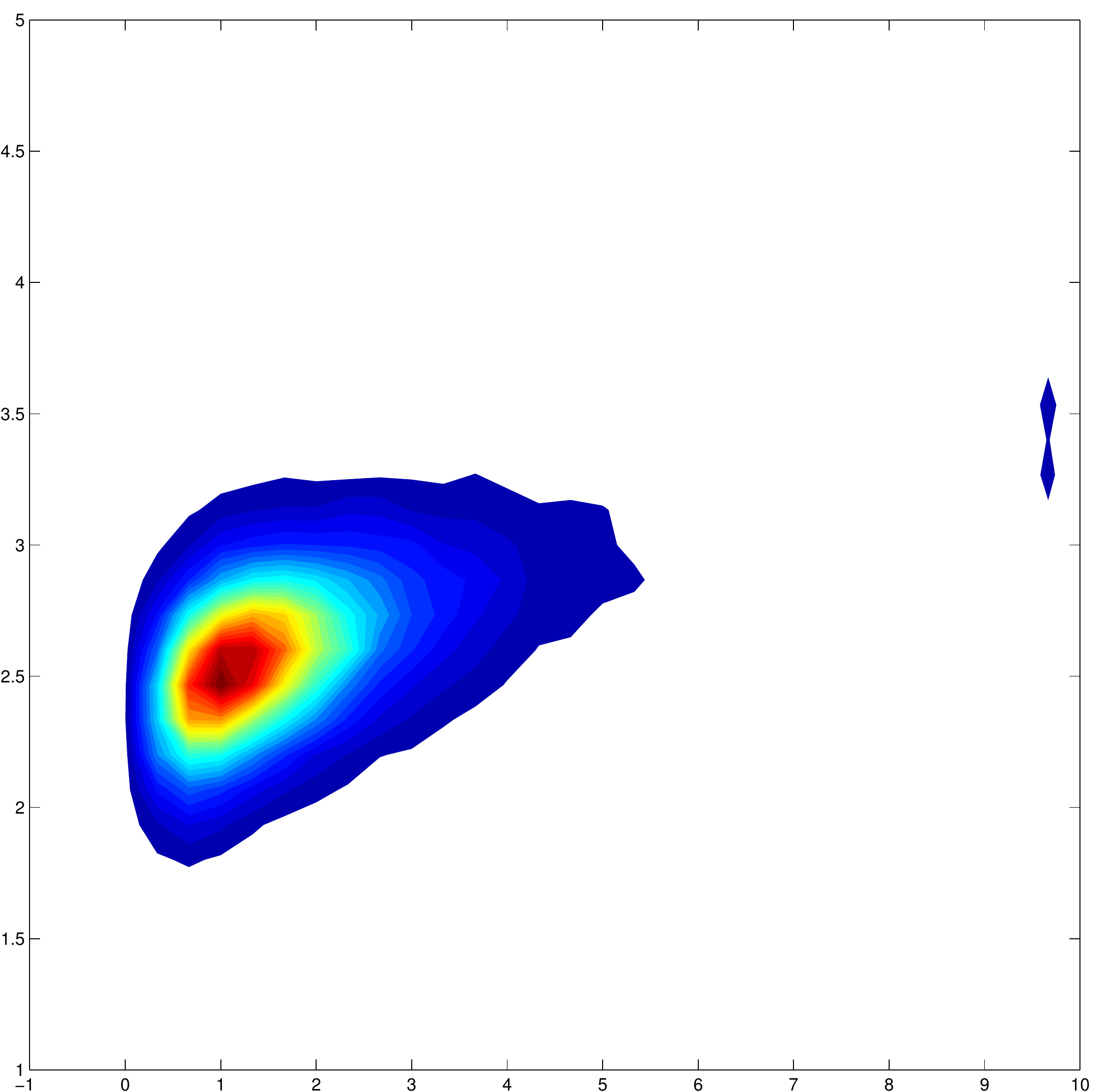}
}
\subfigure[Real data]{
    \includegraphics[width=0.45\columnwidth]{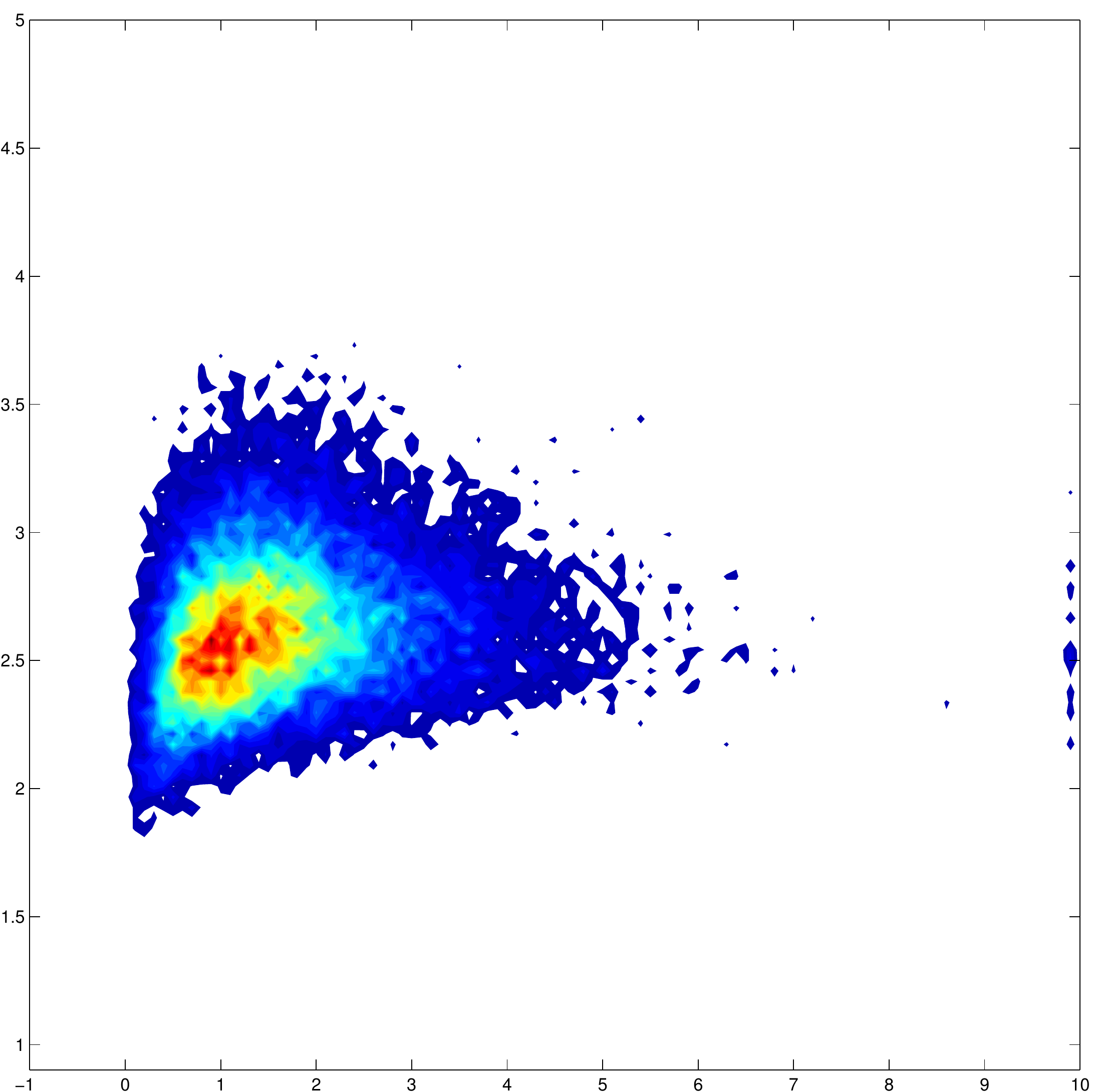}
}
\caption{\textit{\cop matches real data}. (a)-(c): contour plots for \cop (with various $\Ken$). (d): real data. All plots are $\Ratio$ v.s. $\Mean$. In (b), $\Ken = 1.12$, which is the value estimated from the real data. Notice how well (b) matches (d).}
\label{fig:Gumbel_with_diff_Tau}
\end{figure}

The goal of \cop is to model the joint distribution of $\Ratio$ and $\Mean$. As the results presented in Section~\ref{subsec:Marginal distribution of R and M}, their marginals follow \LL. By using Gumbel Copula, we present the definition of the proposed \cop here:
\begin{definition}[\cop] Let $\Ratio$ and $\Mean$ be non-negative random variables following \cop distribution, the CDF of their joint distribution is:
\begin{eqnarray}
 & & F_{\cop}(r,m;\Ken,\alpha_{\Ratio}, \beta_{\Ratio}, \alpha_{\Mean}, \beta_{\Mean}) \nonumber \\
 &=& e^{-([\log(1+(r/\alpha_{\Ratio})^{-\beta_{\Ratio}})]^\Ken+[\log(1+(m/\alpha_{\Mean})^{-\beta_{\Mean}})]^\Ken)^{1/\Ken}} 
\label{eq:cop}
\end{eqnarray}
\noindent where $\ratio, \mean \ge 0$, $\Ken \ge 1$, ($\alpha_{\Ratio}, \beta_{\Ratio}$), ($\alpha_{\Mean}, \beta_{\Mean}$) are the hyper-parameters used in $F_{\mathcal{LL}}(\ratio)$ and $F_{\mathcal{LL}}(\mean)$, respectively.
\end{definition}
In this work, $\Ken$ in Eq\eqref{eq:cop} is estimated by Kendall tau correlation \cite{koutra2013patterns}; the values of ($\alpha_{\Ratio}, \beta_{\Ratio}$), ($\alpha_{\Mean}, \beta_{\Mean}$) are estimated by using MLE as mentioned in Section~\ref{subsec:Marginal distribution of R and M}.

We now show that the proposed \cop distribution preserves the characteristics in the marginal distributions of each random variable:
\begin{lemma}[Marginals of \cop are \LL]
We prove this by taking the limit of $r$ to infinity:
	\begin{eqnarray}
		& & \lim_{r \to \infty} F_{\cop}(r,m) \nonumber \\
		&=& F_{\Mean}(m;\alpha_{\Mean}, \beta_{\Mean}) \nonumber \\
		&=& \frac{1}{1+(m/\alpha_{\Mean})^{-\beta_{\Mean}}}
	\nonumber	
	\end{eqnarray}
Therefore, $\Mean \sim $ \LL($\alpha_{\Mean},\beta_{\Mean}$). We can show $\Ratio\sim$ $\mathcal{LL}(\alpha_{\Ratio},\beta_{\Ratio})$ in a similar manner.\hspace{12pt} 
\end{lemma}

Figure~\ref{fig:Gumbel_with_diff_Tau}(a)(b)(c) illustrate three contour plots of the proposed \cop with setting $\Ken$ to various values, whereas Figure~\ref{fig:Gumbel_with_diff_Tau}(d) provides the contour plot from the empirical data. The contour plot in (b) seems to match the empirical data qualitatively well.

%% file: practice.tex
\pdfoutput=1

We provide the step-by-step guide to apply the proposed \mmm for behavioral modeling and anomaly detection:
\bit
	\item {\bf \mll at user level}: given a user's IAT, use \mll to characterize their marginal IAT distribution with five parameters ($\theta, \alpha_{\IN}, \beta_{\IN}, \alpha_{\OFF}, \beta_{\OFF}$) in Eq\eqref{eq:mll}.
	\item {\bf \cop at group level}: given each user's $\theta$ and $\alpha_{\IN}$ from the previous step, convert them into ratio $\Ratio$, log-median $\Mean$ and then use \cop presented in Eq\eqref{eq:cop} to estimate Copula parameter $\eta$ for the two-dimensional heavy-tail distribution.
	\item {\bf Anomaly detection}: given a user's $\Ratio$ and $\Mean$, calculate its likelihood by using \cop.
\eit

\begin{figure*}[htbp]
\centering
\subfigure[Scatter between $\Ratio$ and $\Mean$]{
    \includegraphics[width=0.30\textwidth]{FIG/ratio_med_cov_noted.pdf}
}
\subfigure[Rank-weirdness plot]{
    \includegraphics[width=0.30\textwidth]{FIG/rank_ll.pdf}
}
\subfigure[Abnormally-active user]{
    \includegraphics[width=0.29\textwidth]{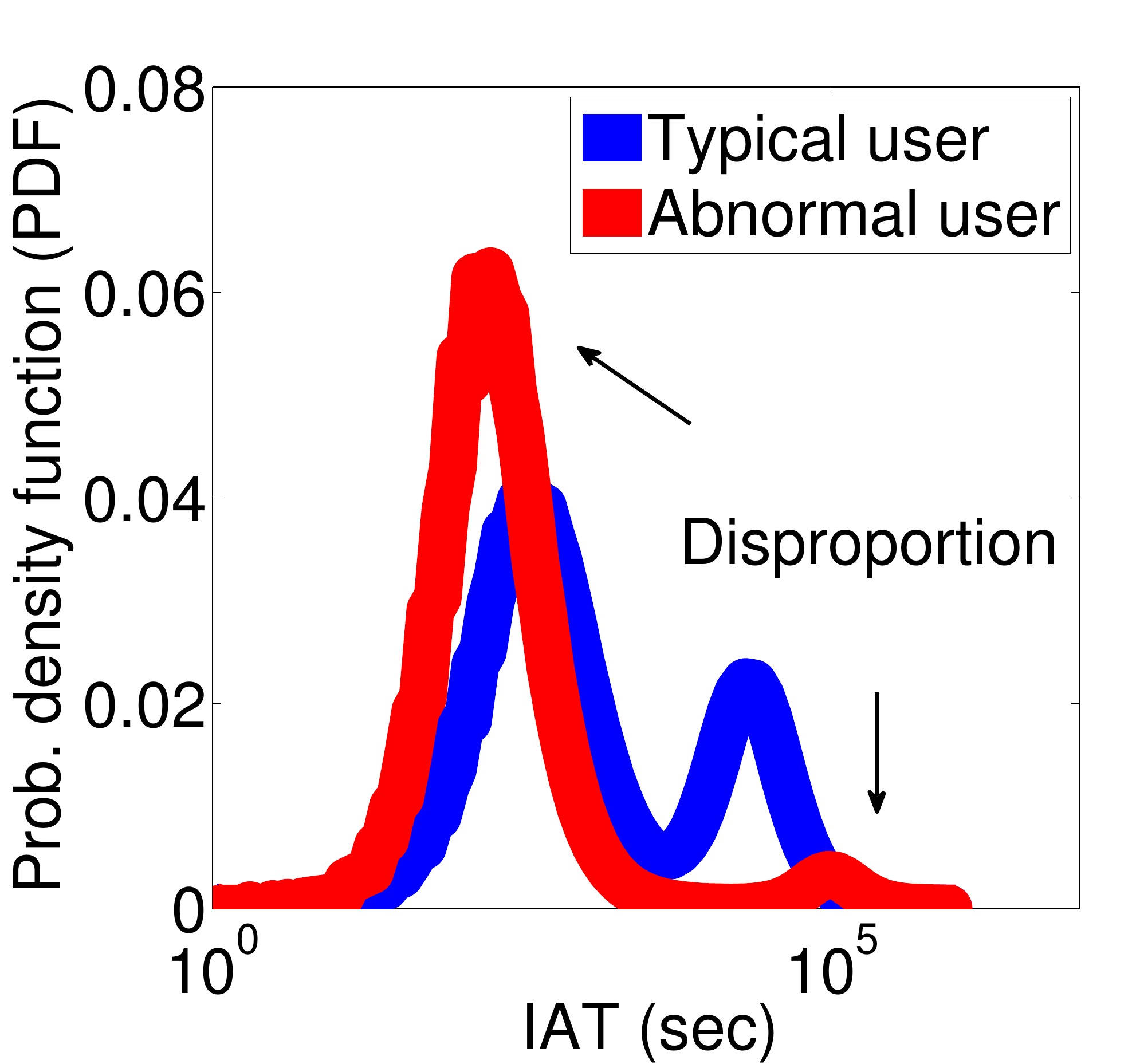}
}
\caption{\textit{\mmm detects anomaly}. In (a), each dot represents a user characterized by $\Ratio$ and $\Mean$ extracted from the \mll distribution. The anomalies spotted in (a) correspond to the few users (marked in red) with the lowest likelihoods in (b). Notice that, compared to the anomalies, the simulated samples with the corresponding ranks have much higher likelihoods (by two orders of magnitude). (c) illustrates the marginal PDF of IAT from an abnormal user detected by \mmm. Notice the disproportion between in-session and take-off: about 30 queries per session, whereas typical users have 3 queries per session.} 
\label{fig:mmm_detects_anomaly}
\end{figure*}

Figure \ref{fig:mmm_detects_anomaly} presents the anomalies detected by \mmm. Figure~\ref{fig:mmm_detects_anomaly}(b) provides ``rank-weirdness'' plot: users are presented in a ``least likely first'' order, by using the likelihood of observing their $\Ratio$ and $\Mean$ calculated by \cop. All users fit on a line, except the first seven users who have tiny likelihoods. As a comparison, the green line shows a synthetic set of users by using Eq\eqref{eq:copula}. Notice that none of the ``green'' users exhibits such tiny likelihoods; further notice that those seven users indeed correspond to outliers in ($\Ratio$, $\Mean$) space, where we enclose them in a red box and two red ellipses for visual clarity in Figure~\ref{fig:mmm_detects_anomaly}(a).

Figure \ref{fig:mmm_detects_anomaly}(c) further illustrates an abnormally-active user detected by \mmm. Notice the disproportion between in-session and take-off (the ratio $\Ratio \approx 30$), which is ten times higher compared to a typical user's (around 3).

%% file: Related_Work.tex
\begin{table*}[tb]
	\centering
	\caption{Metrics of temporal data-mining approaches: \mmm possesses all desired properties}
	\begin{tabular}{P{0.16\textwidth}||P{0.12\textwidth} P{0.12\textwidth} P{0.12\textwidth} P{0.12\textwidth} P{0.08\textwidth}}	
	\hline 
	\textbf{Metrics} & Meiss et al. \cite{meiss2005lack} & M{\"u}nz et al. \cite{munz2007traffic} & Vaz de Melo et al. \cite{vaz2013self} & Liu et al. \cite{liu2010understanding} &\textbf{\mmm} \\
	\hline\hline
	Heavy tail & $\surd$ & & $\surd$ & $\surd$ & $\surd$ \\ \hline
	Bi-modal & & $\surd$ & & & $\surd$ \\ \hline
	IAT modeling & & & $\surd$ & & $\surd$ \\ \hline
	User-level \& group-level modeling & & & & & $\surd$ \\ \hline
	Fits multiple datasets & $\surd$ & $\surd$ & $\surd$ & $\surd$ & $\surd$ \\ \hline
	Anomaly detection & $\surd$ & $\surd$ & $\surd$ & $\surd$ & $\surd$ \\ \hline
	Generative & $\surd$ & & $\surd$ & & $\surd$ \\ \hline
	Interpretable & $\surd$ & $\surd$ & $\surd$ & $\surd$ & $\surd$ \\ \hline
	\end{tabular}	
	\label{tbl:approach_comparison}
\end{table*}

Many prior papers have attempted to model the temporal, Internet-based activities of humans:
\bit
\item \textbf{Internet-based, temporal data.} Vaz de Melo et al. \cite{vaz2013self,de2010surprising} have proposed a self-feeding process to generate IAT following \LL distributions for modeling the Internet-based communications of humans. Becchetti et al. \cite{becchetti2008link} and Castillo et al. \cite{castillo2008query} have proposed novel graph-based algorithms for Web spam detection. Meiss et al. \cite{meiss2005lack} have demonstrated that client-server connections and traffic flows exhibit heavy-tailed probability distributions lacking any typical scale. M{\"u}nz et al. \cite{munz2007traffic} have presented a flow-based anomaly detection scheme based on the K-mean clustering. Gupta et al. \cite{gupta2014outlier} provides a comprehensive survey on outlier detection for temporal data. Veca et al. \cite{vaca2014time} have proposed a time-based collective factorization for monitoring news. Xing et al. \cite{DBLP:conf/sdm/XingPYW11} have proposed to use local shapelets for early classification on time-series data. Ratanamahatana et al. \cite{DBLP:reference/dmkdh/RatanamahatanaLGKVD10} gives a high-level survey of time-series data mining tasks, with an emphasis on time series representations. Furthermore, point processes, time series and inter-arrival time analysis have attracted huge interests, with multiple textbooks (Keogh et al. \cite{DBLP:journals/kais/CamerraSPRK14}).

\item \textbf{Human activities.} Shie et al. \cite{shie2013mining} has proposed a new algorithm (IM-Span) for mining user behavior patterns in mobile commerce environments. Saveski et al. \cite{saveski2011web} has adapted active learning to model the web services. Barabasi \cite{barabasi2005origin} models and explains human dynamics with heavy-tail distributions. Liu et al. \cite{liu2010understanding} have provided a Weibull analysis of Web dwell time, to discover human browsing behaviors. Sarma et al. \cite{das2014commerce} provides a fine tutorial on personalized search.
\eit  
Table~\ref{tbl:approach_comparison} summarizes the comparison among several popular methods. As Table~\ref{tbl:approach_comparison} shows, this is the only work focusing on the surprising pattern of web query IAT: in-session and take-off, and proposing a new framework \mmm to (a) match and explain this pattern, and (b) detect anomaly. To the best of our knowledge, this is the first work to use log-logistic distributions and the Copulas (as a metamodel) to describe the IAT of web queries.

%% file: Conclusion.tex
In this paper, we answer the motivational questions mentioned in the Introduction: `Alice' is submitting one web search per five minutes, for three hours in a row$-$is it normal? How to detect abnormal search behaviors, among Alice and other users? Is there any distinct pattern in Alice's (or other users') search behavior?

We conclude this paper by bringing the answers to these questions:
\bit
    \item{\textbf{A1: \pone.} One key observation of IAT is provided: a bi-modal distribution with the interpretation of in-session and take-off behaviors.
    \item{\textbf{A2: \ptwo.} Specifically, we propose:
	\bit
		\item{``\mll'' to parametrically characterize Alice's (or any person's) IAT by mixturing two log-logistic distributions.}
		\item{``\cop'' to describe the joint probability of two parameters of \mll by using Gumbel Copula.}
	\eit
	}	
	\item{\textbf{A3: \pthree.} \mll generates IAT with the same statistical properties as in the real data, and \cop can detect abnormal users by examining their search behaviors.
	}
	}
\eit
Finally, we provide a practitioners' guide for \mmm, and illustrate its power via ``rank-weirdness'' plot as in Figure~\ref{fig:mmm_detects_anomaly}(b). \mmm exactly pin-points the outliers that a human would spot: the points in red circles/boxes, in Figure~\ref{fig:mmm_detects_anomaly}(a).

%% file: appendix.tex
\subsection*{Kolmogorov-Smirnov (K-S) test}
Kolmogorov-Smirnov test (K-S test) is a non-parametric statistical test for testing the equality of two probability distributions. The null hypothesis assumes the samples are drawn from the given continuous distribution. Mathematically, the Kolmogorov-Smirnov test statistic is defined as:
\begin{equation}
D_n = \sup_{x} |F_n(x) - F(x)|
\nonumber
\end{equation}
\noindent where $F_n(x)$ is the empirical distribution estimated from the sample population, and $F(x)$ is the cumulative distribution function (CDF) of the given probability distribution. Under the null hypothesis, $\sqrt{n} D_n$ converges to the Kolmogorov distribution. Hence, the risk region of Kolmogorov-Smirnov test is $\sqrt{n} D_n > K_{\alpha}$, where $K_{\alpha}$ satisfies that $P(K > K_{\alpha}) = 1-\alpha$, $K$ follows Kolmogorov distribution. 

\subsection*{Bayesian information criterion (BIC)}
Bayesian information criterion(BIC) is a criterion for model selection. In model selection, the criterion
purely based on log-likelihood is likely leading to over-fitting. BIC is a penalized version of log-likelihood.
Mathematically, 
\begin{equation}
BIC = -2 L + k \ln(n)
\nonumber
\end{equation}
\noindent where $L$ is log-likelihood, k is the number of parameters, and n is number of observations.
Hence, minimizing BIC tends to select model with less parameters (parsimony). 

\subsection*{Kendall tau in Gumbel copula}
Kendall tau rank correlation $\Ken$ measures the dependency between two random variables. 
Given random variables $X$, $Y$ and $n$ pairs of their observations, $(x_1, y_1), \dots, (x_n, y_n)$, 
a pair of observations $(x_i, y_i)$ and $(x_j, y_j)$ is called concordant if $(x_i - x_j)(y_i - y_j) > 0$.
Likewise, the pair is called discordant if $(x_i - x_j)(y_i - y_j) < 0$. Hence, $\Ken$ is defined as:
\begin{equation}
\Ken = \frac{(\text{\# of concordant pairs}) - (\text{\# of discordant pairs})}{\frac{1}{2} n (n-1) }
\nonumber
\end{equation}
\noindent Note that $\Ken$ must be in $[-1,1]$. In particular, if $Y$ is rigorously increasing monotone with respect to $X$, $\Ken = 1$, whereas if $Y$ is rigorously decreasing monotone with respect to $X$, then $\Ken = -1$. 
